\pdfoutput=1

\documentclass[manuscript,screen,nonacm]{acmart}
\usepackage{graphicx}
\usepackage{amsmath,amssymb}
\usepackage{booktabs}
\usepackage{url}
\usepackage{multirow}
\usepackage{multicol}
\usepackage{algorithm}
\usepackage{algpseudocode}

\renewcommand\footnotetextcopyrightpermission[1]{}  % no permission footnote
\setcopyright{none}

\graphicspath{{./}{Figures/}{figures/}}

\emergencystretch=\maxdimen
\begin{document}
\raggedbottom

\title{Asking Questions the Right Way: A Multi-Agent Conversational System for Prompt Formulation in Complex Task Resolution}

\author{B. Sankar}
\email{sankarb@iisc.ac.in}
\orcid{0000-0001-5844-6273}
\affiliation{%
  \institution{Department of Mechanical Engineering, Indian Institute of Science (IISc)}
  \city{Bangalore}
  \country{India}
}

\author{Pawni Yadav}
\email{pawniyadav3435@gmail.com}
\affiliation{%
  \institution{Department of Design and Manufacturing, Indian Institute of Science (IISc)}
  \city{Bangalore}
  \country{India}
}

\author{Srinidhi Ranjini Girish}
\email{srinidhirgirish@gmail.com}
\affiliation{%
  \institution{Department of Design and Manufacturing, Indian Institute of Science (IISc)}
  \city{Bangalore}
  \country{India}
}

\author{Amogh A S}
\email{amogh.setty07@gmail.com}
\affiliation{%
  \institution{Department of Design and Manufacturing, Indian Institute of Science (IISc)}
  \city{Bangalore}
  \country{India}
}

\renewcommand{\shortauthors}{B. Sankar et al.}

\begin{abstract}
Large language models (LLMs) have become integral to complex intellectual tasks, yet their output quality remains fundamentally constrained by the quality of user-provided prompts. Users typically resort to iterative, multi-turn prompting that suffers from context degradation, error propagation, and diminishing returns on cognitive investment. This paper presents PAWNI (Prompt Architecture Wizard using Neural Intelligence), an agentic conversational interface comprising eight agents that transforms unstructured user queries into comprehensive, structured prompts through guided question-and-answer dialogue informed by a self-evolving domain knowledge base of expert-derived patterns. Rather than optimising the model's response, PAWNI optimises the question itself by front-loading intent clarification so that a single, well-formed prompt replaces multiple corrective iterations. We also propose a three-tier framework of 18 prompt elements organised into Essential, Enhancement, and Elevation categories, derived from a systematic review of existing prompting guidelines. The system's primary contribution is architectural; to probe whether it behaves as intended, and to validate a measurement protocol for future work, we conducted an exploratory within-subjects study ($N=4$) across four domain-diverse complex tasks, combining 32-channel EEG, NASA-TLX workload assessment, and behavioural interaction metrics. All four participants produced more structurally complete prompts with PAWNI (42\% to 91\% of assessed elements), rated the resulting LLM outputs higher on every quality dimension, and reported lower subjective workload (39.6 to 21.7 on NASA-TLX). Every participant reached a satisfactory output in a single LLM turn, against one to twelve turns unaided. Effect sizes are large but, at this sample size, unstable and not generalisable; we report them with small-sample corrections and bootstrap intervals, and treat direction consistency as the primary evidence. These results are exploratory. They motivate, rather than establish, the hypothesis that asking the right questions---not merely generating better answers---is the more effective lever for improving human-AI collaboration on complex tasks.
\end{abstract}



\maketitle

%% ---------------------------------------------------------------------------
%% MAIN BODY
%%
%% arXiv always compiles from the root of the uploaded archive, so this path
%% is relative to the root. Keep 1_PAWNI_main_body.tex at the top level of the
%% tarball, alongside this file.
%% ---------------------------------------------------------------------------
% ============================================================

% BATCH 1: §1 Introduction + §2 Related Work
% Paper: Asking Questions the Right Way
% Target: CUI 2026 (ACM Conference)
% ============================================================

\section{Introduction}
\label{sec:introduction}

Large Language Models (LLMs) have become deeply embedded in everyday workflows, from drafting professional documents to generating full-scale software applications~\cite{brown2020language, ouyang2022training, zhao2023survey}. Recent traffic statistics indicate that conversational AI platforms such as ChatGPT surpass several hundred million monthly active users, and an increasing number of individuals now prefer these systems over traditional search engines for personalised, multi-step tasks~\cite{schulhoff2024prompt, sahoo2024systematic}. The appeal is straightforward: human-like interaction, rapid information retrieval, and the capacity to generate contextualised responses in natural language.

The primary interface through which users communicate with LLMs is the \textit{prompt}---a natural language instruction that specifies the task, provides context, and constrains the expected output. Prompt engineering, the practice of designing and refining such instructions, has emerged as a critical determinant of output quality~\cite{sahoo2024systematic, chen2024taxonomy}. A well-structured prompt that clearly articulates the user's intent, domain context, and desired output format elicits substantially better responses than a vague or underspecified one~\cite{sinha2025effects, giray2025knowledge}. In essence, to get the right answers from an LLM, one must ask the right questions.

The problem, however, is that users frequently fail to do so. The human mind is inherently unstructured in how it articulates requirements, particularly for complex, multi-faceted tasks such as designing a software system, planning a curriculum, or drafting a clinical research protocol. Users tend to provide short, context-deficient prompts and then rely on multi-turn conversations to iteratively refine the output~\cite{li2025llms, bai2025hidden}. This trial-and-error approach suffers from well-documented limitations: LLMs exhibit a 39\% accuracy drop in multi-turn conversations compared to exhaustive single-shot prompts~\cite{li2025llms}; context rot causes models to lose critical nuances as conversations lengthen; and error propagation means that incorrect assumptions from early turns compound in subsequent responses~\cite{wang2025cognitive}. Recent evidence suggests that prolonged reliance on LLM-generated content can also induce cognitive offloading and reduced engagement~\cite{kosmyna2025brain, georgiou2025chatgpt}.

Existing prompting frameworks---such as Google's five-step approach, OpenAI's best practices, and Microsoft's Prompt-Evaluate-Optimise cycle---offer useful heuristics but remain generic checklists that are neither model-aware nor task-adaptive~\cite{schulhoff2024prompt}. They do not account for task-specific best practices, do not adapt to individual user contexts, and critically, they place the entire burden of prompt construction on the user. More recent work on proactive AI systems has explored models that ask clarifying questions rather than generating immediate responses~\cite{deng2025reasoning, li2023thinking, hu2025autonomous}. However, these systems treat question-asking as an isolated capability rather than integrating it within an end-to-end pipeline that also provides domain knowledge, evaluates prompt quality, and ensures structural completeness.

This paper presents PAWNI (\textbf{P}rompt \textbf{A}rchitecture \textbf{W}izard with \textbf{N}eural \textbf{I}ntelligence), an eight-agent conversational AI system that transforms unstructured user queries into comprehensive, structured prompts for complex task resolution. PAWNI operates on a central thesis: \textit{asking the right questions leads to better answers}. Rather than optimising the LLM's response generation, PAWNI optimises the question itself---front-loading the cognitive effort into structured intent clarification through an AI-guided question-and-answer process. The system introduces three key innovations: (i) a three-tier prompt element framework comprising 18 elements organised into Essential, Enhancement, and Elevation categories, derived through a systematic analysis of existing prompting guidelines; (ii) a self-evolving pattern file system that accumulates and weights domain-specific best practices through semantic comparison; and (iii) a reverse-reasoning evaluation loop where the system generates an ideal response from its own prompt draft, evaluates it against the original requirements using multiple LLM judges, and iteratively refines the prompt based on the feedback.

We evaluate PAWNI through a within-subjects pilot study ($N=4$) comparing naive LLM interaction against PAWNI-assisted interaction across four domain-diverse complex tasks. The evaluation employs a convergent multi-method approach combining 32-channel EEG-based neurophysiological measurement of cognitive load, NASA-TLX subjective workload assessment, system usability scales, and behavioural metrics including conversation turns, time-on-task, and prompt structural analysis. This paper addresses three primary and two secondary research questions:

\begin{itemize}
    \item \textbf{RQ1 (Output Quality):} Does using PAWNI-generated structured prompts result in LLM outputs that are more complete and comprehensive compared to outputs from iterative naive prompting?
    \item \textbf{RQ2 (Cognitive Load):} How does the temporal distribution and magnitude of cognitive load differ between naive LLM interaction and PAWNI-assisted prompt formulation?
    \item \textbf{RQ3 (Interaction Efficiency):} Does PAWNI-assisted prompt formulation reduce the number of conversational turns and iterative corrections required to achieve task completion?
    \item \textbf{RQ4 (Prompt Quality):} How do the structural and linguistic properties of PAWNI-generated prompts differ from user-authored prompts? \textit{(Secondary)}
    \item \textbf{RQ5 (Cognitive Topology):} Do the topographic distributions of cortical activation during the two conditions differ in a manner consistent with established neurocognitive models? \textit{(Secondary)}
\end{itemize}

The remainder of this paper is organised as follows. Section~\ref{sec:related} reviews related work on prompt engineering techniques, multi-turn conversation limitations, and proactive AI systems. Section~\ref{sec:contributions} presents our theoretical contributions: the three-tier prompt element framework and task complexity taxonomy. Section~\ref{sec:system} describes the PAWNI system architecture and its eight agents. Section~\ref{sec:methodology} details the experimental methodology, including the EEG-based cognitive load measurement protocol. Section~\ref{sec:results} presents the pilot study results across all research questions. Section~\ref{sec:discussion} discusses implications, and Section~\ref{sec:conclusion} concludes with future directions.

% ============================================================
\section{Related Work}
\label{sec:related}

% ------------------------------------------------------------
\subsection{Prompt Engineering Techniques and Frameworks}
\label{subsec:rw_techniques}

The development of prompting techniques has progressed substantially since the introduction of in-context learning~\cite{chen2024taxonomy, liu2023pretrain}. Zero-shot prompting relies entirely on the model's pre-trained knowledge and instructions~\cite{kojima2022large}, whereas few-shot prompting provides a limited set of input-output exemplars to guide pattern recognition~\cite{sahoo2024systematic, vatsal2024survey}. Chain-of-Thought (CoT) prompting~\cite{wei2022chain} marked a significant advance by encouraging models to generate intermediate reasoning steps, thereby improving performance on multi-step arithmetic and commonsense tasks. Self-consistency decoding can further strengthen CoT reasoning~\cite{wang2023selfconsistency}, and Least-to-Most prompting decomposes complex queries into a sequence of simpler subproblems~\cite{zhou2023least}. CoT was further generalised by Tree-of-Thought (ToT)~\cite{yao2023tree}, which explores multiple reasoning paths using tree-based search, and Graph-of-Thought~\cite{besta2024graph}, which allows non-linear, interconnected reasoning structures; recent work has also accelerated graph-based LLM reasoning over knowledge graphs~\cite{jiang2025fast}. The ReAct framework~\cite{yao2022react} interleaves reasoning with external tool use, enabling models to query databases and search engines mid-response.

Frameworks for organising prompts have also emerged. The Prompt Report~\cite{schulhoff2024prompt} catalogued 58 distinct prompting techniques and proposed a structured vocabulary. White et al.~\cite{white2023prompt} structured prompting techniques into a pattern catalogue analogous to software design patterns, including the Persona, Reflection, and Recipe patterns. Markup-based approaches such as Microsoft's Prompt Orchestration Markup Language (POML) decouple content from instructions using XML-like tags. Despite this proliferation, a critical limitation persists: all these techniques assume the user already knows what to ask. They operate within a \textit{blind-thinking} paradigm~\cite{deng2025reasoning} where the model reasons extensively using its own internal knowledge, even when the input prompt is ambiguous or critically incomplete.

% ------------------------------------------------------------
\subsection{Prompt Structure and Elements}
\label{subsec:rw_elements}

Research on the architecture of effective prompts has identified several recurring components. A well-structured prompt typically comprises a Directive that defines the core task, a Persona that sets the model's perspective, Examples that demonstrate the expected pattern, additional Context, and Output Format specifications~\cite{giray2024systematic, chen2024taxonomy}. Task-specific constraints such as domain, intent, and precise requirements are essential in specialised applications~\cite{giray2025knowledge}. Empirical studies indicate that prompt specificity---achieved through technical keywords and conditional phrasing---and contextual richness are key determinants of output quality~\cite{giray2025knowledge}. Furthermore, prompt length itself influences performance: shorter, vague prompts produce simplified responses, whilst longer prompts with extensive domain context improve model performance on domain-specific tasks~\cite{sinha2025effects}.

Conversely, ineffective prompting is characterised by knowledge gaps: missing context, missing specifications, multiple intents in the same thread, and unclear instructions~\cite{giray2025knowledge}. Research on neural transparency has shown that prompts can activate latent behavioural trait vectors within the model, altering the response according to the user's framing~\cite{chen2025neural}. The ordering of elements within a prompt also matters---providing examples and background information before the directive ensures that the model processes context before instructions, leading to better responses~\cite{giray2024systematic}. Despite this growing understanding of what constitutes an effective prompt, no existing system translates these findings into an automated pipeline that ensures all necessary elements are present for a given task.

% ------------------------------------------------------------
\subsection{Multi-Turn Conversations and Their Limitations}
\label{subsec:rw_multiturn}

Multi-turn conversations are the default interaction mode for commercial LLMs, yet they introduce fundamental reliability problems. Laban et al.~\cite{li2025llms} demonstrated that LLMs suffer measurable performance degradation across turns, with models making assumptions about vague initial prompts, propagating errors by treating incorrect previous outputs as ground truth, and failing to self-correct. This degradation is compounded by positional sensitivity: models exhibit reduced accuracy when critical information is scattered within a larger context of distractors~\cite{bai2025hidden}. Adapala~\cite{wang2025cognitive} adapted principles from Human Cognitive Load Theory to LLMs, identifying \textit{context saturation}---the overload of task-irrelevant information---and \textit{attentional residue} from constant task switching as key factors in hallucination and reasoning degradation.

These findings have a direct implication for complex task resolution. In human-to-human conversations, context and nuances are naturally retained, corrections to earlier statements propagate through subsequent exchanges, and an implicit shared understanding grounds the interaction. In human-to-LLM conversations, none of these properties hold reliably. Small ambiguities compound over turns, and there is no guarantee that the model will maintain consistency with its own prior statements.

This observation motivates the core design principle of PAWNI: for complex tasks, it is preferable to invest effort in constructing one comprehensive, well-structured prompt rather than engaging in iterative multi-turn refinement. This parallels how human teams approach complex projects in practice. Before the AI era, accomplishing a complex task began with a planning phase where team members collectively defined goals, deliverables, and constraints. Domain expertise was acquired, conflicts in approach were resolved, and a detailed project scope document was drafted, reviewed, and refined before execution commenced. This front-loaded planning process minimized trial-and-error during execution. PAWNI translates this human-collaborative model into a human-AI collaboration system, using specialized agents to emulate each stage of the collaborative planning process.

% ------------------------------------------------------------
\subsection{Question-Asking and Proactive AI Systems}
\label{subsec:rw_proactive}

A nascent but growing body of work explores AI systems that ask rather than answer. Chen et al.~\cite{deng2025reasoning} introduced the paradigm of Proactive Interactive Reasoning, where models identify ambiguous or missing information and ask clarifying questions before generating a response. Their work reports improved accuracy and reduced reasoning overhead on under-specified inputs across mathematical reasoning, code generation and document editing. Li et al.~\cite{li2023thinking} developed Thinking Assistants---LLM-based conversational agents that ask domain-based clarifying questions instead of providing direct answers, fostering deeper user understanding of the task at hand. Su~\cite{hu2025autonomous} proposed Autonomous Question Formation systems that prioritize question generation over task execution in the decision process. Advanced multi-agent architectures using questioner-judge models have been explored for generating and evaluating critical questions~\cite{garcia2025ellis}. In the multi-agent domain, recent surveys~\cite{guo2024multiagent, tran2025multiagent, wang2024survey, chen2024llmmas, plaat2025agentic, yue2025survey} document the rapid growth of LLM-based multi-agent systems for complex problem-solving (e.g., Generative Agents~\cite{park2023generative} and MetaGPT~\cite{hong2024metagpt}), though these have primarily focused on task execution rather than prompt formulation.

The Universal Self-Consistency framework~\cite{chen2024usc} demonstrated that self-evaluation across multiple reasoning paths reduces errors in free-form generation. Self-Refine~\cite{madaan2023selfrefine} introduced iterative self-feedback loops for output improvement. Robust conversational question answering through reinforced reformulation has been explored in dialogue systems~\cite{wu2023robust}. However, none of these systems combine dynamic knowledge acquisition, structured question-asking, multi-model evaluation, and iterative feedback into a single pipeline specifically designed for prompt formulation.

PAWNI addresses this gap by providing an end-to-end system that: (a) classifies task complexity and acquires domain-specific knowledge dynamically, (b) conducts structured question-and-answer sessions to elicit comprehensive user requirements, (c) generates and evaluates prompt drafts through a reverse-reasoning workflow with multi-model judges, and (d) outputs a structurally complete prompt containing all 18 identified prompt elements. To the best of our knowledge, this is the first system that integrates agentic question-asking with a self-evolving knowledge base and multi-model prompt evaluation for the specific purpose of prompt optimisation.

% ------------------------------------------------------------
\subsection{Automated Prompt Optimisation and Built-in Query Rewriting}
\label{subsec:rw_autoprompt}

A parallel line of work automates prompt construction without involving the user. Automatic Prompt Engineering (APE) treats instruction generation as black-box search, using an LLM to propose candidate instructions and selecting those that maximise held-out task accuracy~\cite{zhou2023ape}. OPRO extends this by placing previous candidates and their scores in context, allowing the optimiser to refine proposals across iterations~\cite{yang2024optimizers}. Prompt optimisation with textual ``gradients'' applies a descent-like procedure over natural-language feedback~\cite{pryzant2023apo, yuksekgonul2024textgrad}, and DSPy compiles declarative pipeline specifications into optimised instructions and demonstrations against a programmatic metric~\cite{khattab2024dspy}. In deployed systems, comparable transformations occur invisibly: queries are expanded before retrieval, system prompts are auto-tuned, and vendor tooling increasingly offers one-click prompt improvement.

These methods are effective and, for their intended setting, largely solve the problem of instruction phrasing. They share a precondition, however, that does not hold for the scenario addressed in this paper: they require a task distribution and an automatically computable success metric. A single user with a single novel complex task supplies neither. More fundamentally, every method in this family operates on the information already present in the user's input. They can reformulate, expand, decompose and reorder; they cannot supply a constraint the user never stated. Where the deficiency is a \emph{knowledge gap} rather than a phrasing problem---missing context, missing specification, unstated deliverable~\cite{giray2025knowledge}---optimisation over phrasings cannot close it, and a system that attempts to close it by inference is hallucinating requirements.

The alternative is to ask. Clarifying-question generation is well established in conversational information seeking~\cite{zamani2020clarifying}, and recent work trains language models to request missing information before acting, whether through simulated future turns~\cite{andukuri2024stargate}, unclear-instruction detection~\cite{wang2024learningtoask}, proactive planning agents~\cite{zhang2024askbeforeplan}, information-gain objectives~\cite{mazzaccara2024informative}, or domain-specific question quality~\cite{li2025clinicalquestions}. This is the tradition in which PAWNI sits. What distinguishes it is that asking is embedded in a full pipeline: questions are grounded in retrieved and weighted task-specific knowledge rather than generated ad hoc, they are enumerated against an explicit completeness target, and the resulting prompt is independently evaluated before delivery. Automated optimisation and elicitation are complementary---the former improves how a specification is expressed, the latter improves what the specification contains---and Section~\ref{subsec:positioning} returns to how they might be combined.

% ============================================================
% BATCH 2: §3 Contributions + §4 System Design
% Paper: Asking Questions the Right Way
% Target: CUI 2026 (ACM Conference)
% ============================================================

\section{Contributions: Prompt Elements and Task Taxonomy}
\label{sec:contributions}

% ------------------------------------------------------------
\subsection{Redefining Prompts: From Questions to Structured Directives}
\label{subsec:redefining}

A prompt, as defined in the context of generative AI, is the input instruction given to an LLM to guide it in producing a specific, desired output. It functions as a set of directions---much like a stage prompt that cues an actor to deliver a performance aligned with the script and audience expectations. The clarity, specificity, and structure of the prompt directly affect the quality and usefulness of the model's output.

A question, by contrast, is a simple interrogative statement that seeks information on a specific topic. A question is therefore a \textit{subset} of a prompt: while every question can serve as a prompt, a prompt entails far more than a question. It includes specifications, directions, constraints, persona assignments, and output format requirements that collectively guide an LLM toward a detailed, actionable response.

This distinction matters because users, trained from childhood to ask direct questions of teachers and mentors, tend to carry the same habit into LLM interactions. A generic question addressed to a teacher who knows the student's background yields a reasonable answer; the same generic question addressed to a model that lacks any such context yields a generic, often inadequate response. PAWNI bridges this gap by providing an autonomous pipeline to convert human questions into model-optimised prompts.

Common deficiencies in user-authored prompts include: \textit{missing context}---the absence of essential details such as goals, prior attempts, or project information; \textit{multiple intents}---the introduction of several issues without clear separation; \textit{unclear instructions}---vague or ambiguous directives open to multiple interpretations; and \textit{missing specifications}---the absence of critical technical requirements~\cite{giray2025knowledge}. To address these systematically, we developed a comprehensive set of prompt elements through an empirical methodology described below.

% ------------------------------------------------------------
\subsection{A Three-Tier Prompt Element Framework}
\label{subsec:tiers}

We derived our prompt element framework using a PRISM-based methodology: we systematically reviewed existing prompting guidelines from leading technology organisations (Google, OpenAI, Microsoft, Anthropic) and research literature~\cite{schulhoff2024prompt, sahoo2024systematic, chen2024taxonomy, giray2024systematic}, identified their structural limitations and recurring components, and consolidated the essential prompt constituents into a unified superset. The resulting 18 elements were then organised into three hierarchical tiers based on their functional role in prompt construction.

\textbf{Tier~1: Essential Core Attributes.} These eight elements define the structural foundation of a prompt and directly determine output alignment. Their absence leads to ambiguity and misaligned responses.

\textbf{Tier~2: Enhancement Elements.} These seven elements improve precision, output quality, and audience alignment. They refine \textit{how} the task is executed and optimise the response for specific contexts.

\textbf{Tier~3: Elevation (Control) Elements.} These three elements govern evaluation, safety, and model behaviour. They are optional and highly technical, but they ensure reliability and consistency.

Table~\ref{tab:prompt_elements} presents the complete framework. During the PAWNI Architect agent's question-and-answer session (Section~\ref{subsec:architect}), missing Essential elements must be mandatorily clarified by the user, whereas the system makes reasonable assumptions for Enhancement and Elevation elements when the user does not specify them.

\begin{table*}[!htbp]
\centering
\caption{Three-tier prompt element framework. Elements are grouped by functional tier. Essential elements are mandatory; Enhancement and Elevation elements are recommended but can be inferred.}
\label{tab:prompt_elements}
\small
\begin{tabular}{p{0.8cm} p{3.2cm} p{9.5cm}}
\toprule
\textbf{Tier} & \textbf{Element} & \textbf{Description} \\
\midrule
\multirow{8}{*}{\rotatebox{90}{\textbf{Essential (8)}}}
& Role / Persona & Professional expertise, identity assignment, knowledge boundaries, user expertise level \\
& Context & Background information, situational setup, temporal context, prerequisites \\
& Goal / Objective & Overarching goal, what success looks like, desired end state \\
& Task / Instructions & Specific actions, task decomposition, sub-task sequencing, deliverable specifications \\
& Input Type / Format & Raw data specification, references, examples, input language and boundaries \\
& Output Format & Format (JSON, markdown, etc.), length, layout, section organisation, template structure \\
& Final Deliverable & The exact artefact the model must produce \\
& Constraints & Explicit limitations, negative prompting, scope boundaries, ethical and safety guidelines \\
\midrule
\multirow{7}{*}{\rotatebox{90}{\textbf{Enhancement (7)}}}
& Reference Data & External sources, citation requirements, authority references, documentation links \\
& Examples & Zero/one/few-shot demonstrations, input-output pairs, edge cases, counter-examples \\
& Style / Tone & Writing style, emotional register, formality level, language complexity \\
& Target Audience & Reader demographics, knowledge level, accessibility needs \\
& Reasoning Mode & Chain-of-thought, step-by-step, problem decomposition, self-verification requirements \\
& Formatting Rules & Header hierarchy, table structures, citation formats, visual elements, desired detail level \\
& Task-Specific Elements & Domain-dependent requirements unique to the task type \\
\midrule
\multirow{3}{*}{\rotatebox{90}{\textbf{Elev. (3)}}}
& Best Practices & Safety, error handling, delimiters, grammar checking, action verbs, labelling \\
& Confidence Level & Certainty thresholds, uncertainty handling, ambiguity acknowledgement \\
& Model Settings & Temperature, top-p, top-k, maximum new tokens \\
\bottomrule
\end{tabular}
\end{table*}

% ------------------------------------------------------------
\subsection{Task Complexity Classification}
\label{subsec:task_classification}

Not all user queries benefit from the full PAWNI pipeline. A factual lookup such as ``What is the capital of Germany?'' requires no structured prompt engineering, whereas designing a 16-week university curriculum demands substantial context, constraints, and formatting. We therefore classify tasks into two categories, summarised in Table~\ref{tab:task_complexity}.

\begin{table}[!htbp]
\centering
\caption{Simple vs.\ Complex task classification criteria.}
\label{tab:task_complexity}
\small
\begin{tabular}{p{2.8cm} p{2.2cm} p{2.5cm}}
\toprule
\textbf{Property} & \textbf{Simple} & \textbf{Complex} \\
\midrule
Answer type & Single factual & Multi-part, structured \\
Domain knowledge & General & Best practices needed \\
Creativity required & Minimal & High \\
Constraints & Few or none & Multiple \\
Iteration needed & No & Yes \\
Output length & Short ($<$200 words) & Long ($>$200 words) \\
Skill dependency & None & Domain-specific \\
\bottomrule
\end{tabular}
\end{table}

Simple tasks bypass the agentic pipeline entirely and are passed directly to the LLM. Complex tasks are further classified into a Master Task Taxonomy of 13 categories: Content Creation, Code Generation, Creative Writing, Strategic Planning, Education \& Curriculum, Research \& Analysis, Technical Documentation, Communication, Design \& UX, Data \& Analytics, Operations, Legal \& Compliance, and Multimedia. Each category contains 5--11 subtasks (approximately 90 subtasks in total). This taxonomy is self-evolving: if a prompt cannot be classified into any existing category, a new entry is created and, after repeated invocations, promoted to the official taxonomy.

% ============================================================
\section{System Design: PAWNI}
\label{sec:system}

% ------------------------------------------------------------
\subsection{Design Philosophy}
\label{subsec:philosophy}

PAWNI's design is grounded in an analogy to pre-AI human collaboration on complex projects. Before AI, accomplishing a complex task began with a planning phase: team members collectively defined end goals, expected outcomes, and deliverables. Domain expertise was then acquired---through hiring specialists or knowledge transfer sessions. Conflicts in approach were resolved through discussion. An initial project scope document was drafted, reviewed by stakeholders, and iteratively refined before execution. This front-loaded planning process ensured that execution involved minimal trial-and-error.

PAWNI translates each stage of this collaborative process into a specialised AI agent. The user states their requirements (initial input). A classification agent identifies the task and cleans the narrative. A knowledge agent acquires domain expertise. A questioner agent aligns direction and resolves ambiguities through structured inquiry. A drafting agent produces a preliminary output. An evaluation agent scrutinises it. A feedback agent translates the evaluation into actionable improvements. And a final agent prepares the comprehensive prompt document. This separation of concerns ensures loose coupling between agents, enabling independent utilisation and clear accountability for each transformation step.

% ------------------------------------------------------------
\subsection{Pipeline Architecture}
\label{subsec:architecture}

\begin{figure*}[!htbp]
\centering
\includegraphics[width=\columnwidth]{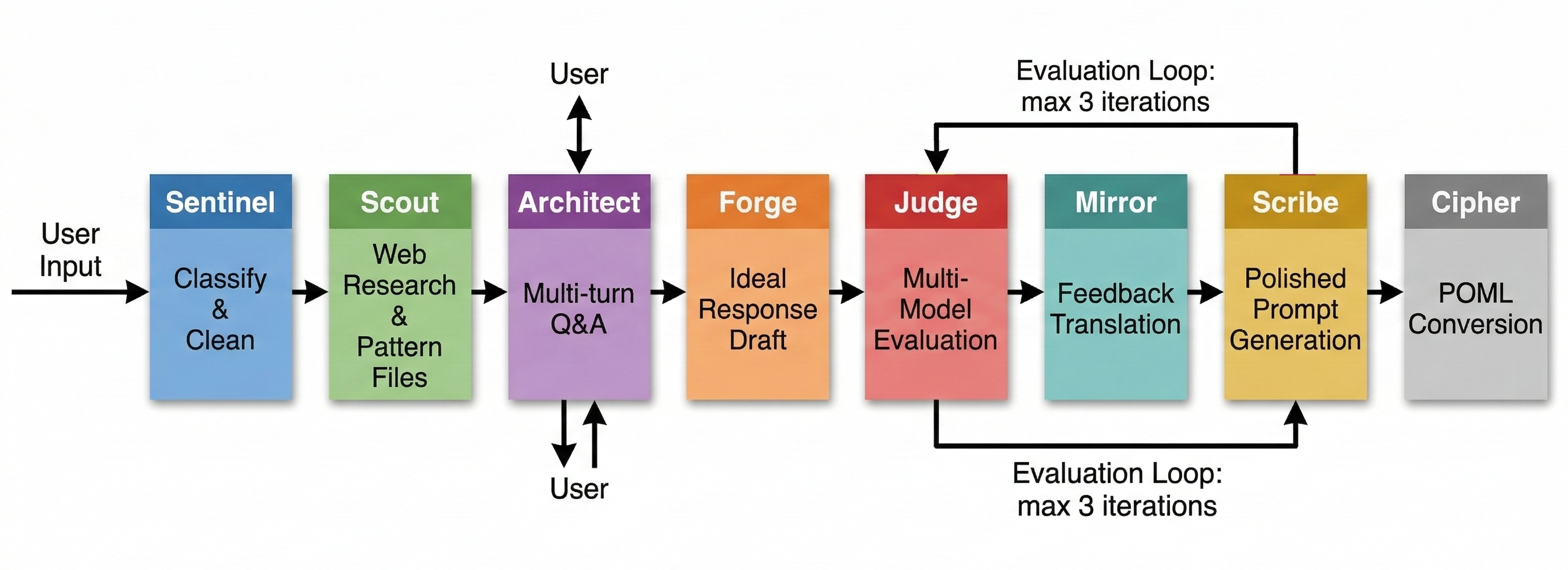}
\caption{PAWNI pipeline architecture. Eight agents process the user's raw input sequentially. The Architect agent conducts multi-turn Q\&A with the user. The Forge--Judge--Mirror--Scribe loop iteratively refines the prompt until a quality threshold is met.}
\Description{Block diagram of the PAWNI pipeline showing eight sequential agents from user input to final prompt. The Architect stage includes a multi-turn Q\&A with the user, followed by an iterative Forge--Judge--Mirror--Scribe loop that refines the prompt until a quality threshold is reached.}
\label{fig:architecture}
\end{figure*}

Figure~\ref{fig:architecture} illustrates the complete PAWNI pipeline. The system comprises eight sequentially orchestrated agents that transform an ambiguous, unstructured user prompt into a comprehensive structured prompt through multiple processing stages: dynamic task-based assumption clarification, detailed prompt generation with the defined prompt elements, self-evaluation via reverse reasoning, iterative feedback loops, and consistent structural formatting.

% ------------------------------------------------------------
\subsection{Agent Descriptions}
\label{subsec:agents}

\subsubsection{Sentinel Agent --- The Preprocessor}
\label{subsec:sentinel}

The Sentinel is the pipeline's entry point, responsible for two preprocessing tasks: prompt cleaning and intent classification. Prompt cleaning corrects grammatical and typographical errors, removes filler words, and reduces the input to its core intent. The classifier then operates at two levels. First, it determines whether the task is simple or complex using the criteria in Table~\ref{tab:task_complexity}. Simple tasks bypass the pipeline and receive only a cleaned, paraphrased version of the original prompt. For complex tasks, a second classification determines the task type and subtask from the Master Task Taxonomy. Compound tasks (e.g., ``build me a website and write its marketing copy'') are decomposed into atomic subtasks, and the user is asked which to address first.

Task matching against the existing taxonomy uses embedding-based semantic similarity via \texttt{pgvector}. If the cosine similarity exceeds 0.85, the existing canonical task entry is used automatically. For similarities between 0.70 and 0.85, a confirmation step verifies the match. Below 0.70, a new taxonomy entry is created.

\subsubsection{Scout Agent --- Dynamic Subject Matter Expert}
\label{subsec:scout}

The Scout acquires structured domain knowledge for the identified task. It makes two parallel calls to a web-search API: (i) a \textit{task-level} call that fetches generic best practices, dos and don'ts, and current trends for the task type (e.g., ``curriculum design''), and (ii) a \textit{prompt-specific} call that fetches knowledge relevant to the user's particular topic (e.g., ``quantum computing for engineering undergraduates''). Both calls return structured JSON with five categories: best practices, dos, don'ts, additional details, and current trends.

The task-level knowledge is stored in a persistent \textit{pattern file} in the database (Section~\ref{subsec:patternfiles}), whilst the prompt-specific knowledge is stored ephemerally for the current session. By providing both stable task-level patterns and contextually rich prompt-specific knowledge, the Scout enables the downstream Architect to ask informed, domain-relevant questions.

\subsubsection{Architect Agent --- The Questioner}
\label{subsec:architect}

The Architect is the QA specialist of the pipeline and embodies PAWNI's core contribution: asking the right questions. It receives the cleaned prompt, task classification, and both pattern file snapshots (task-level and prompt-specific). It first analyses the cleaned prompt against the 18-element framework (Table~\ref{tab:prompt_elements}) to identify missing critical elements. It then generates targeted multiple-choice questions---approximately nine per turn---informed by the pattern file knowledge and prioritised by the weight of the corresponding best practices.

High-weight best practices (those confirmed across multiple prior invocations) are prioritised in early questions to capture the user's preferences on the most established domain conventions first. The QA process proceeds iteratively until all required Essential elements have been clarified. The user may type \texttt{/done} at any point to signal that sufficient information has been provided. Upon completion, the Architect produces a structured paragraph---a dense, information-rich summary of all gathered requirements---which is passed to the Forge agent.

\subsubsection{Forge Agent --- Ideal Response Generator}
\label{subsec:forge}

The Forge implements the first phase of PAWNI's \textit{reverse-reasoning} workflow. Traditional prompt engineering follows a \texttt{User Input $\rightarrow$ Prompt Structuring $\rightarrow$ Output} pipeline. PAWNI instead follows:

\begin{center}
\texttt{User Input $\rightarrow$ Prompt Structuring $\rightarrow$ Ideal Response $\rightarrow$ Evaluation $\rightarrow$ Feedback $\rightarrow$ Re-Prompt Structuring $\rightarrow$ Output}
\end{center}

Rather than directly polishing the prompt, the Forge first drafts the best ideal response that could be generated from the Architect's structured paragraph. This response exposes the implicit assumptions the model makes and highlights gaps in task comprehension. The draft is then forwarded to the Judge for quality evaluation. This reverse approach---generating an answer first and then improving the question based on what the answer reveals---is central to PAWNI's quality assurance mechanism.

\subsubsection{Judge Agent --- Multi-Model Evaluator}
\label{subsec:judge}

The Judge evaluates the Forge's output against a defined rubric that assesses four criteria: (i) the presence of all 18 prompt elements, (ii) alignment with user requirements, (iii) clarity and absence of ambiguity, and (iv) structural completeness. The evaluation is performed independently by four different LLM models (GPT, Claude, GLM, and Grok in the current implementation). Each model returns a score on a 0--10 scale along with lists of positives, negatives, and recommendations. The results are aggregated: a mean score is computed, and the lists are deduplicated and merged. If the score falls below a configurable threshold (default: 8/10), the output proceeds to the Mirror for feedback incorporation.

\subsubsection{Mirror Agent --- Feedback Translator}
\label{subsec:mirror}

The Mirror translates the Judge's evaluation into improved instructions for the next iteration. It cross-references the evaluation feedback with the Architect's original structured paragraph to ensure that no original requirement is lost or distorted during refinement. All recommendations from the Judge are incorporated, and the Mirror explicitly prevents the introduction of new tasks or requirements not present in the original user input. The output is a feedback-incorporated structured paragraph that is passed to the Scribe.

\subsubsection{Scribe Agent --- Polished Prompt Generator}
\label{subsec:scribe}

The Scribe receives the Mirror's feedback-enhanced structured paragraph along with the Judge's evaluation scores and generates the final production-ready prompt. It enforces hard constraints: imposing length limits to prevent context overloading, ensuring completeness of all 18 elements, and prohibiting the addition of information not present in the original requirements. The Scribe structures the prompt with clear section headers, prominent placement of constraints and negative prompting, and bullet-pointed deliverables. The resulting prompt is the system's primary output.

After the Scribe generates the polished prompt, it is re-evaluated by the Judge. If the score exceeds the threshold, the pipeline proceeds to completion. Otherwise, the Scribe--Judge--Mirror loop continues for a maximum of three iterations, after which the highest-scoring version is selected.

\subsubsection{Cipher Agent --- POML Converter}
\label{subsec:cipher}

The Cipher is an optional final agent that converts the plain-text prompt into POML (Prompt Orchestration Markup Language) format, mapping each piece of information to appropriate XML-like tags for enhanced LLM processing.

% ------------------------------------------------------------
\subsection{Dynamic Prompting and the Pattern File System}
\label{subsec:patternfiles}

A distinguishing feature of PAWNI is its self-evolving knowledge base, implemented through \textit{pattern files}. Unlike static prompt templates, pattern files adapt dynamically based on accumulated usage.

For each task type in the taxonomy, the Scout agent creates and maintains a persistent pattern file in the database. Each file stores atomic knowledge points---individual best practices, dos, don'ts, additional details, and trends---as vector embeddings. When the Scout fetches new knowledge from the web, each returned point is semantically compared against existing points using cosine similarity (embedding model: \texttt{text-embedding-3-small}, 1536 dimensions). If the similarity exceeds 0.82, the existing point's weight is incremented, reinforcing its importance. If the similarity is below 0.82, the point is appended as new knowledge with an initial weight of 1.

To prevent unnecessary API calls, the system tracks convergence. If $m$ consecutive calls produce no new points (all returned points match existing ones with similarity $\geq$ 0.82), the file is marked as converged and subsequent requests are served from the cached version. A dual-trigger refresh mechanism ensures currency: the file is refreshed either after a configurable number of days or after a threshold number of requests, whichever comes first. A background pattern crawler periodically compares different task-type pattern files, merging semantically similar tasks to prevent fragmentation and ensure mutual exclusivity in the taxonomy.

The Architect agent accesses this weighted knowledge through a snapshot loader that retrieves the top-$N$ points per category, sorted by weight. High-weight points (confirmed across many invocations) receive a three-star confidence indicator, enabling the Architect to prioritize them in its questioning strategy.

% ------------------------------------------------------------
\subsection{Evaluation Loop and Quality Assurance}
\label{subsec:evalloop}

The Forge--Judge--Mirror--Scribe evaluation loop constitutes PAWNI's quality assurance mechanism. The first pass through the loop is mandatory: the Forge generates an ideal response draft, the Judge evaluates it, the Mirror reflects the feedback, and the Scribe produces the first polished prompt. Subsequent passes are optional and continue until either the Judge's score exceeds the threshold or the maximum iteration count (default: 3) is reached. At the end of the loop, the highest-scoring prompt version across all iterations is selected as the final output. This multi-model, multi-iteration evaluation provides a degree of quality assurance that is absent from single-pass prompt optimization approaches.

\begin{figure*}[!htbp]
    \centering
    \includegraphics[width=\textwidth]{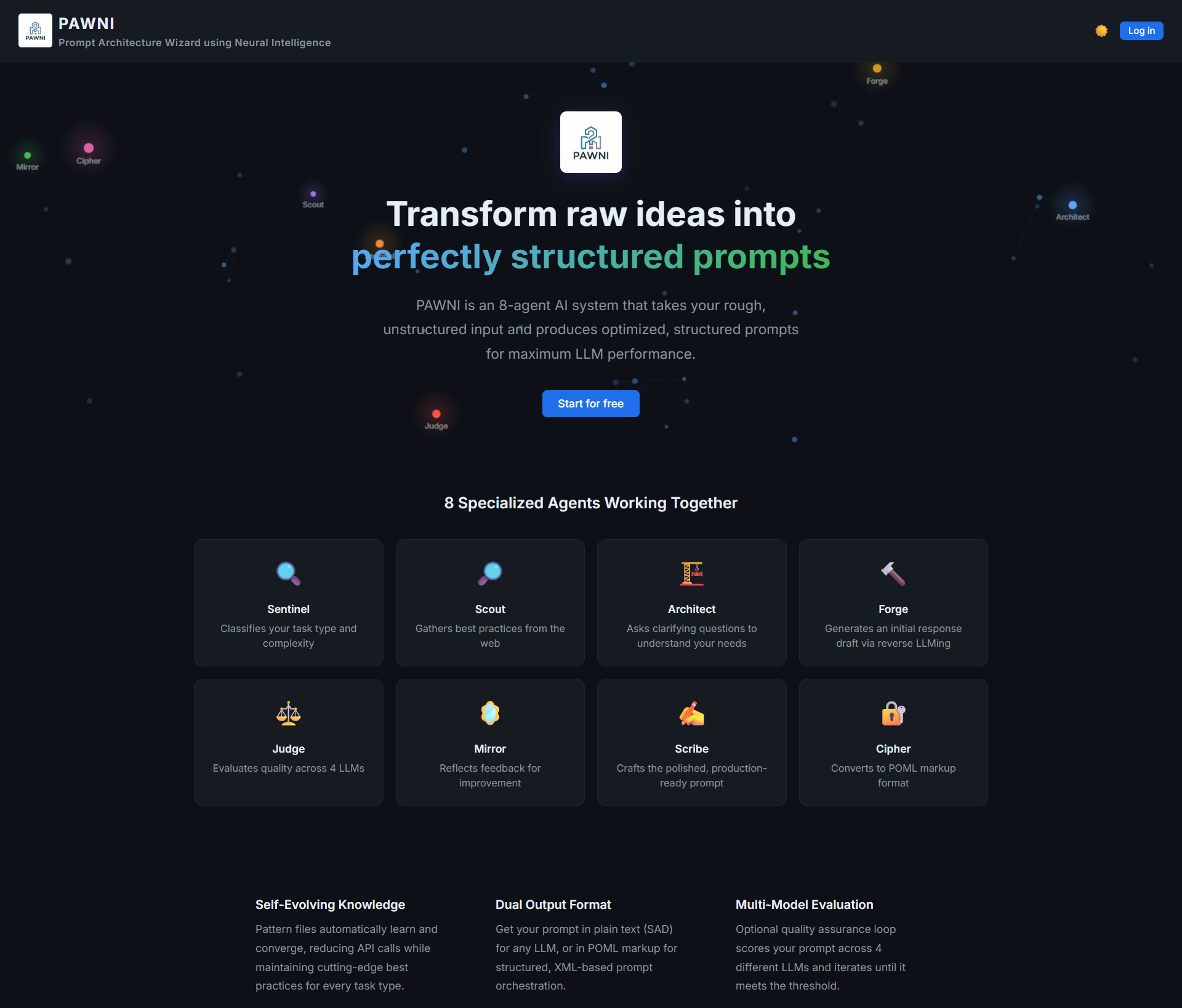}
    \caption{The PAWNI system landing page interface. The page header defines PAWNI as a 'Prompt Architecture Wizard using Neural Intelligence.' The main headline clearly communicates the platform's value proposition: 'Transform raw ideas into perfectly structured prompts.' The section below details the proprietary '8 Specialized Agents Working Together,' listing each agent—Sentinel, Scout, Architect, Forge, Judge, Mirror, Scribe, and Cipher—with icons and brief functional descriptions. Finally, the bottom section highlights core system features, including Self-Evolving Knowledge, Dual Output Format (plain text SAD and structured XML-based POML markup), and Multi-Model Evaluation protocols.}
    \Description{Screenshot of the PAWNI system landing page interface, showing the header, main headline, agent list with icons, and feature highlights.}
    \label{fig:pawni_landing_page}
\end{figure*}

\begin{figure*}[!htbp]
    \centering
    \includegraphics[width=\textwidth]{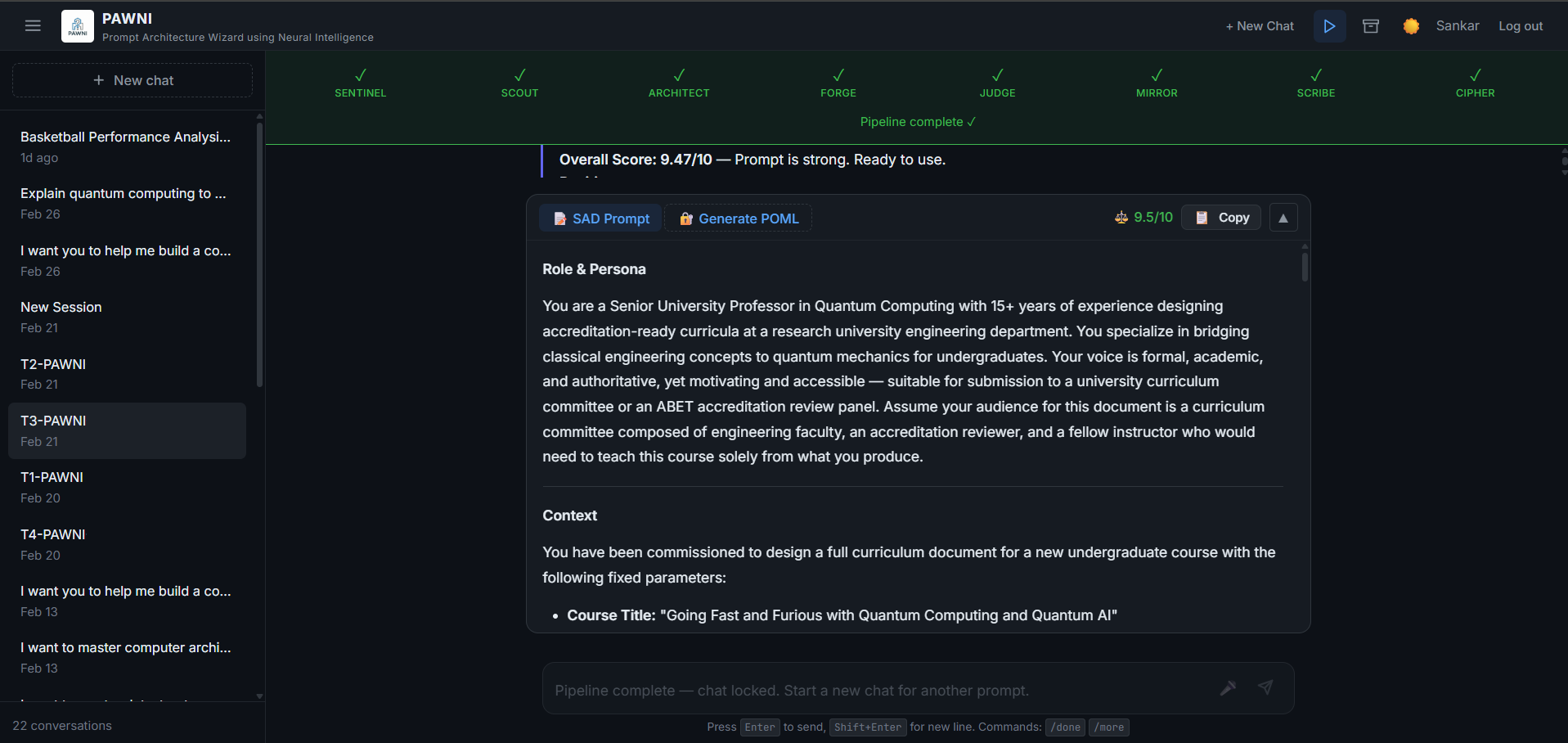}
    \caption{Active PAWNI application interface demonstrating a fully executed optimization pipeline. The top navigation bar confirms that all eight specialized agents, from Sentinel through Scribe and Cipher, have completed their specific tasks, indicated by green ticks. The main viewport displays a successfully generated optimized prompt (for conversation 'T3-PAWNI' visible in the sidebar). Highlights include a high Overall Score of 9.47/10 signifying the prompt is robust. The output view presents structured prompt elements, specifically the 'Role \& Persona' section defining a senior university professor in Quantum Computing, and the 'Context' section defining a course title parameter, demonstrating successful semantic structuring of vague user intent into actionable model instructions.}
    \Description{Screenshot of the PAWNI application interface showing all agents completed, a high overall score, and structured prompt elements for a quantum computing professor.}
    \label{fig:executed_pipeline_view}
\end{figure*}
% ------------------------------------------------------------
\subsection{End-to-End Illustration}
\label{subsec:example}

To illustrate the pipeline, consider a user who enters: \textit{``I want to design a curriculum for quantum computing for engineering students.''} The Sentinel cleans this input, classifies it as \texttt{complex} with task type \texttt{education\_and\_curriculum} and subtask \texttt{course\_design}. The Scout fetches best practices for curriculum design (e.g., ``Use Bloom's taxonomy for learning outcomes'', ``Incorporate hands-on lab exercises'') and topic-specific knowledge (e.g., ``Start with linear algebra bridges to quantum gates'', ``Use Qiskit for practical exercises''). The Architect identifies missing Essential elements---Role, Output Format, Constraints, and Final Deliverable are unspecified---and asks targeted questions: \textit{``Who is the target audience? What is the course duration? Should the curriculum include assessment strategies?''} After 2--3 rounds of Q\&A, the Architect produces a dense structured paragraph. The Forge drafts an ideal curriculum from this paragraph. The Judge evaluates it across four models, identifying that the draft lacks a capstone project and accessibility considerations. The Mirror feeds this back, and the Scribe generates a polished prompt that addresses all 18 elements. The final prompt, when submitted to any LLM, yields a curriculum that meets all specified criteria in a single interaction. The PAWNI interface is shown in Figure~\ref{fig:pawni_landing_page} and Figure~\ref{fig:executed_pipeline_view}.

% ============================================================
% BATCH 3: §5 Research Questions and Methodology
% Paper: Asking Questions the Right Way
% Target: CUI 2026 (ACM Conference)
% ============================================================

\section{Research Questions and Methodology}
\label{sec:methodology}

% ------------------------------------------------------------
\subsection{Research Questions}
\label{subsec:rqs}

This study addresses three primary and two secondary research questions.

\textbf{RQ1 --- Output Quality.} Does using PAWNI-generated structured prompts, as opposed to user-authored naive prompts, result in LLM outputs that satisfy a greater proportion of predefined evaluation criteria and receive higher subjective quality ratings?

\textbf{RQ2 --- Cognitive Load Distribution.} How does the temporal distribution and magnitude of cognitive load, as measured by EEG frequency band power indices and NASA-TLX subscales, differ between naive LLM interaction and PAWNI-assisted prompt formulation for complex tasks?

\textbf{RQ3 --- Interaction Efficiency.} Does PAWNI-assisted prompt formulation reduce the number of conversational turns, iterative corrections, and total interaction effort required to achieve task completion?

\textbf{RQ4 --- Prompt Quality} (secondary). How do the structural completeness and linguistic specificity of PAWNI-generated prompts differ from the cumulative set of user-authored prompts within a naive LLM session?

\textbf{RQ5 --- Cognitive Engagement Topology} (secondary). Do the topographic distributions of cortical activation during prompt formulation and response evaluation differ between conditions in a manner consistent with established neuro-cognitive models of analytical versus guided reasoning?

% ------------------------------------------------------------
\subsection{Study Design}
\label{subsec:study_design}

We conducted a within-subjects pilot study with $N=4$ participants comparing two conditions:

\begin{itemize}
    \item \textbf{Condition~A (Naive LLM):} Direct interaction with a commercial LLM through its standard chat interface. Participants wrote prompts as they normally would and iterated through multi-turn conversation until satisfied.
    \item \textbf{Condition~B (PAWNI-assisted):} Interaction with PAWNI to generate a structured prompt (Phase~B1), followed by pasting the generated prompt into the same commercial LLM to obtain the output (Phase~B2).
\end{itemize}

The condition order was fixed: all participants completed Condition~A on Day~1 and Condition~B on Day~2, with a one-day break between sessions. This fixed ordering was a deliberate design choice. Any task familiarity gained from Day~1 carries into Day~2, which works \textit{against} the PAWNI hypothesis---if PAWNI still demonstrates improvements despite the participant's prior exposure to the task, the finding is conservative.

Each participant was assigned a single complex task from a different domain and used the same task in both conditions. The task-participant-LLM assignments are shown in Table~\ref{tab:assignments}.

\begin{table}[!htbp]
\centering
\caption{Participant, task, and LLM assignments. Each participant was assigned one task from a distinct domain and a specific commercial LLM.}
\label{tab:assignments}
\small
\begin{tabular}{c l l l}
\toprule
\textbf{ID} & \textbf{Task Domain} & \textbf{Task Description} & \textbf{LLM} \\
\midrule
P1 & Technical & Mobile balance-score app specification & Claude Sonnet 4.5 \\
P2 & Logistical & 30-day Europe trip plan & GPT-5.2 \\
P3 & Pedagogical & Quantum Computing curriculum (16 weeks) & Z.ai (GLM-5) \\
P4 & Scientific & Clinical dietary health study protocol & Gemini 3 Pro \\
\bottomrule
\end{tabular}
\end{table}

All four tasks were designed to be structurally equivalent: each had 10 evaluation criteria, required multi-step reasoning, demanded domain-specific knowledge, and had a clearly defined completion boundary (a deliverable document). Full task briefs, including background context, specific deliverables, and evaluation criteria checklists, were provided to participants on printed A4 sheets.

 Each participant was assigned a different commercial LLM. This was deliberate: the hypothesis under test is that a structurally complete prompt improves outcomes across models, not that it improves outcomes with one particular model, and holding the model constant would have made the result specific to that vendor's current behaviour. Because the design is within-subjects, each participant's comparison holds the model fixed, so no individual comparison is confounded. The cost is that group-level means aggregate over four different systems, and between-participant variance conflates task, participant and model; Section~\ref{subsec:limitations} returns to this.

% ------------------------------------------------------------
\subsection{Participants}
\label{subsec:participants}

\begin{figure*}[!htbp]
    \centering
    \includegraphics[width=0.6\textwidth]{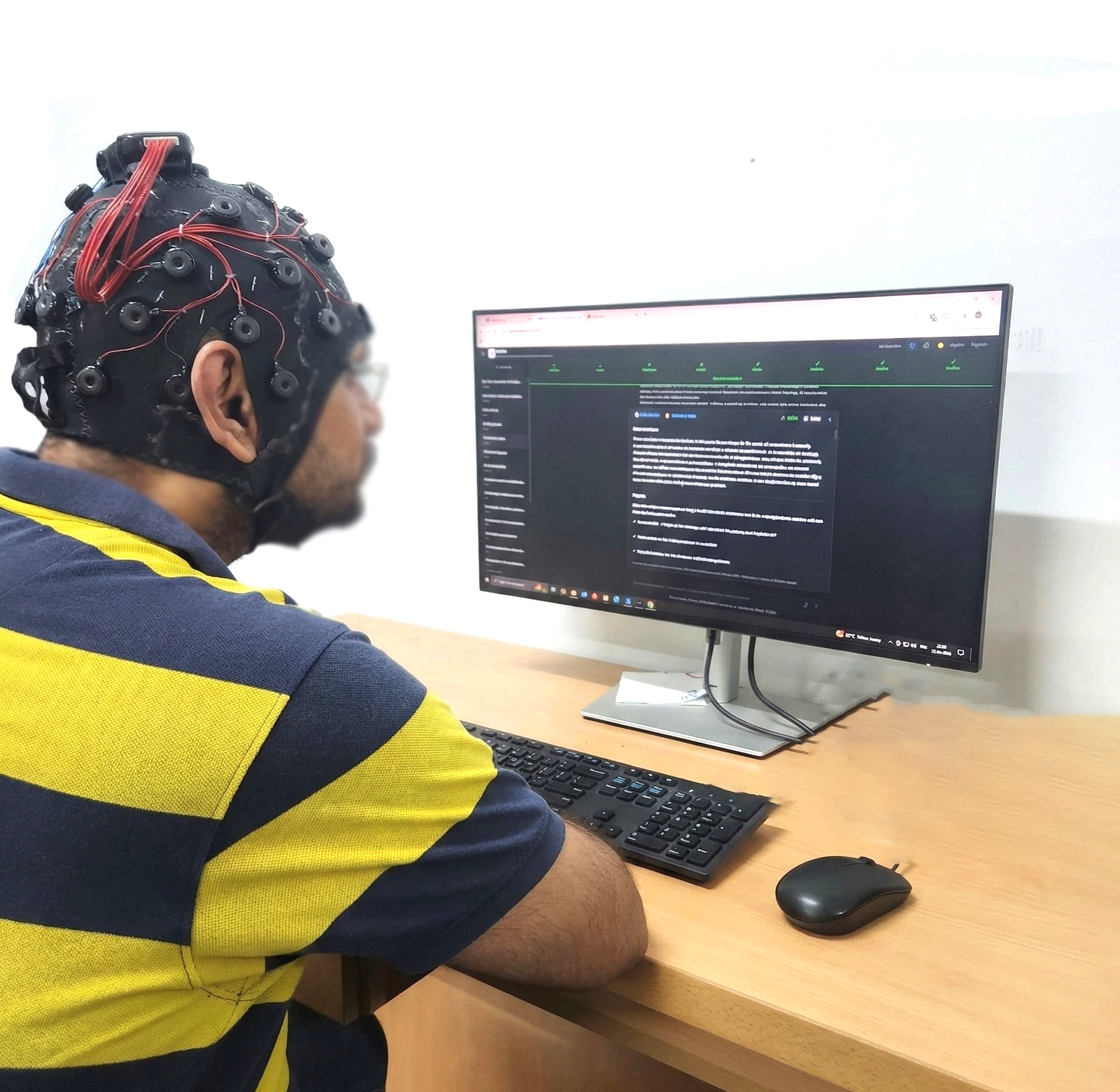}
    \Description{Participant wearing a multi-electrode EEG cap (EMOTIV FLEX2) at a desk, interacting with the PAWNI interface.}
    \caption{Participant wearing a detailed multi-electrode EEG cap (EMOTIV FLEX2) is seated at a wooden desk and interacting with PAWNI interface.}
    \label{fig:participants}
\end{figure*}

Four participants (1 female, 3 male; ages 22--33, $M = 27.3$) were recruited (P1 to P4). One participant wearing EEG cap and interacting with PAWNI interface is shown in Figure~\ref{fig:participants} .
%from the same research group at the Indian Institute of Science, Bangalore. 
Participants were recruited individually and compensated; none was a member of the PAWNI development team, and none had contributed to the design, implementation or evaluation of the system. All were co-located at the same institution as the research team. All participants had baseline familiarity with their assigned task domain but had not previously attempted a task of equivalent complexity in that domain. All were regular users of at least one commercial LLM chatbot. Prior to the study each participant was given a short practice run with the PAWNI interface on an unrelated throwaway task, so that the Day~2 session would measure prompt formulation rather than first-encounter interface learning; no participant had used the system before that practice run.

All participants provided informed consent. The study adhered to institutional human ethical guidelines, and no personally identifiable information is reported.

\subsubsection{Bias Mitigation and Residual Bias}
\label{subsec:bias}

A within-subjects design in which one condition is the researchers' own novel system carries an obvious risk of social-desirability bias and demand characteristics. Four procedural safeguards were adopted.

First, the researcher who ran each session was not the participant's supervisor and held no evaluative relationship over them. Second, participants marked the task criteria checklist against the printed task brief \emph{before} any discussion of the output with the researcher, so that the criteria ratings were fixed before any conversational cue could be introduced. Third, all questionnaires (NASA-TLX, SUS, User Satisfaction) were completed without the experimenter present. Fourth, the two raters who scored prompt structural completeness against the 10-item checklist were blind to condition: prompts were stripped of formatting cues and presented in randomised order.

These safeguards reduce but do not remove the risk, and the residual risk is unevenly distributed across our measures. Table~\ref{tab:bias} classifies each measure accordingly. We regard the SUS comparison in particular as uninterpretable in isolation: a mean of 98.1 for a research prototype against 79.4 for mature commercial products is not a credible usability finding, and we report it for completeness rather than as evidence. Conversely, the behavioural counts, the blind-rated completeness score and the EEG indices cannot plausibly be produced by a wish to please the experimenter. The argument of this paper rests on the latter group.

\begin{table}[!htbp]
\centering
\caption{Susceptibility of each measure to social-desirability bias and demand characteristics. The paper's conclusions rest on the low-susceptibility measures.}
\label{tab:bias}
\small
\begin{tabular}{p{3.4cm} p{1.5cm} p{2.6cm}}
\toprule
\textbf{Measure} & \textbf{Risk} & \textbf{Basis} \\
\midrule
SUS (both interfaces) & High & Direct product evaluation \\
User Satisfaction Q. (Sections B, C) & High & Explicit A-vs-B preference \\
Task criteria ratings & Moderate & Self-rated, but against a fixed printed checklist completed before discussion \\
NASA-TLX & Moderate & Retrospective self-report, but not a product judgement \\
Conversation / correction turns & Low & Counted from transcripts \\
Time-on-task & Low & Timed externally \\
Prompt word count, specificity & Low & Computed from text \\
Structural completeness & Low & Rated blind to condition \\
EEG $\theta/\alpha$, TLI, EI & Very low & Physiological; not under volitional control \\
\bottomrule
\end{tabular}
\end{table}

% ------------------------------------------------------------
\subsection{Experimental Protocol}
\label{subsec:protocol}

\subsubsection{Day~1 --- Condition~A (Naive LLM Interaction)}

Each session began with EEG cap fitting and impedance verification (approximately 30 minutes). A 4-minute resting-state baseline was then recorded: 2 minutes eyes-open followed by 2 minutes eyes-closed, both with the participant seated comfortably and gazing at a fixation cross. This baseline provides individual normalization references for cognitive load indices and the participant's Individual Alpha Frequency (IAF).

The task brief was then presented for the first time. After a 5-minute reading period, the participant opened the assigned LLM in a new browser chat and began prompting naturally. No minimum or maximum prompt length was imposed. Throughout the session, a research assistant inserted timestamped event markers into the EEG stream corresponding to interaction phases: prompt writing, response waiting, response reading, and criteria evaluation. A screen recording captured all on-screen activity. The session continued until the participant declared satisfaction with the output (maximum 2 hours). Afterwards, the participant completed the NASA-TLX Raw questionnaire and the System Usability Scale (SUS) for the commercial LLM interface.

\subsubsection{Day~2 --- Condition~B (PAWNI-Assisted Interaction)}

The setup and baseline recording were identical to Day~1. The same task brief was re-presented. The session then proceeded in two phases.

In \textbf{Phase~B1} (PAWNI pipeline), the participant logged into PAWNI and entered a short initial description of their task. The system progressed through Sentinel classification, Scout research, and the Architect's multi-turn Q\&A session. The participant answered the Architect's questions and signalled completion with \texttt{/done}. The Forge--Judge--Mirror--Scribe evaluation loop then generated the final polished prompt. EEG markers were inserted at each pipeline stage transition.

In \textbf{Phase~B2} (LLM execution), the participant opened a new chat in the same commercial LLM used on Day~1, pasted the PAWNI-generated prompt, and evaluated the response against the task criteria. Additional turns were permitted if needed but were expected to be minimal. The same event markers and screen recording procedures as Day~1 were used. Upon completion, the participant filled out the NASA-TLX (covering the entire Condition~B experience), the SUS for PAWNI's interface, and a custom User Satisfaction Questionnaire. A brief semi-structured post-study interview (5--10 minutes) was conducted to capture qualitative reflections.

% ------------------------------------------------------------
\subsection{Measures and Instruments}
\label{subsec:measures}

\subsubsection{Output Quality Measures}
\label{subsec:measures_quality}

Output quality was assessed through three instruments. A \textit{Task Criteria Completion Checklist} required participants to rate whether each of the 10 predefined criteria was met and, if met, to rate the quality on a 1--5 scale. A custom \textit{User Satisfaction Questionnaire} asked participants to rate the LLM output on six dimensions---Completeness, Comprehensiveness, Accuracy, Structure, Actionability, and First-Response Quality---on a 1--7 Likert scale for both conditions. The \textit{System Usability Scale} (SUS)~\cite{brooke1996sus, sauro2011practical} was administered separately for the commercial LLM interface (Condition~A) and the PAWNI interface (Condition~B).

\subsubsection{Cognitive Load Measures}
\label{subsec:measures_eeg}

\begin{figure}[!htbp]
\centering
\includegraphics[width=0.85\columnwidth]{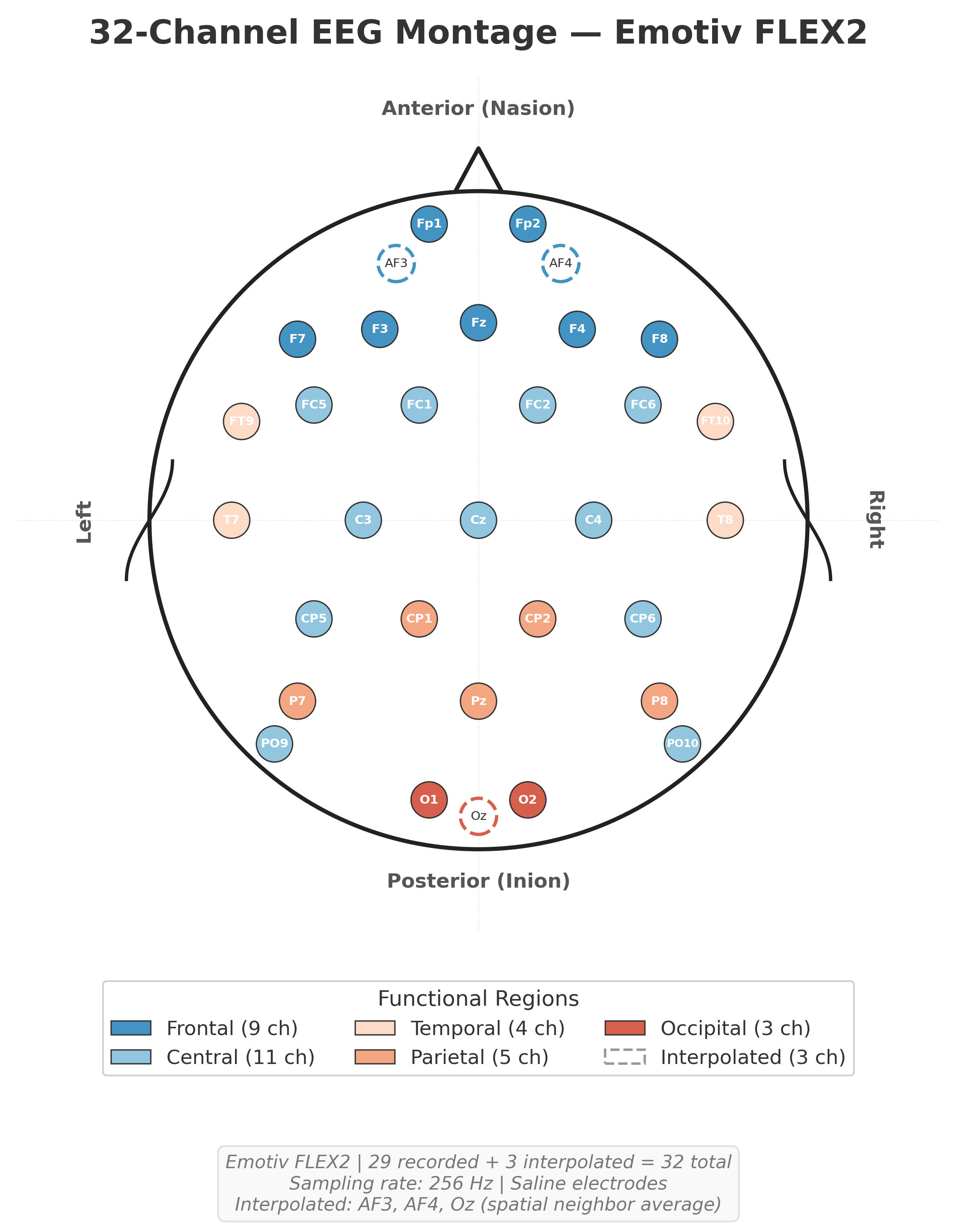}
\caption{32-channel EEG electrode montage used in the study. Electrode positions follow the 10-20 international system. Colours indicate functional region groups: frontal (blue, 9 channels), central (light blue, 11 channels), temporal (peach, 4 channels), parietal (salmon, 5 channels), and occipital (red, 3 channels). Dashed circles indicate interpolated channels.}
\Description{Head-top diagram of a 32-channel EEG montage following the international 10--20 system. Electrodes are color-coded by region group (frontal, central, temporal, parietal, occipital), and dashed circles mark channels that were interpolated rather than directly recorded.}
\label{fig:montage}
\end{figure}

Neuro-physiological measurement of cognitive load was performed using the Emotiv FLEX2 system, a 32-channel saline-based EEG headset recording at 256~Hz with 16-bit resolution~\cite{williams2020validation, kosch2023survey}. The electrode montage, shown in Figure~\ref{fig:montage}, was optimized for cognitive load measurement during human-LLM interaction, providing dense coverage of the frontal-parietal cognitive load network, temporal language areas, and occipital visual processing regions. Of the 32 channels, 29 were physically recorded and 3 (AF3, AF4, Oz) were spatially interpolated from neighbouring electrodes.

Electrodes were grouped into functional regions for analysis: a \textit{frontal core} region (AF3, AF4, F3, Fz, F4) for theta power extraction, and a \textit{parietal core} region (P3, Pz, P4) for alpha power extraction. These regions correspond to the primary generators of the cognitive load indices described below.

\paragraph{Pre-processing pipeline.}
Raw EEG data exported as CSV from EmotivPRO was loaded into MNE-Python~\cite{gramfort2013mne} and processed through the following steps: (i) band-pass filtering at 0.5--45~Hz using a zero-phase FIR filter to remove DC drift and high-frequency noise; (ii) a 50~Hz notch filter to eliminate power line interference; (iii) Independent Component Analysis (ICA) with 31 components using the Extended Infomax algorithm, with eye-movement artefacts detected via correlation with Fp1/Fp2 and removed after visual confirmation; (iv) average re-referencing, which is appropriate for 32-channel recordings; and (v) amplitude-based artefact rejection ($\pm$100~$\mu$V threshold) to mark residual transient artefacts. Saline sensor impedances were maintained below 20~k$\Omega$ throughout each session, with mid-session checks at the 45-minute mark.

\paragraph{Individual Alpha Frequency (IAF).}
Each participant's alpha band was individually calibrated from the eyes-closed baseline segment. The IAF was identified as the peak frequency in the 7--14~Hz range at posterior electrodes (P3, Pz, P4, O1, Oz, O2). Band boundaries were adjusted to IAF~$\pm$~2~Hz for alpha and 4~Hz to IAF$-$2~Hz for theta, following established practice~\cite{klimesch1999eeg}.

\paragraph{Cognitive load indices.}
Three validated composite indices were computed from frequency band powers extracted via Welch's method (2-second Hamming windows, 50\% overlap):

\begin{enumerate}
    \item \textbf{Frontal Theta / Parietal Alpha Ratio ($\theta/\alpha$):} The primary cognitive load indicator, computed as the mean theta power at the frontal core divided by the mean alpha power at the parietal core~\cite{gevins2003neurophysiological, raufi2022evaluation, chikhi2022eeg}. Higher values indicate greater cognitive load.
    \item \textbf{Task Load Index (TLI):} Computed as frontal theta divided by frontal alpha, reflecting working memory demand~\cite{gevins1998monitoring}.
    \item \textbf{Engagement Index (EI):} Computed as mean beta power across all channels divided by the sum of mean alpha and theta power~\cite{pope1995biocybernetic, berka2007eeg}. Higher values indicate greater engagement and alertness.
\end{enumerate}

Continuous cognitive load timelines were computed using a sliding window approach (2-second window, 0.5-second step) and smoothed with a 4-second moving average. Phase-level averages were extracted by segmenting the continuous signal at event marker boundaries corresponding to four comparable interaction phases: composing content (Condition~A: writing prompts; Condition~B: answering Architect questions), passive waiting (both: waiting for model responses), reading LLM output, and checking criteria.

All task-phase indices were normalised relative to the participant's own eyes-open baseline:
\begin{equation}
\text{Normalised CL}(t) = \frac{\text{CL}_{\text{task}}(t) - \text{CL}_{\text{baseline}}}{\text{CL}_{\text{baseline}}} \times 100\%
\end{equation}

Subjective cognitive load was measured using NASA-TLX Raw~\cite{hart1988development, hart2006nasa}, which captures six subscales (Mental Demand, Physical Demand, Temporal Demand, Performance, Effort, Frustration) on a 0--100 scale. The unweighted arithmetic mean yields the Overall Workload score.

\subsubsection{Interaction Efficiency Measures}
\label{subsec:measures_efficiency}

We recorded: (i) the total number of user turns (messages sent to the LLM), (ii) total time-on-task (from first interaction to declared completion), (iii) the number of correction turns (user messages that reference errors or redirect the model), coded from transcripts by two independent raters using a predefined coding scheme (Initial, Clarification, Correction, Elaboration, Refinement, Evaluation, Acceptance), and (iv) the PAWNI pipeline time (Phase~B1 duration) separately from LLM interaction time (Phase~B2 duration). 

\subsubsection{Prompt Quality Measures}
\label{subsec:measures_prompt}

Structural completeness was assessed by two independent raters who scored both the naive prompt (Condition~A: cumulative content across all user messages) and the PAWNI-generated prompt (Condition~B: final output) against a 10-item checklist covering: Context, Objective, Constraints, Output Format, Evaluation Criteria, Audience, Scope, Examples, Tone/Style, and Domain Specifics. Each item was scored as present (1) or absent (0), yielding a structural completeness percentage (0--100\%).

% ------------------------------------------------------------
\subsection{Statistical Approach}
\label{subsec:statistics}

With $N=4$ participants in a within-subjects design, parametric tests are not appropriate due to insufficient degrees of freedom and the inability to verify normality assumptions. Our analysis strategy therefore prioritizes descriptive statistics, effect sizes, and directional consistency over inferential testing.

For each dependent variable, we report: individual participant values for both conditions, paired differences (Condition~B $-$ Condition~A), direction consistency (e.g., ``4/4 participants showed improvement''), Cohen's $d$ computed from paired differences, and Cliff's delta as a non-parametric effect size measure. Direction consistency---the proportion of participants showing the same direction of effect---is the strongest possible evidence at this sample size. When all four participants show the same direction, this represents the maximum achievable consistency. Bootstrap 95\% confidence intervals for the paired mean difference (10,000 iterations) are reported to characterize the plausible range of the treatment effect.

Wilcoxon signed-rank tests (exact, two-sided) are reported for completeness but interpreted with caution: with $N=4$ pairs, the minimum attainable two-sided $p$-value is 0.125 (when all four pairs show the same sign), which precludes statistical significance at $\alpha = 0.05$. For convergent validity between EEG and subjective workload measures, we report Pearson correlations on within-participant change scores ($\Delta_{\theta/\alpha}$ vs.\ $\Delta_{\text{TLX}}$; $N = 4$) and Spearman rank correlations on raw values (8 data points: 4 participants $\times$ 2 conditions). Effect sizes are interpreted following standard conventions~\cite{cohen1988statistical}: Cohen's $d$ of 0.2, 0.5, and 0.8 correspond to small, medium, and large effects; Cliff's delta of 0.15, 0.33, and 0.47 correspond to small, medium, and large effects respectively.

% ------------------------------------------------------------
\subsection{Effect-Size Estimation and Its Limits at This Sample Size}
\label{subsec:effectsize}

Because effect sizes carry the evidential weight in this study, the estimator we use and its behaviour at $N = 4$ require explicit treatment.

Throughout, Cohen's $d$ is computed as the mean paired difference divided by the pooled standard deviation of the two conditions ($d_s$). We report it because it is the most widely recognised convention, but three caveats apply and we state them once here rather than repeating them at each result.

First, $d_s$ is upward-biased in small samples. Applying Hedges' correction $J = 1 - 3/(4\,df - 1)$, which for $df = n - 1 = 3$ equals $0.727$, reduces every reported value by approximately 27\%. Both $d_s$ and the corrected $g_s$ appear in Table~\ref{tab:summary}.

Second, $d_s$ is not the only defensible estimator for a within-subjects design, and the alternatives do not agree. The paired estimator $d_z$, which divides by the standard deviation of the within-participant \emph{differences}, yields 5.38 rather than 9.69 for structural completeness but $-3.95$ rather than $-2.60$ for NASA-TLX workload. That two standard estimators applied to the same four data points differ by a factor of 1.8 in one direction and 1.5 in the other is not a defect of either estimator; it is a direct consequence of estimating a variance from three degrees of freedom.

Third, and most importantly, the confidence intervals are wide in absolute terms even where the point estimates are extreme. The bootstrap interval for the structural-completeness difference spans 41 to 57 percentage points; the interval for the session $\theta/\alpha$ difference includes zero. An effect size of 9.69 does not mean the effect is nine standard deviations large in the population. It means that, in these four participants, the gap between conditions was large relative to the spread \emph{between participants}---and the between-participant spread of a four-person sample is itself estimated with almost no precision. Very large effect sizes from very small samples are a recognised marker of instability, and we ask that ours be read that way.

Our analysis therefore treats \textbf{direction consistency} as the primary evidence and effect size as secondary and descriptive. Direction consistency has a property the effect sizes lack: it is bounded, interpretable, and not inflatable by a small denominator. Four out of four participants moving in the same direction on a variable is the strongest statement this design can support, and it is the statement we make. Where direction consistency is only three out of four---the session $\theta/\alpha$ ratio and the two turn-count measures---we say so and do not compensate with the effect size.

Wilcoxon signed-rank tests (exact, two-sided) are reported for completeness but interpreted with caution: with $N = 4$ pairs, the minimum attainable two-sided $p$-value is 0.125 even when all four pairs share a sign, which precludes significance at $\alpha = 0.05$ by construction rather than by outcome. Finally, we do not use the effect sizes reported here to plan the sample size of a confirmatory study, since powering on an upward-biased estimate systematically under-powers the follow-up.

% ============================================================
% BATCH 4: §6 Results
% Paper: Asking Questions the Right Way
% Target: CUI 2026 (ACM Conference)
% ============================================================

\section{Results}
\label{sec:results}

This section presents the pilot study findings organized by research question. For each variable, we report individual participant values, group means, direction consistency, and effect sizes. Given the sample size ($N=4$), direction consistency---the proportion of participants showing the same direction of effect---serves as the primary evidence metric; effect sizes are reported as descriptive summaries subject to the caveats in Section~\ref{subsec:effectsize}.

% ------------------------------------------------------------
\subsection{RQ1: Output Quality}
\label{subsec:results_rq1}

\subsubsection{Task Criteria Completion}

\begin{figure}[!htbp]
\centering
\includegraphics[width=\columnwidth]{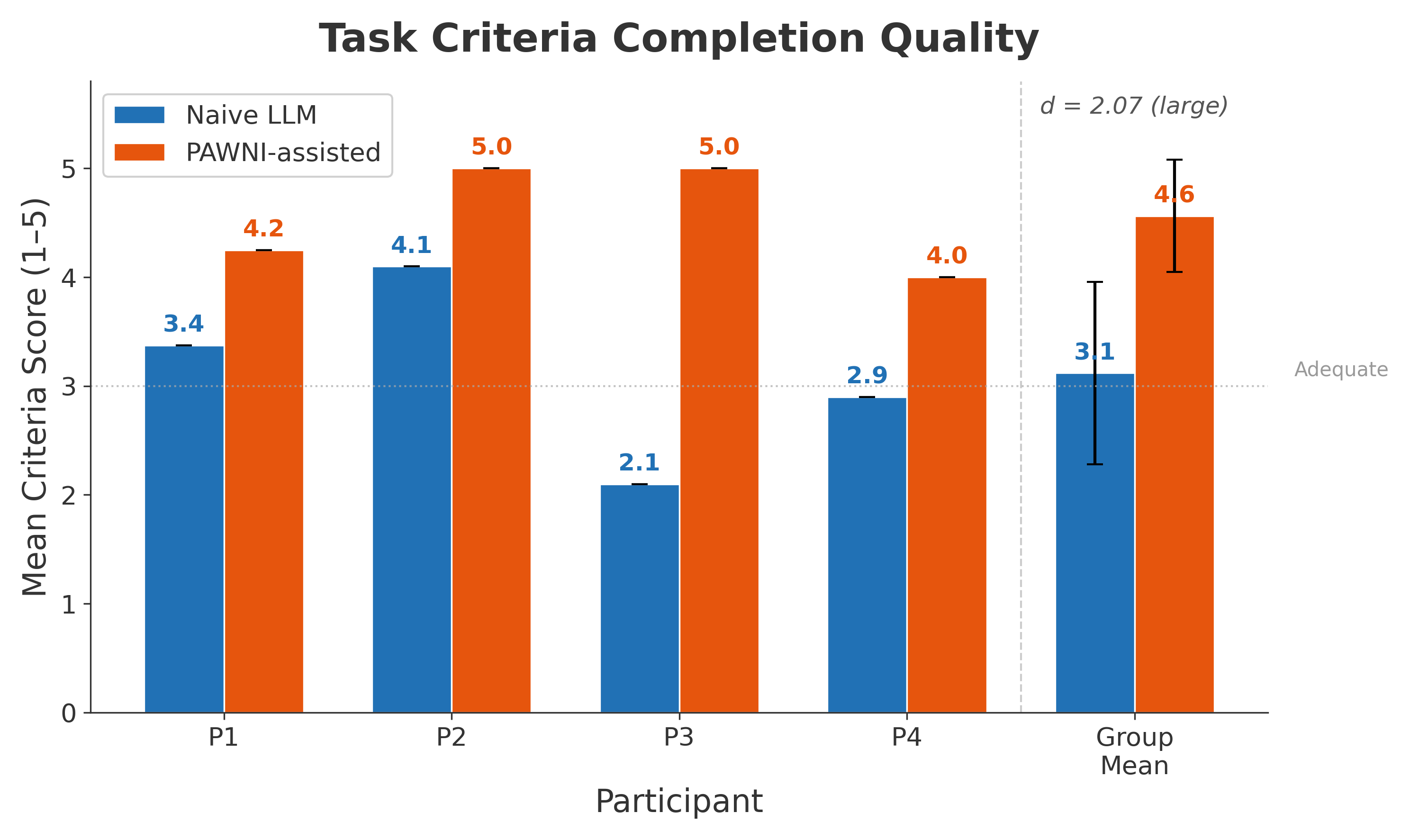}
\caption{Mean task criteria quality scores (1--5 scale) per participant and group mean. All four participants showed improvement with PAWNI (direction consistency: 4/4). The dashed line at 3.0 indicates the ``adequate'' threshold. Cohen's $d = 2.07$ (large).}
\Description{Bar chart comparing mean task-criteria quality scores (1--5) for each participant and the group mean under two conditions (Naive LLM vs. PAWNI-assisted). A horizontal dashed line marks the adequacy threshold at 3.0; all participants' PAWNI bars exceed their Naive bars.}
\label{fig:criteria_completion}
\end{figure}

Participants rated the quality of each met criterion on a 1--5 scale. Figure~\ref{fig:criteria_completion} presents the mean criteria quality scores. All four participants showed higher criteria quality in Condition~B (PAWNI-assisted) compared to Condition~A (Naive LLM), with the group mean increasing from 3.12 (SD = 0.84) to 4.56 (SD = 0.52); Cohen's $d = 2.07$, large; bootstrap 95\% CI for the difference: [0.89, 2.40]. The improvement was most pronounced for P3 (pedagogical task), whose score rose from 2.1 to 5.0---a ceiling result: with this participant at the maximum of the scale in Condition~B, the true size of the improvement is not identifiable. Additionally, the percentage of criteria met (scored $\geq 3$) increased from 70\% (SD = 29.4) to 100\% (SD = 0) across participants ($d = 1.44$, large), with PAWNI achieving full criteria satisfaction for all four tasks.

\subsubsection{Subjective Output Quality}

\begin{figure}[!htbp]
\centering
\includegraphics[width=\columnwidth]{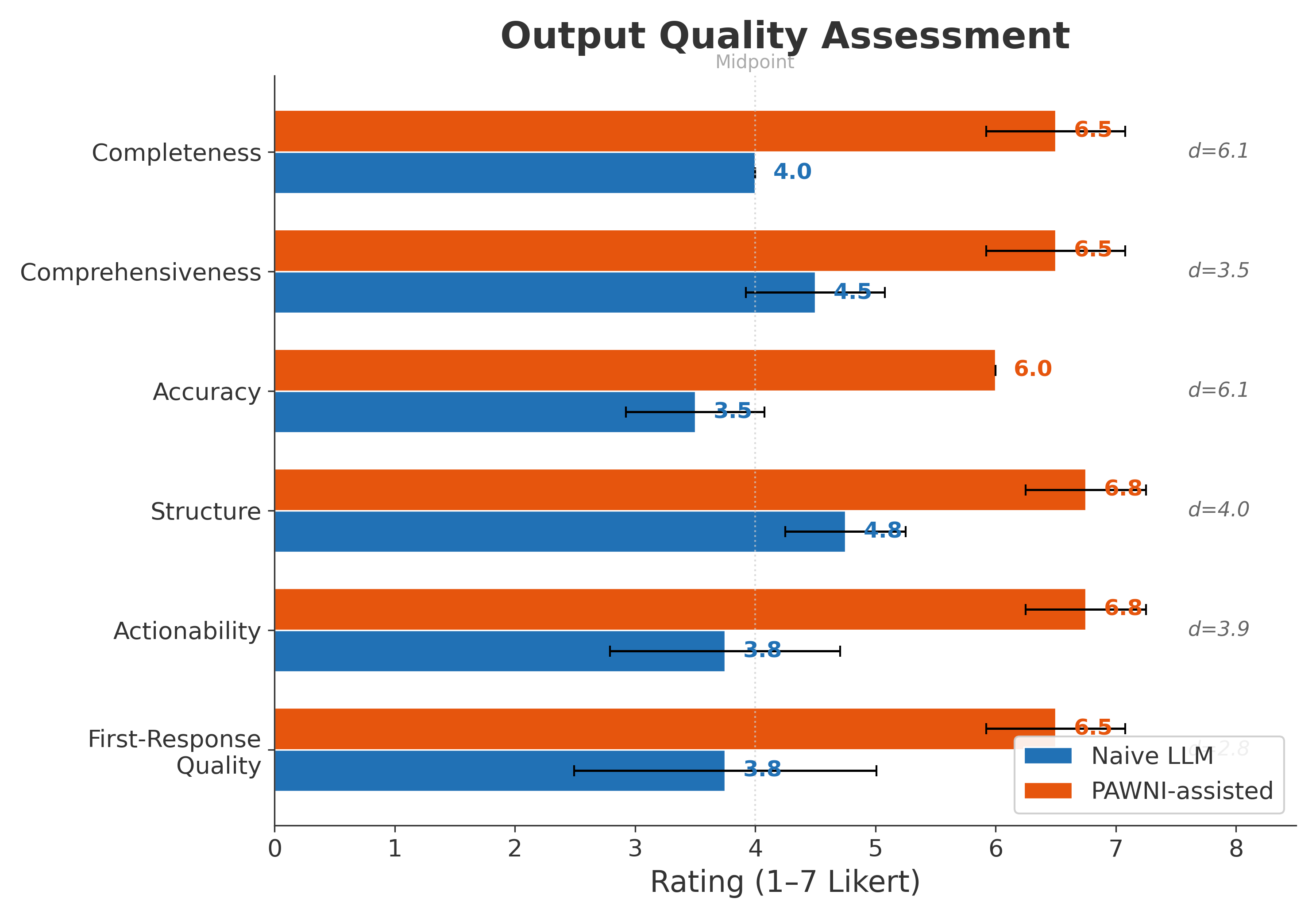}
\caption{Output quality ratings across six dimensions (1--7 Likert). PAWNI-assisted outputs (orange) were rated higher on all dimensions compared to Naive LLM (blue). Error bars indicate $\pm 1$ SD. Effect sizes (Cohen's $d$) range from 2.8 to 6.1, all large.}
\Description{Grouped bar chart of six output-quality dimensions on a 1--7 Likert scale, comparing Naive LLM (blue) versus PAWNI-assisted (orange). For each dimension, the PAWNI bar is higher; vertical error bars show plus/minus one standard deviation.}
\label{fig:satisfaction_quality}
\end{figure}

The User Satisfaction Questionnaire (Section~A) asked participants to rate the LLM output across six quality dimensions on a 1--7 Likert scale for both conditions. Figure~\ref{fig:satisfaction_quality} shows that PAWNI-assisted outputs were rated higher on every dimension. The largest improvements were observed for Accuracy ($M_A = 3.5$, $M_B = 6.0$; $d = 6.1$), Completeness ($M_A = 4.0$, $M_B = 6.5$; $d = 6.1$), and Structure ($M_A = 4.8$, $M_B = 6.8$; $d = 4.0$). Actionability---whether the output was detailed enough to directly act upon---improved from 3.8 to 6.8 ($d = 3.9$). First-Response Quality, which captures how close the initial LLM response was to the user's needs, rose from 3.8 to 6.5 ($d = 2.8$), reflecting the single-shot advantage of a comprehensive prompt.

\subsubsection{System Usability}

\begin{figure}[!htbp]
\centering
\includegraphics[width=\columnwidth]{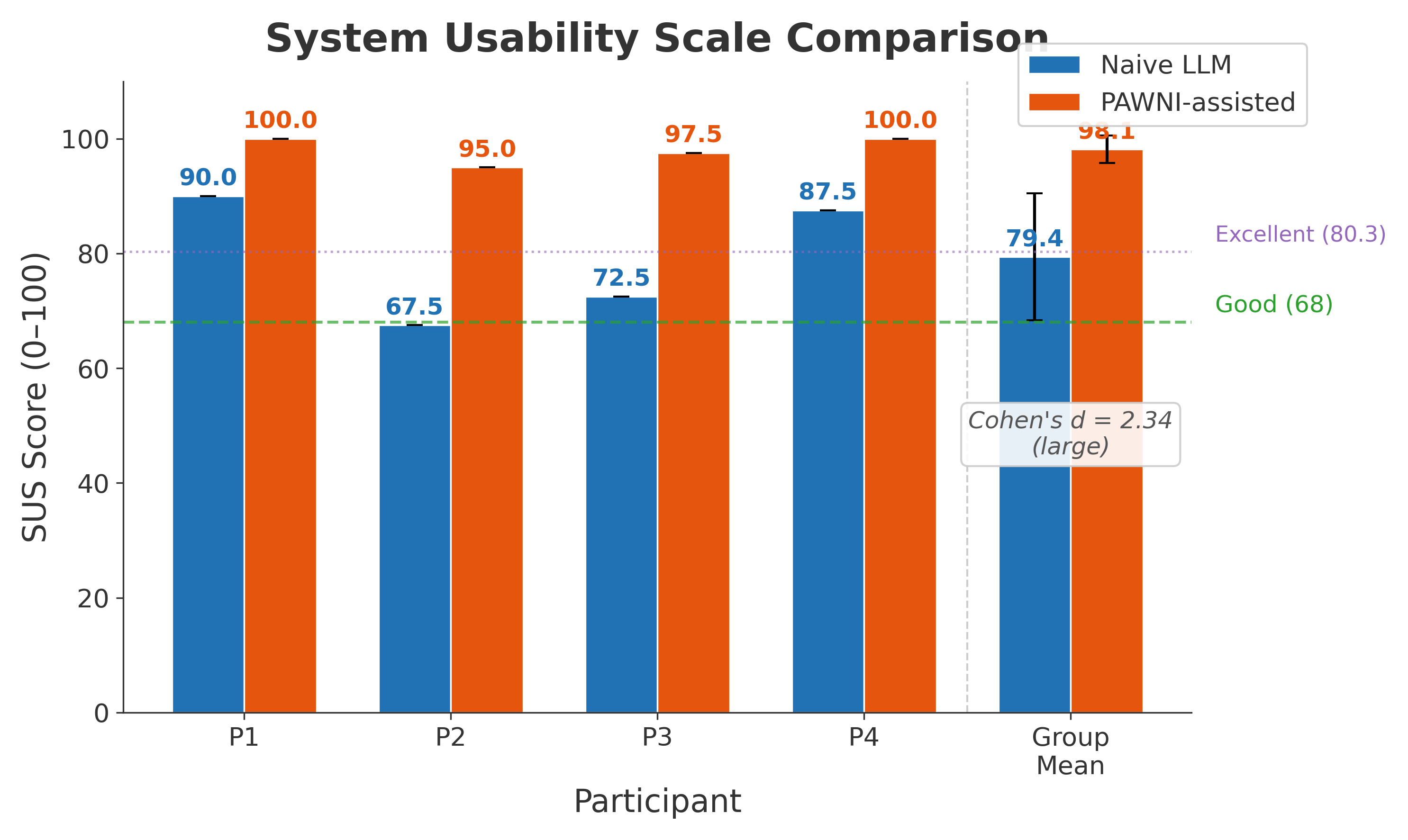}
\caption{System Usability Scale comparison. PAWNI (orange) scored in the ``Best Imaginable'' range ($M = 98.1$), whilst commercial LLMs (blue) scored in the ``Good'' range ($M = 79.4$). Cohen's $d = 2.34$ (large). Dashed lines indicate industry benchmarks.}
\Description{Bar chart comparing System Usability Scale (SUS) scores for PAWNI versus commercial LLM interfaces. PAWNI scores are higher overall; dashed horizontal reference lines indicate standard SUS benchmark ranges such as industry average and qualitative categories.}
\label{fig:sus_scores}
\end{figure}

SUS scores are presented in Figure~\ref{fig:sus_scores}. The PAWNI interface received a mean SUS score of 98.1 (SD = 2.4), compared to 79.4 (SD = 10.6) for the commercial LLM interfaces. All four participants rated PAWNI higher, with P2 and P4 assigning perfect scores of 100. Cohen's $d = 2.34$ (large). The commercial LLM interfaces scored in the ``Good'' range (mean 79.4, above the industry average of 68), whilst PAWNI scored in the ``Best Imaginable'' range. We do not regard this comparison as interpretable. A mean of 98.1 for a research prototype against 79.4 for mature commercial interfaces is more plausibly an artefact of novelty and demand characteristics than a usability finding; see Section~\ref{subsec:bias}. It is reported for completeness and no conclusion rests on it.

% ------------------------------------------------------------
\subsection{RQ2: Cognitive Load}
\label{subsec:results_rq2}

\subsubsection{Subjective Workload (NASA-TLX)}

\begin{figure}[!htbp]
\centering
\includegraphics[width=0.85\columnwidth]{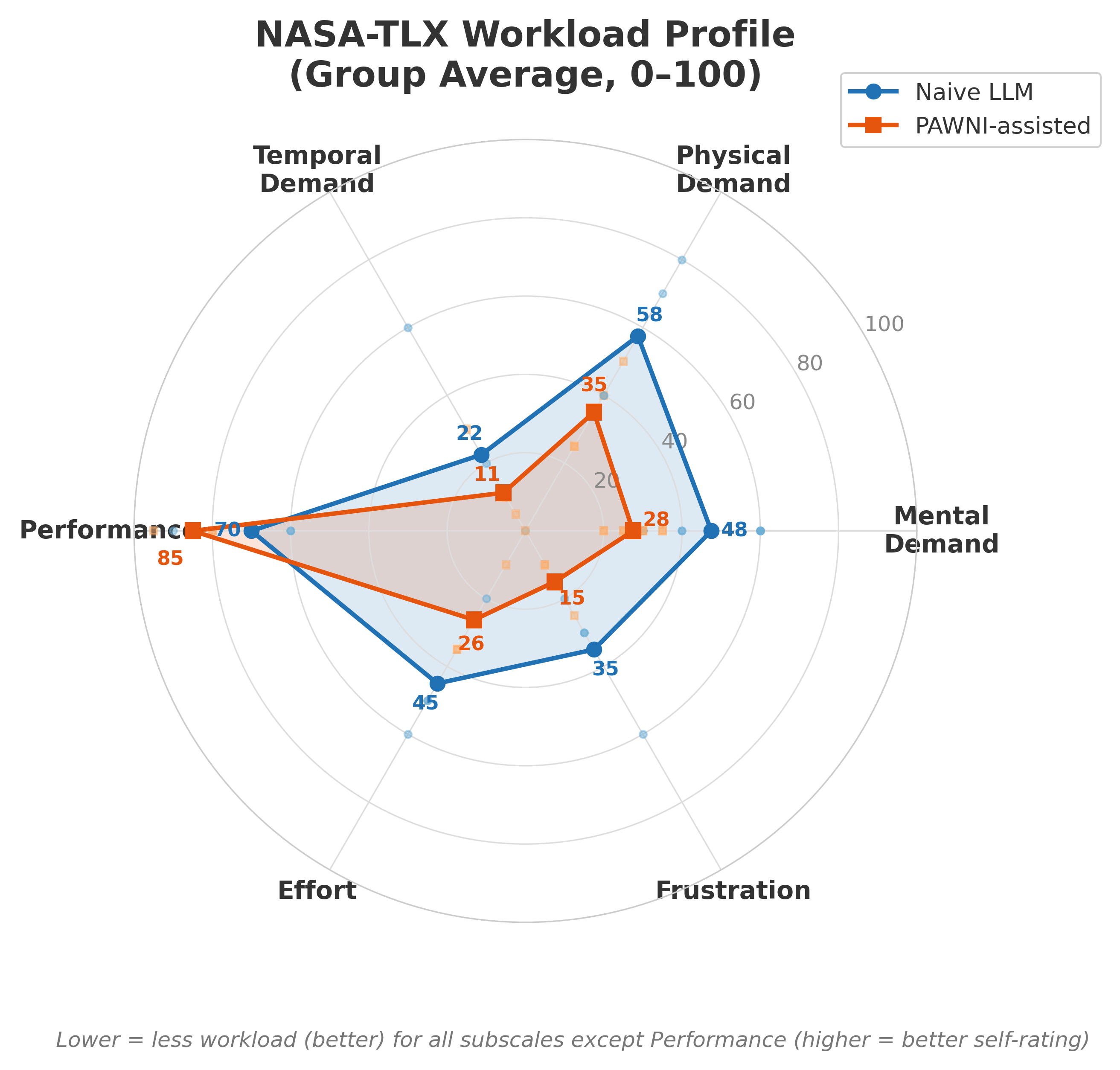}
\caption{NASA-TLX workload profile (group average, 0--100). For all demand sub-scales, lower values indicate less workload (better). For Performance, higher values indicate better self-rated performance. PAWNI (orange) shows a uniformly smaller workload footprint than Naive LLM (blue).}
\Description{Radar (spider) chart of group-average NASA-TLX sub-scale scores on a 0--100 scale, comparing Naive LLM (blue polygon) to PAWNI-assisted (orange polygon). The PAWNI polygon is smaller on demand-related axes and higher on the Performance axis, indicating lower workload and better perceived performance.}
\label{fig:nasa_radar}
\end{figure}

Figure~\ref{fig:nasa_radar} presents the group-average NASA-TLX workload profile. PAWNI-assisted interaction produced lower scores on all five demand subscales. The largest reductions were observed for Physical Demand ($57.5 \rightarrow 35.0$; $d = 1.33$), Mental Demand ($47.5 \rightarrow 27.5$; $d = 1.73$), and Effort ($45.0 \rightarrow 26.3$; $d = 1.27$). Performance self-rating (inverted scale: higher = better self-rated performance) improved from 70 to 85 ($d = 1.34$), indicating that participants felt more successful with PAWNI. Frustration decreased from 35.0 to 15.0 ($d = 1.51$), representing a 57\% reduction. Temporal Demand showed the smallest effect ($22.5 \rightarrow 11.3$; $d = 0.54$, medium), which is expected since PAWNI's pipeline introduces its own time demands even as it reduces iterative correction time.

\begin{figure}[!htbp]
\centering
\includegraphics[width=\columnwidth]{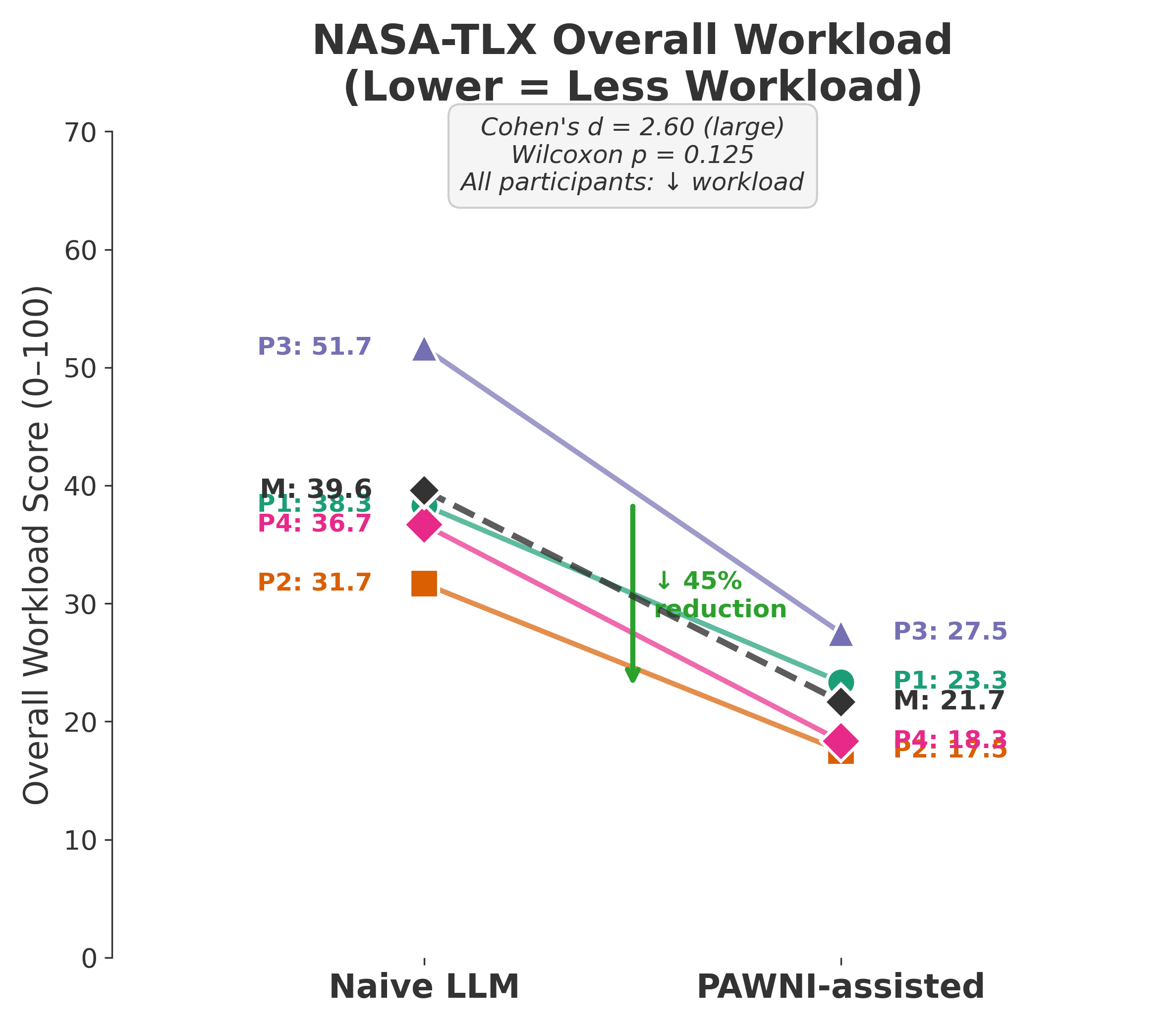}
\caption{NASA-TLX Overall Workload paired dot plot. Every participant showed reduced workload with PAWNI. Group mean decreased from 39.6 to 21.7 (45\% reduction). Cohen's $d = 2.60$ (large).}
\Description{Paired dot plot of NASA-TLX Overall Workload for each participant under Naive LLM and PAWNI-assisted conditions. Each participant has two points connected by a line; all lines slope downward from Naive to PAWNI, indicating reduced workload with PAWNI.}
\label{fig:nasa_dotplot}
\end{figure}

Figure~\ref{fig:nasa_dotplot} shows the Overall Workload score (unweighted mean of all six subscales) per participant. All four participants reported lower workload with PAWNI, with the group mean decreasing from 39.6 to 21.7---a 45\% reduction (Cohen's $d = 2.60$, large; Wilcoxon $p = 0.125$, the minimum achievable with $N = 4$ when all pairs show the same direction).

\subsubsection{EEG-Based Cognitive Load}

\begin{figure*}[!htbp]
\centering
\includegraphics[width=\textwidth]{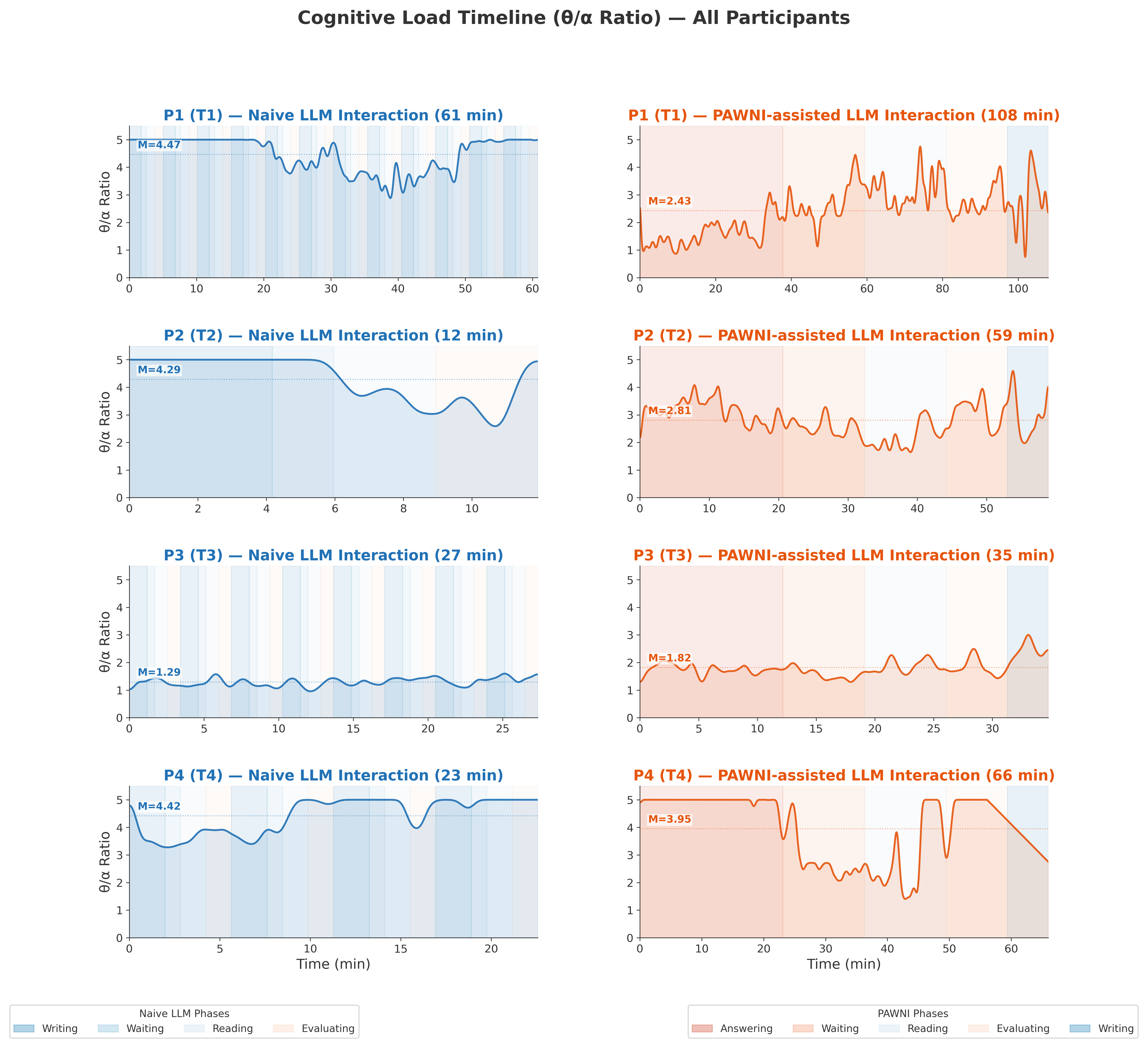}
\caption{Continuous cognitive load timelines ($\theta/\alpha$ ratio) for all participants. Left column: Condition~A (Naive LLM). Right column: Condition~B (PAWNI-assisted). Background colours indicate interaction phases (writing, waiting, reading, evaluating for Condition~A; answering, waiting, reading, evaluating for Condition~B). Mean $\theta/\alpha$ values (M) are shown per panel.}
\Description{Eight time-series panels (four participants by two conditions) showing the EEG cognitive-load index $\theta/\alpha$ over time. The left column shows Naive LLM sessions and the right column shows PAWNI-assisted sessions. Background shading marks interaction phases (e.g., writing/answering, waiting, reading, evaluating), and each panel reports its mean value.}
\label{fig:cl_timeline}
\end{figure*}

Figure~\ref{fig:cl_timeline} presents the continuous $\theta/\alpha$ ratio timelines for all four participants across both conditions. Condition~A exhibits sustained high cognitive load with oscillatory patterns corresponding to the write--wait--read--evaluate cycle of multi-turn conversation. Condition~B shows a qualitatively different profile: a moderate, variable cognitive load during the PAWNI Architect Q\&A phase, followed by reduced load during the LLM execution phase. Three of four participants (P1, P2, P4) showed lower mean $\theta/\alpha$ ratios in the PAWNI condition. P3 was the exception, showing an increase ($M = 1.55 \rightarrow 2.14$), which may reflect the more engaged cognitive processing elicited by the Architect's structured questions on a topic (quantum computing) where the participant had limited prior knowledge.

\begin{figure*}[!htbp]
\centering
\includegraphics[width=\textwidth]{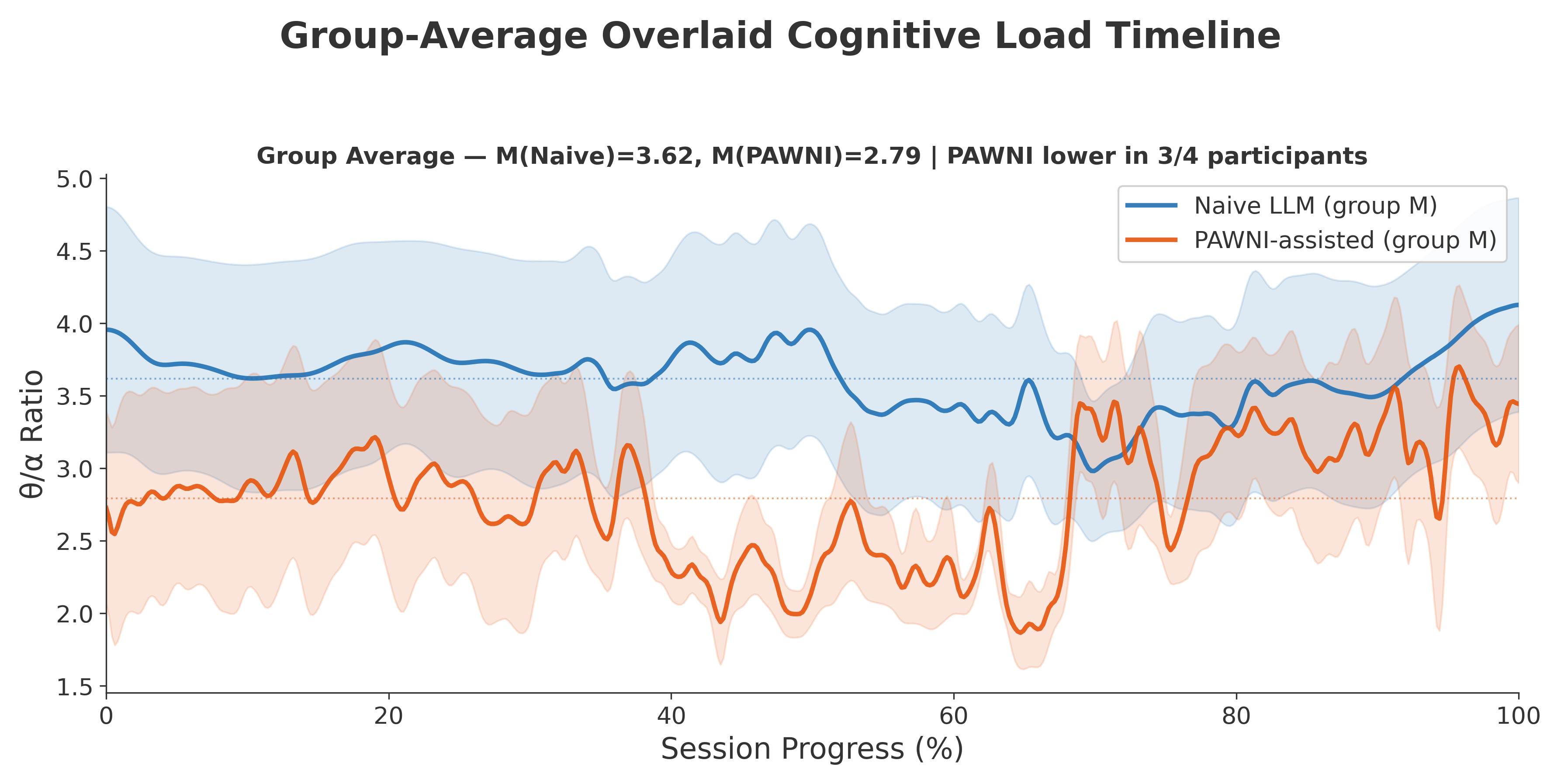}
\caption{Group-average overlaid cognitive load timeline. Both conditions are mapped to normalised session progress (0--100\%). The Naive LLM condition (blue) shows consistently higher $\theta/\alpha$ ratios. Shaded regions represent $\pm 1$ SD across participants.}
\Description{Line plot of group-average EEG cognitive load ($\theta/\alpha$ ratio) over normalized session progress from 0 to 100 percent. Two lines are overlaid: Naive LLM in blue and PAWNI-assisted in orange. Shaded bands around each line indicate plus/minus one standard deviation; the blue line remains higher across most of the session.}
\label{fig:cl_overlay_agg}
\end{figure*}

The group-average overlaid timeline (Figure~\ref{fig:cl_overlay_agg}) provides a direct visual comparison on a normalised session-progress axis. The Naive LLM condition (blue) maintains a consistently higher $\theta/\alpha$ ratio compared to the PAWNI condition (orange) across most of the session. At the session level, the mean $\theta/\alpha$ ratio was 3.39 (SD = 1.23) for Condition~A and 2.91 (SD = 0.76) for Condition~B, representing a 14\% reduction (Cohen's $d = 0.48$, small; Wilcoxon $p = 0.375$) \footnote{Two $\theta/\alpha$ summaries appear in this paper and are computed differently. \emph{ROI $\theta/\alpha$} (reported in the text and in Table~\ref{tab:summary}) is the ratio of mean frontal-core theta power to mean parietal-core alpha power over the whole session, computed once per session. \emph{Timeline $\theta/\alpha$} (annotated on Figures~\ref{fig:cl_timeline}, \ref{fig:cl_overlay_agg} and~\ref{fig:app_overlaid_cl}) is the mean of the continuous sliding-window index. The two differ because the sliding-window index is an average of ratios whilst the ROI index is a ratio of averages. Group means are 3.39 vs.\ 2.91 (ROI) and 3.62 vs.\ 2.79 (timeline); both show the same direction and comparable magnitude. All statistics in this paper use the ROI values.}. The modest session-level effect size reflects the influence of P3, who showed an increase rather than a decrease. Excluding P3, the remaining three participants exhibited a mean reduction of 21\% in their $\theta/\alpha$ ratios. The temporal visualisation reveals a pattern that session-level means obscure: the gap between conditions widens progressively towards the later portions of the session, as multi-turn fatigue accumulates in Condition~A whilst Condition~B transitions into the lower-demand LLM execution phase.

\begin{figure*}[!htbp]
\centering
\includegraphics[width=\textwidth]{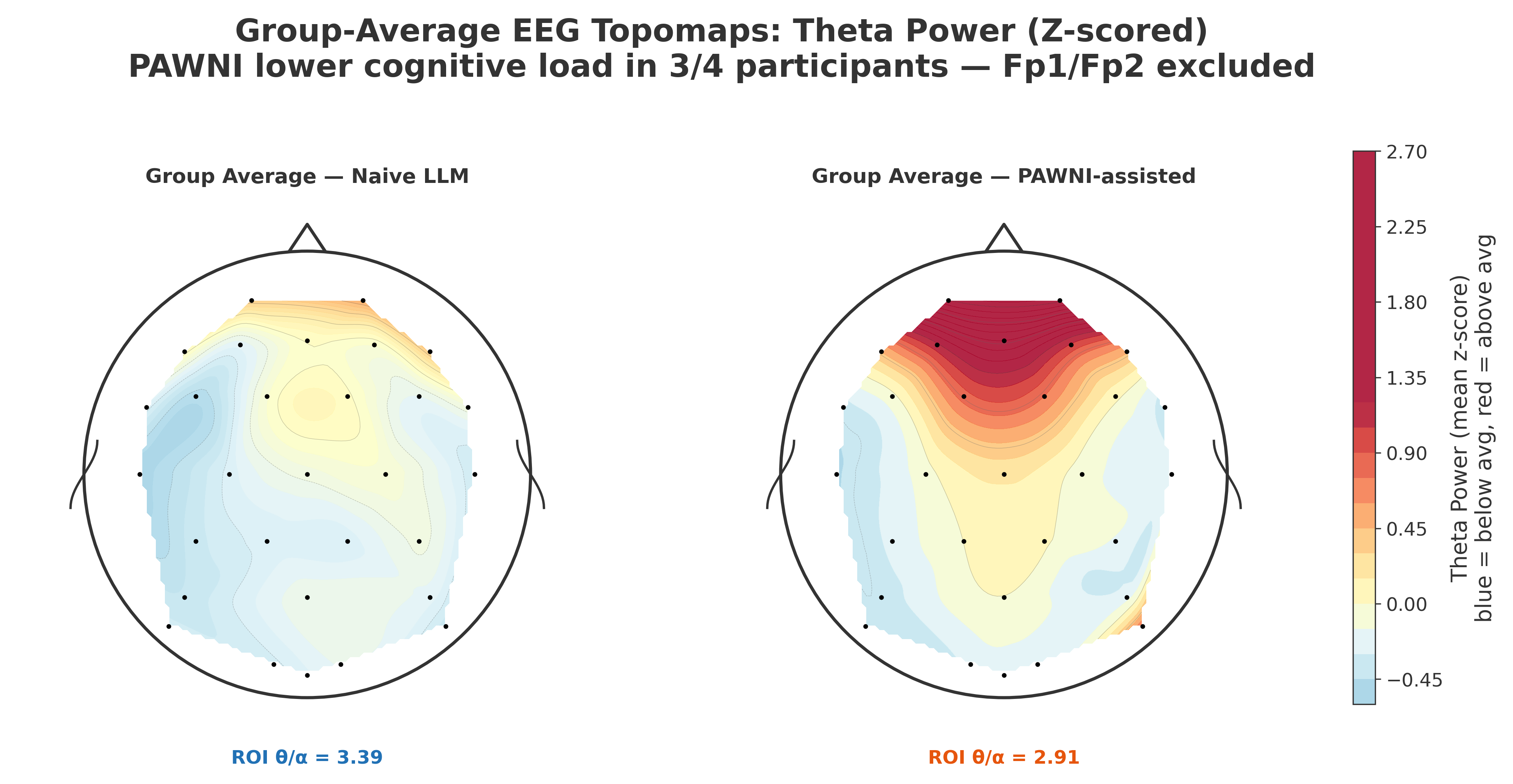}
\caption{Group-average EEG topographic maps of theta power (z-scored within participant). Left: Naive LLM. Right: PAWNI-assisted. Fp1/Fp2 channels excluded to avoid eye-movement artefacts. The PAWNI condition shows more prominent frontal theta concentration. ROI $\theta/\alpha$: 3.39 (Naive) vs.\ 2.91 (PAWNI).}
\Description{Two scalp topography heatmaps showing group-average theta-band power (z-scored within participant) for Naive LLM (left) and PAWNI-assisted (right) conditions. Frontal electrodes show relatively stronger theta activity in the PAWNI map; Fp1/Fp2 are omitted to reduce eye-movement artefact influence.}
\label{fig:topomaps_agg}
\end{figure*}

Topographic analysis (Figure~\ref{fig:topomaps_agg}) reveals the spatial distribution of theta power across the scalp. In the PAWNI condition, theta activity is concentrated in the frontal region, consistent with focused, task-directed cognitive processing. In the Naive LLM condition, frontal theta is relatively lower and more diffusely distributed, suggesting less organised cognitive engagement. The group-average ROI $\theta/\alpha$ ratio was 3.39 for Condition~A and 2.91 for Condition~B.

\begin{figure*}[!htbp]
\centering
\includegraphics[width=\textwidth]{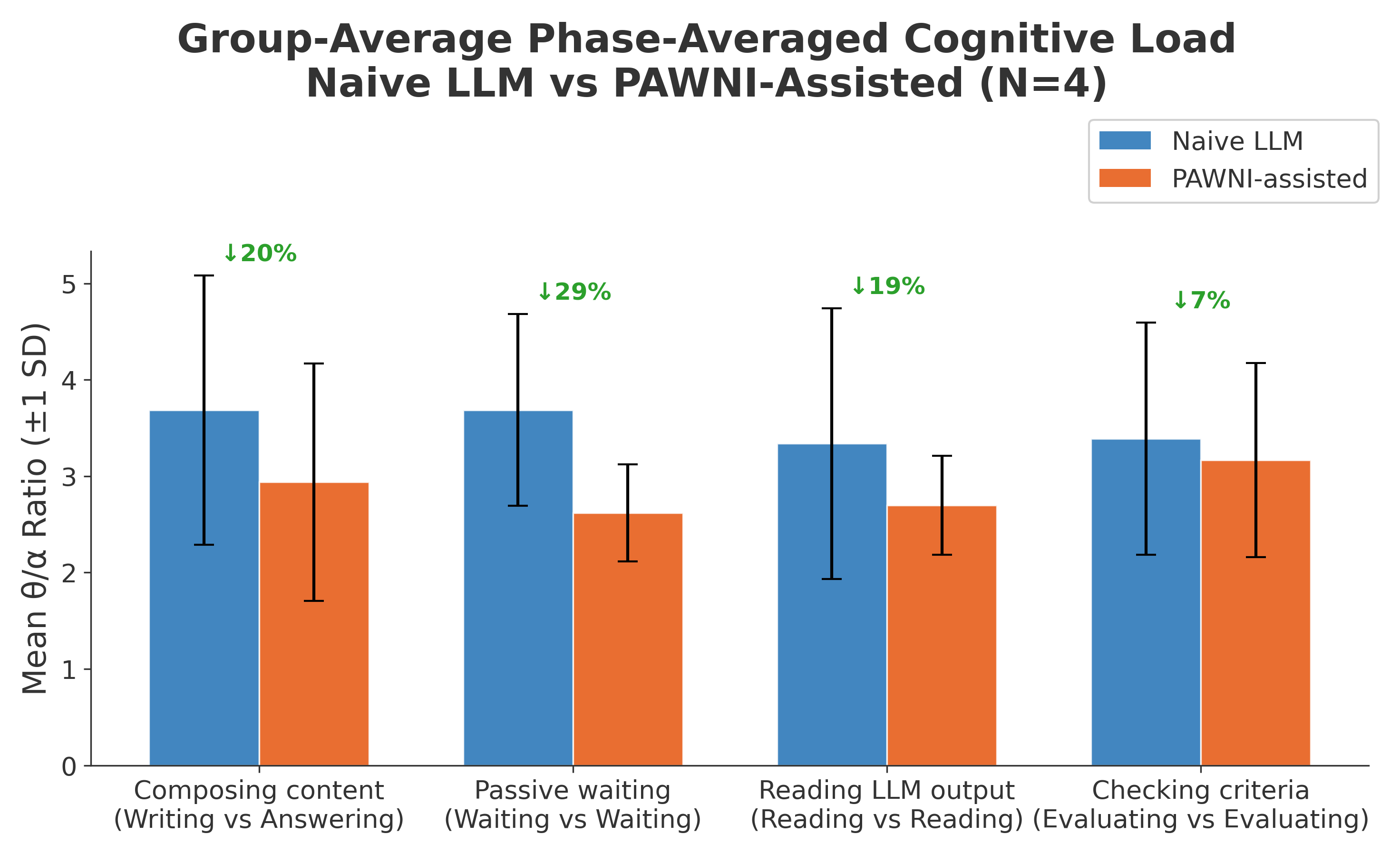}
\caption{Group-average phase-averaged cognitive load ($\theta/\alpha$ ratio, $\pm 1$ SD). Four functionally comparable phases are compared between conditions. Percentage labels indicate the reduction from Naive LLM to PAWNI. The largest reduction occurred during passive waiting and composing content.}
\Description{Grouped bar chart of phase-averaged EEG cognitive load ($\theta/\alpha$ ratio) across four interaction phases for Naive LLM and PAWNI-assisted conditions. Error bars show plus/minus one standard deviation, and percentage annotations indicate the reduction from Naive to PAWNI for each phase.}
\label{fig:phase_cl_agg}
\end{figure*}

Phase-averaged cognitive load (Figure~\ref{fig:phase_cl_agg}) compares the $\theta/\alpha$ ratio across four functionally comparable interaction phases. The largest reduction occurred during passive waiting ($\downarrow$29\%), followed by composing content ($\downarrow$20\%) and reading LLM output ($\downarrow$19\%). The checking criteria phase showed the smallest reduction ($\downarrow$7\%), which is expected since evaluating output against criteria requires similar cognitive effort regardless of how the prompt was generated.

\subsubsection{EEG--NASA-TLX Cross-Validation}

\begin{figure}[!htbp]
\centering
\includegraphics[width=\columnwidth]{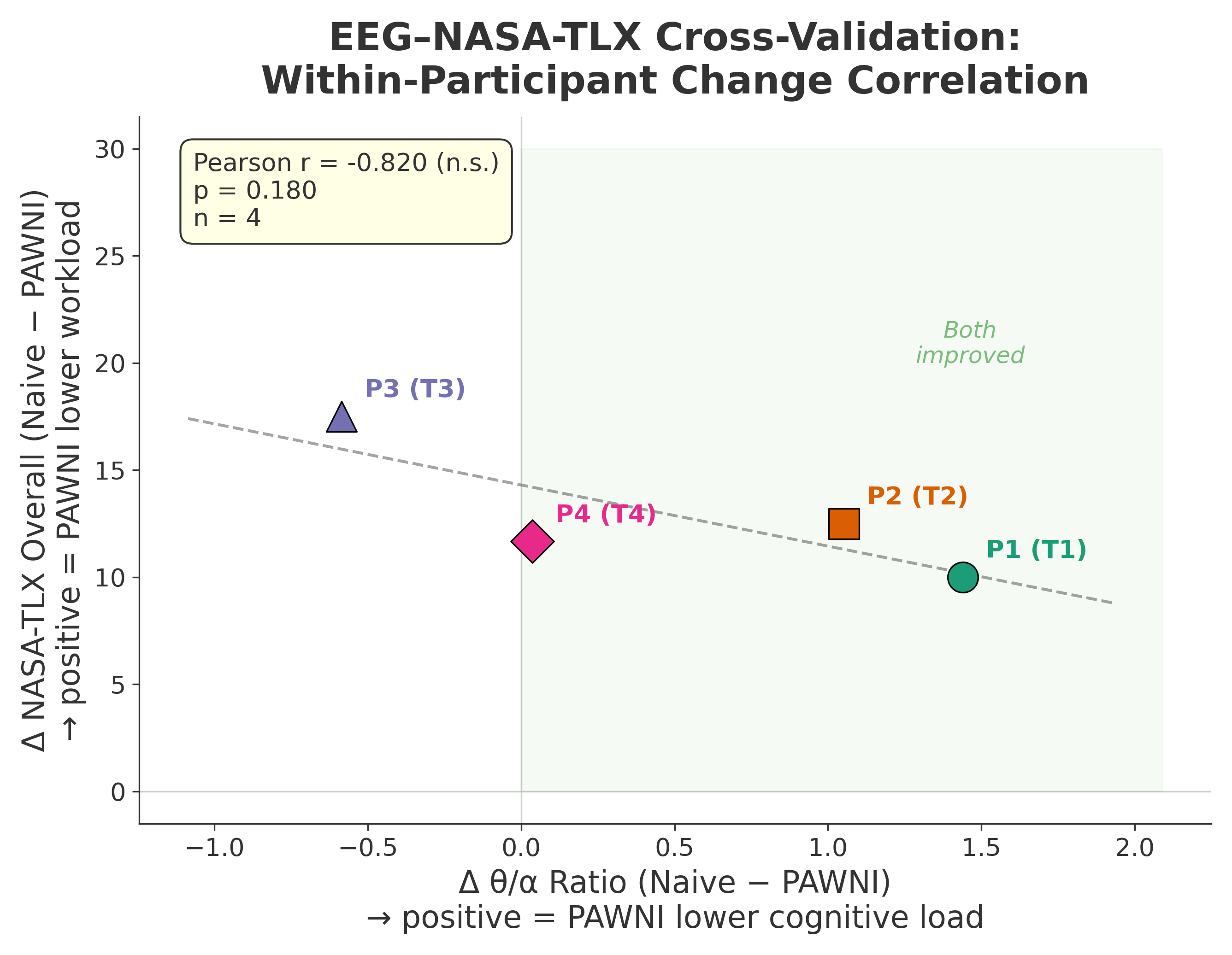}
\caption{EEG--NASA-TLX cross-validation scatter plot. Each point represents one participant. The $x$-axis shows the change in mean $\theta/\alpha$ ratio (positive = PAWNI lower cognitive load). The $y$-axis shows the change in NASA-TLX Overall Workload (positive = PAWNI lower workload). The upper-right quadrant (green shading) indicates improvement on both measures.}
\Description{Scatter plot with one point per participant showing the relationship between change in EEG cognitive load ($\theta/\alpha$, x-axis; positive means PAWNI reduced load) and change in NASA-TLX overall workload (y-axis; positive means PAWNI reduced workload). A shaded upper-right region highlights participants who improved on both measures.}
\label{fig:eeg_nasa}
\end{figure}

To assess convergent validity between objective (EEG) and subjective (NASA-TLX) measures of cognitive load, we examined both within-participant change scores and raw value correlations. Figure~\ref{fig:eeg_nasa} plots the change in $\theta/\alpha$ ratio ($\Delta_{\theta/\alpha}$ = Naive $-$ PAWNI; positive = PAWNI lower) against the change in NASA-TLX Overall Workload ($\Delta_{\text{TLX}}$ = Naive $-$ PAWNI; positive = PAWNI lower). All four participants fall in the upper half of the plot ($\Delta_{\text{TLX}} > 0$), confirming that every participant reported lower subjective workload with PAWNI. Three of four participants (P1, P2, P4) also show $\Delta_{\theta/\alpha} > 0$, placing them in the upper-right quadrant where both measures agree. The Pearson correlation between the two change scores is $r = -0.82$ ($p = 0.18$, non-significant at $N = 4$). The negative correlation indicates that participants with the largest EEG-measured reductions tended to show comparatively smaller NASA-TLX reductions, and vice versa. A secondary Spearman correlation on raw values (8 data points: 4 participants $\times$ 2 conditions) yielded $\rho = -0.08$ ($p = 0.84$), indicating no linear relationship between absolute $\theta/\alpha$ levels and absolute NASA-TLX scores. This dissociation between objective and subjective measures is not uncommon in cognitive load research: EEG captures moment-to-moment neural resource allocation, whereas NASA-TLX reflects a retrospective, holistic assessment that integrates both effort magnitude and perceived productivity~\cite{antonenko2010using, raufi2022evaluation}.

% ------------------------------------------------------------
\subsection{RQ3: Interaction Efficiency}
\label{subsec:results_rq3}

\subsubsection{Conversational Turns}

\begin{figure}[!htbp]
\centering
\includegraphics[width=\columnwidth]{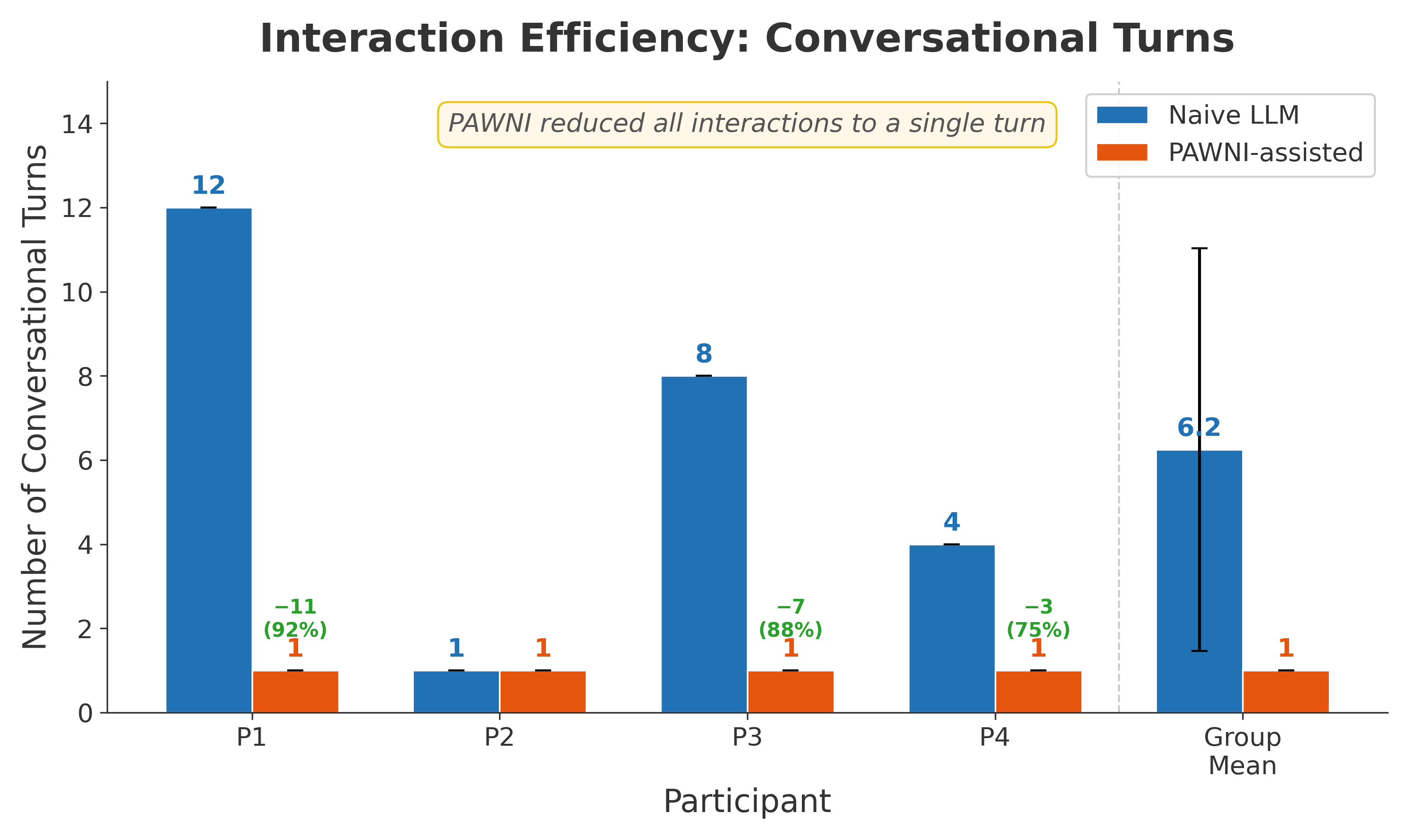}
\caption{Number of conversational turns per participant. PAWNI reduced all interactions to a single LLM turn. Percentage labels indicate the turn reduction for each participant.}
\Description{Bar chart showing the number of conversational turns per participant for Naive LLM versus PAWNI-assisted prompting. PAWNI bars are all at one turn, while Naive LLM bars vary across participants; annotations show percent reduction in turns.}
\label{fig:turns}
\end{figure}

Figure~\ref{fig:turns} shows that PAWNI reduced all interactions to a single LLM turn. In Condition~A, participants used between 1 and 12 turns ($M = 6.25$, SD = 4.79); in Condition~B, every participant required exactly 1 turn to obtain a satisfactory output ($d = 1.55$, large). P1, who used 12 turns naively (including 5 correction turns), exemplifies the trial-and-error pattern that PAWNI eliminates. Across participants, correction turns---messages explicitly addressing model errors or redirecting the output---averaged 2.75 (SD = 2.22) in Condition~A and 0 in Condition~B ($d = 1.75$, large). P2 was the only participant who did not require correction turns in either condition, having submitted a single lengthy prompt in Condition~A; P2 therefore contributes no within-participant improvement to the turn-count measures, which is why their direction consistency is 3/4 rather than 4/4.

\subsubsection{Time-on-Task}

\begin{figure*}[!htbp]
\centering
\includegraphics[width=\textwidth]{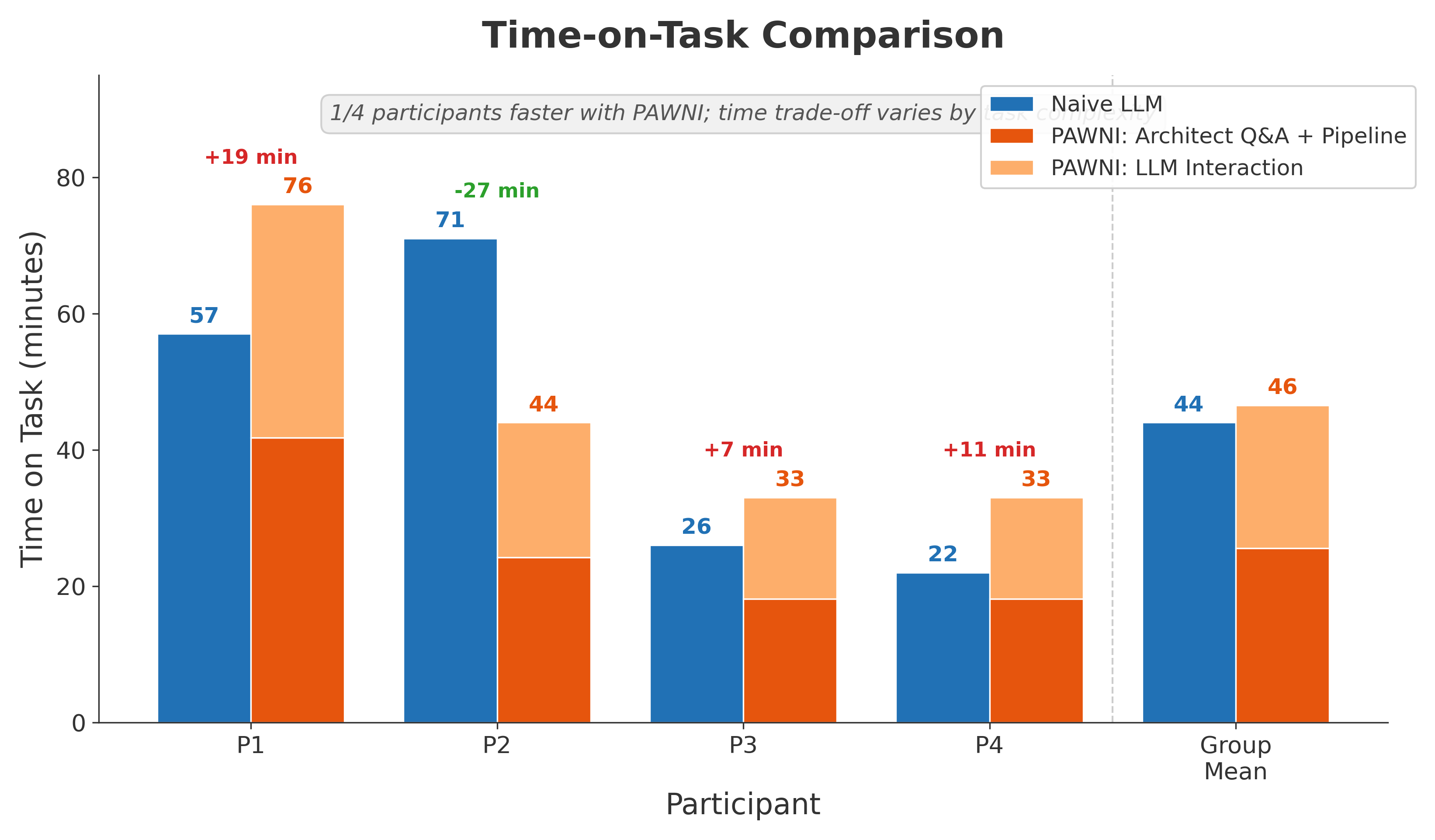}
\caption{Time-on-task comparison. Blue bars: Condition~A total time. Orange (dark): PAWNI pipeline time (Architect Q\&A + agent processing). Orange (light): PAWNI LLM interaction time. Red/green labels indicate time difference. P2 saved 27 minutes; P1 spent 19 minutes more.}
\Description{Stacked bar chart comparing time-on-task by participant. For Naive LLM, a single blue bar shows total time. For PAWNI, the time is split into dark-orange pipeline time and light-orange LLM interaction time. Text labels indicate the net time difference between conditions for each participant.}
\label{fig:time}
\end{figure*}

Time-on-task results (Figure~\ref{fig:time}) reveal a nuanced pattern. The group-average total time was comparable: 44.0 minutes (SD = 23.9, Condition~A) versus 46.5 minutes (SD = 20.3, Condition~B); Cohen's $d = -0.11$, negligible. However, these averages mask substantial individual variation. P2, who had the most complex task (30-day travel itinerary) and used only 1 naive turn (submitting one long prompt), spent 71 minutes in Condition~A but only 44 minutes with PAWNI---a 27-minute saving. P1, conversely, spent 19 minutes longer with PAWNI (76 vs.\ 57 minutes), attributable to an extended Architect Q\&A session on a technically demanding task (mobile app specification).

The stacked bars in Figure~\ref{fig:time} decompose PAWNI time into pipeline time (Architect Q\&A + agent processing) and LLM interaction time. Across all participants, the LLM interaction phase in Condition~B was substantially shorter than the total Condition~A time, consistent with the PAWNI-generated prompt reducing the need for iterative refinement. The pipeline overhead represents an upfront investment in intent clarification that is repaid through reduced downstream iteration.

\subsubsection{Prompt Structural Completeness}

\begin{figure}[!htbp]
\centering
\includegraphics[width=\columnwidth]{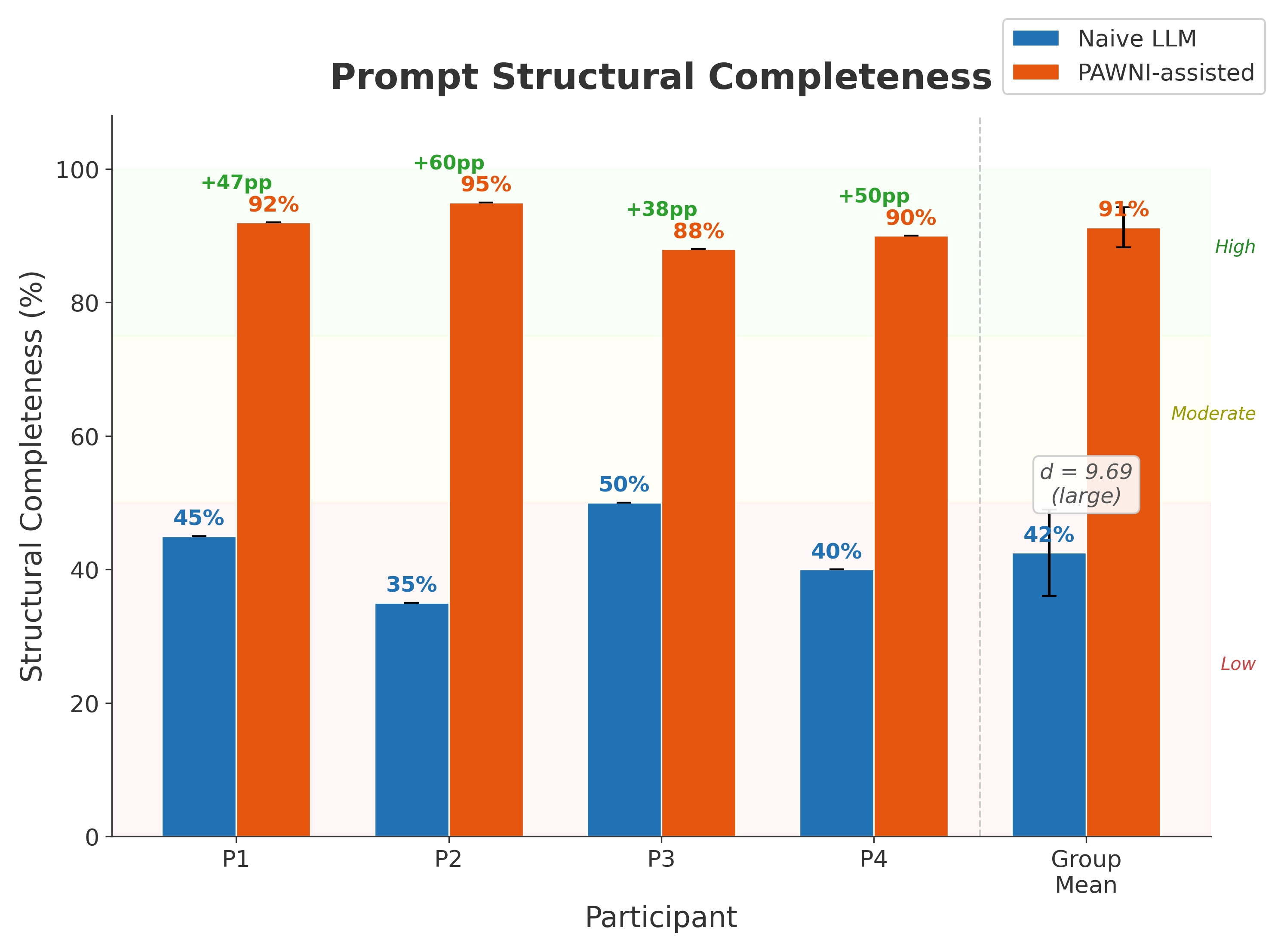}
\caption{Prompt structural completeness (percentage of 10 assessed elements present). Naive prompts (blue) averaged 42\%; PAWNI-generated prompts (orange) averaged 91\%. Cohen's $d = 9.69$ (very large). Shaded bands indicate Low, Moderate, and High completeness zones.}
\Description{Bar chart comparing prompt structural completeness (percent of assessed elements present) for Naive prompts versus PAWNI-generated prompts, shown per participant and/or as summary values. Background shading indicates qualitative zones (low, moderate, high completeness), and PAWNI bars fall in the high zone while Naive bars remain low to moderate.}
\label{fig:completeness}
\end{figure}

Figure~\ref{fig:completeness} presents the structural completeness of prompts in both conditions. Naive prompts scored a mean of 42.5\% completeness (SD = 6.5, range: 35--50\%), whilst PAWNI-generated prompts scored a mean of 91.3\% (SD = 3.0, range: 88--95\%). The improvement ranged from +38 to +60 percentage points across participants (Cohen's $d = 9.69$, very large; bootstrap 95\% CI: [41.0, 56.8]). In this sample, user-authored prompts remained structurally incomplete even after multiple turns of refinement---lacking elements such as explicit constraints, audience specification, and output format---whilst PAWNI systematically ensures the presence of nearly all 18 prompt elements.

Two additional prompt quality metrics corroborate this finding. Counting all user-authored text in Condition~A against the final Scribe output in Condition~B, prompt length increased from a mean of 538 words (SD = 553, range 138--1{,}353) to 2{,}251 words (SD = 287, range 1{,}998--2{,}530); $d_s = 3.89$, bootstrap 95\% CI [1{,}011, 2{,}234]. The large Condition~A standard deviation is itself informative: P1 wrote 1{,}353 words across twelve turns whilst P2 wrote 138 in one, yet both produced prompts scoring below 50\% on structural completeness. Length and completeness are evidently not the same thing---a point that argues against reading the PAWNI prompts' greater length as the operative variable. Prompt specificity---the ratio of domain-specific terminology to total terms, computed via TF-IDF against a general-language corpus---increased from 0.29 (SD = 0.05) to 0.79 (SD = 0.03); $d = 11.50$, very large. This indicates that PAWNI prompts contain markedly more domain-relevant vocabulary, which narrows the model's search space and reduces the likelihood of generic or off-target responses.

% ------------------------------------------------------------
\subsection{Supplementary Findings}
\label{subsec:results_supplementary}

\subsubsection{Topographic Difference Maps (RQ5)}

\begin{figure*}[!htbp]
\centering
\includegraphics[width=\textwidth]{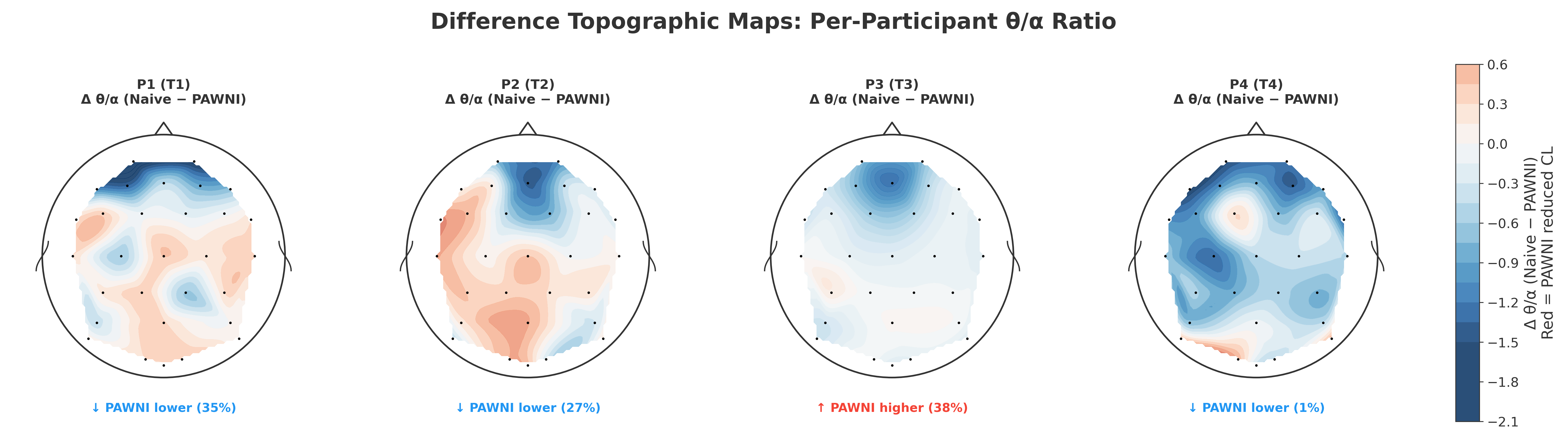}
\caption{Per-participant difference topographic maps ($\Delta\,\theta/\alpha$ ratio, Naive $-$ PAWNI). Blue indicates PAWNI reduced cognitive load; red indicates increase. Three of four participants show predominant frontal reductions.}
\Description{Set of four scalp topography difference maps, one per participant, showing the change in $\theta/\alpha$ ratio between conditions (Naive minus PAWNI). Blue regions indicate lower cognitive load with PAWNI and red regions indicate higher load with PAWNI; most maps show blue in frontal areas.}
\label{fig:diff_topomaps}
\end{figure*}

Per-participant difference topographic maps (Figure~\ref{fig:diff_topomaps}) visualise the spatial distribution of $\theta/\alpha$ ratio change (Naive $-$ PAWNI) across the scalp. Blue regions indicate lower cognitive load in the PAWNI condition; red regions indicate higher. P1 and P2 show broad frontal reductions (35\% and 27\% respectively). P3 shows a 38\% increase, consistent with the elevated engagement observed in the continuous timeline data. P4 shows a marginal 1\% reduction. The predominant pattern across participants is a frontal reduction in cognitive load with PAWNI, consistent with reduced demand on the dorsolateral prefrontal cortex during structured (as opposed to open-ended) prompt formulation.

% ------------------------------------------------------------
\subsection{User Experience and Process Assessment}
\label{subsec:results_ux}

\begin{figure}[!htbp]
\centering
\includegraphics[width=\columnwidth]{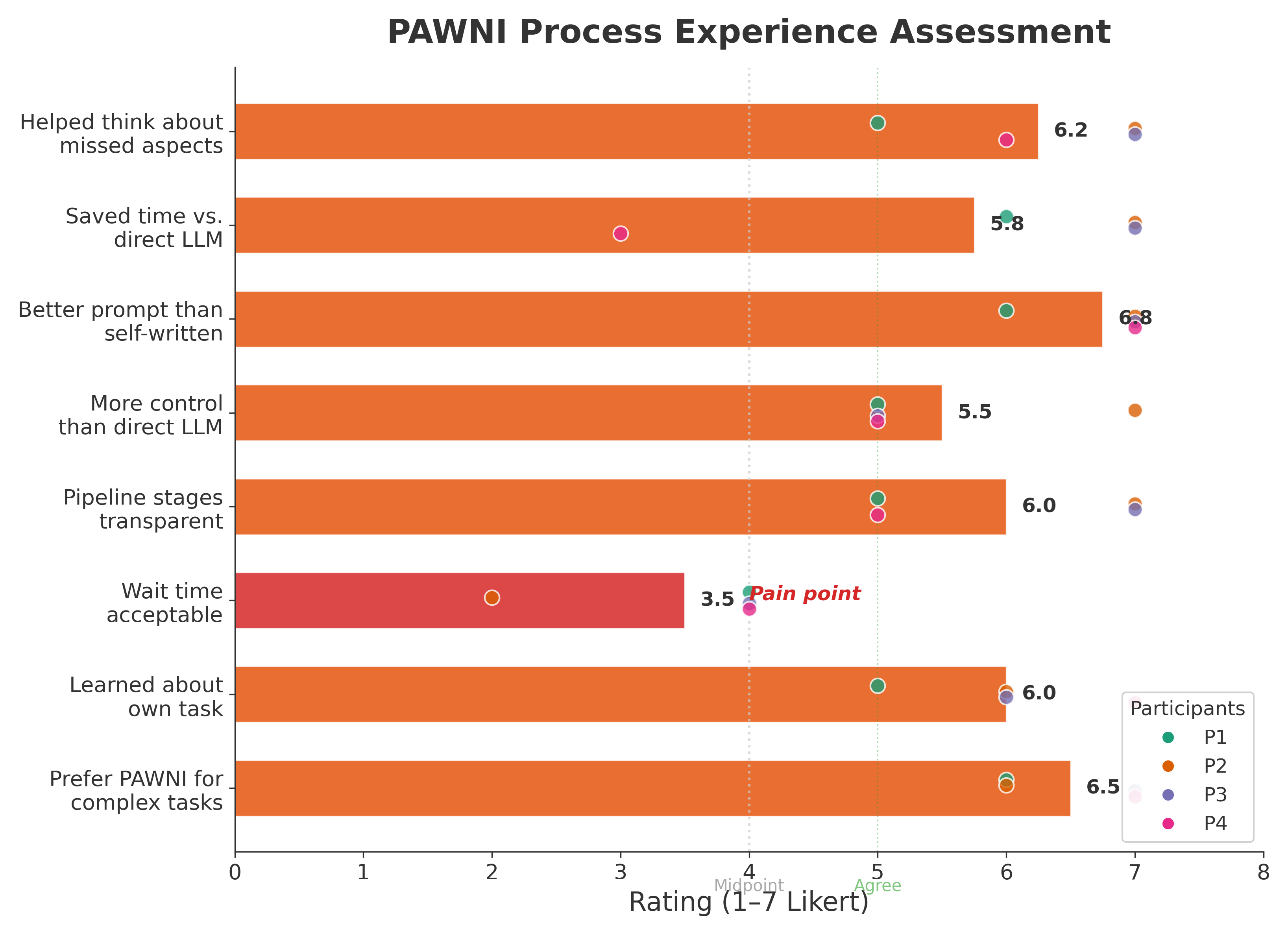}
\caption{PAWNI process experience assessment (1--7 Likert, higher = more agreement). Individual participant dots are overlaid. Wait time acceptability ($M = 3.5$) was the only item below the midpoint, indicating pipeline latency as a key improvement area.}
\Description{Likert-scale summary plot of participants' ratings (1--7) for multiple PAWNI process-experience statements. Mean markers are shown for each item with individual participant points overlaid; the wait-time acceptability item is visibly lower than the others.}
\label{fig:process_experience}
\end{figure}

The User Satisfaction Questionnaire (Section~B) assessed PAWNI-specific process experience. Figure~\ref{fig:process_experience} shows that participants rated PAWNI favourably across most dimensions. The highest-rated item was ``Better prompt than self-written'' ($M = 6.8$/7), followed by ``Prefer PAWNI for complex tasks'' ($M = 6.5$/7) and ``Helped think about missed aspects'' ($M = 6.2$/7). The lowest-rated item was ``Wait time acceptable'' ($M = 3.5$/7), identifying pipeline latency as the primary pain point for user experience improvement.

\begin{figure}[!htbp]
\centering
\includegraphics[width=\columnwidth]{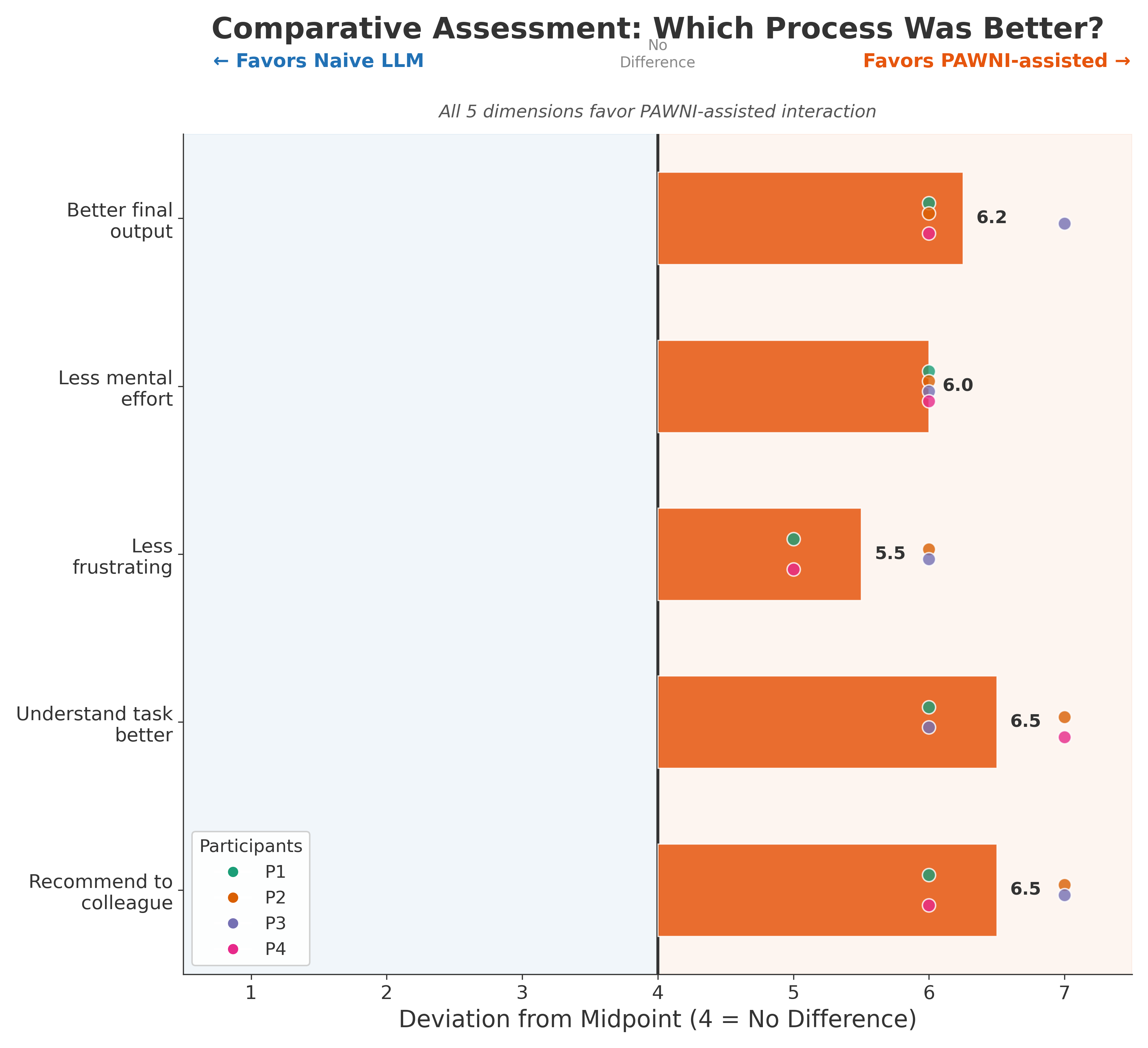}
\caption{Comparative assessment: which process was better? (1 = Naive LLM much better, 4 = no difference, 7 = PAWNI much better). All five dimensions favour PAWNI. The vertical line at 4 indicates the no-difference midpoint.}
\Description{Comparative Likert-scale plot (1--7) where values above 4 favor PAWNI and values below 4 favor Naive LLM. Points or bars for five comparison dimensions all lie to the right of a vertical reference line at 4, indicating participants preferred PAWNI across all dimensions.}
\label{fig:comparative}
\end{figure}

The comparative assessment (Section~C of the questionnaire) asked participants to directly compare the two conditions on a 1--7 scale where 4 indicates no difference, values above 4 favour PAWNI, and values below 4 favour the Naive LLM. Figure~\ref{fig:comparative} shows that all five comparison dimensions favoured PAWNI. The strongest preferences were for ``Helped understand task better'' ($M = 6.5$) and ``Would recommend to colleague'' ($M = 6.5$). ``Less frustrating'' received the lowest comparative advantage ($M = 5.5$), likely reflecting the pipeline wait times noted above.

% ------------------------------------------------------------
\subsection{Summary of Key Findings}
\label{subsec:results_summary}

Table~\ref{tab:summary} consolidates the primary metrics across all research questions. Of the nine key variables (excluding time-on-task, which showed no consistent direction), six improved in the same direction for all four participants and three did so for three of four; all nine had effect sizes in the conventionally large range ($d > 0.8$) with the exception of the session-level EEG $\theta/\alpha$ ratio. The three variables at 3/4 rather than 4/4 are the session $\theta/\alpha$ ratio and the two turn-count measures, the latter because P2 required only a single turn in \emph{both} conditions and therefore could not improve on them. The session-level EEG $\theta/\alpha$ ratio showed a small effect ($d = 0.48$) with 3/4 directional consistency, though the temporal and phase-level analyses reported above reveal clearer separation. Time-on-task was the only variable without a consistent directional effect, reflecting PAWNI's design trade-off: upfront pipeline investment reduces downstream iteration, but the net time impact depends on task complexity and individual prompting behaviour.

\begin{table*}[!htbp]
\centering
\caption{Summary of key findings across all research questions. Direction indicates how many of four participants improved with PAWNI. $\uparrow$ = higher is better; $\downarrow$ = lower is better. $d_s$ is Cohen's $d$ computed from the pooled standard deviation (the convention used throughout the text); $g_s$ applies the small-sample correction $J = 1 - 3/(4\,df - 1) = 0.727$ for $df = 3$; $d_z$ is the paired-samples effect size computed from the standard deviation of the within-participant differences. Bootstrap 95\% CIs (10{,}000 resamples) are for the paired mean difference. The three estimators disagree by factors of up to 1.8 on the same data; this divergence is itself a property of $N = 4$ and is discussed in Section~\ref{subsec:effectsize}.}
\label{tab:summary}
\small
\begin{tabular}{l c c c c c c c}
\toprule
\textbf{Variable} & \textbf{Naive} & \textbf{PAWNI} & \textbf{$\Delta$} & \textbf{Dir.} & \textbf{$d_s$} & \textbf{$g_s$} & \textbf{Bootstrap 95\% CI} \\
\midrule
Task criteria quality (1--5) $\uparrow$ & 3.12 $\pm$ 0.84 & 4.56 $\pm$ 0.52 & +1.44 & 4/4 & 2.07 & 1.51 & [0.85, 2.40] \\
Output quality (1--7 avg) $\uparrow$ & 4.04 $\pm$ 0.44 & 6.50 $\pm$ 0.21 & +2.46 & 4/4 & 4.24 & 3.08 & [2.00, 2.90] \\
SUS score (0--100) $\uparrow$ & 79.4 $\pm$ 11.1 & 98.1 $\pm$ 2.4 & +18.8 & 4/4 & 2.34 & 1.70 & [11.3, 26.3] \\
NASA-TLX workload (0--100) $\downarrow$ & 39.6 $\pm$ 8.5 & 21.7 $\pm$ 4.7 & $-$17.9 & 4/4 & 2.60 & 1.89 & [$-$21.9, $-$14.6] \\
Session $\theta/\alpha$ ratio (EEG) $\downarrow$ & 3.39 $\pm$ 1.23 & 2.91 $\pm$ 0.76 & $-$0.48 & 3/4 & 0.48 & 0.35 & [$-$1.24, 0.28] \\
LLM turns (count) $\downarrow$ & 6.25 $\pm$ 4.79 & 1.0 $\pm$ 0.0 & $-$5.25 & 3/4 & 1.55 & 1.13 & [$-$9.00, $-$1.50] \\
Correction turns $\downarrow$ & 2.75 $\pm$ 2.22 & 0.0 $\pm$ 0.0 & $-$2.75 & 3/4 & 1.75 & 1.27 & [$-$4.50, $-$1.00] \\
Time-on-task (min) $\downarrow$ & 44.0 $\pm$ 23.9 & 46.5 $\pm$ 20.3 & +2.5 & 1/4 & $-$0.11 & $-$0.08 & [$-$17.5, 16.0] \\
Prompt completeness (\%) $\uparrow$ & 42.5 $\pm$ 6.5 & 91.3 $\pm$ 3.0 & +48.8 & 4/4 & 9.69 & 7.05 & [41.0, 56.8] \\
Prompt specificity (0--1) $\uparrow$ & 0.29 $\pm$ 0.05 & 0.79 $\pm$ 0.03 & +0.50 & 4/4 & 11.50 & 8.36 & --- \\
\bottomrule
\end{tabular}
\end{table*}

% ------------------------------------------------------------
\subsection{Component-Level Evidence from Pipeline Logs}
\label{subsec:component}

The measures reported so far characterise the pipeline as a whole. To provide some visibility into the individual agents, we additionally extracted structured records from the complete interaction logs of all four PAWNI sessions. These are process measures rather than outcome measures: they describe what each agent did, not what difference it made. Table~\ref{tab:component} summarises them.

Three observations follow. First, the Scout's behaviour was stable across four unrelated domains---code generation, travel logistics, curriculum design and clinical protocol design---returning between 54 and 63 task-level knowledge points and between 62 and 77 prompt-specific points, always distributed across the same five categories (best practices, dos, don'ts, additional details, current trends). This consistency is what the pattern-file design assumes but had not previously been demonstrated outside development.

Second, the Architect's question load scaled with the under-specification of the initial request rather than being fixed. The two participants whose seed prompts were already detailed (P1, 835 words; P4, 310 words) completed in three rounds, whilst the participant whose seed prompt contained an internally infeasible constraint (P2) required five rounds, two of which were spent resolving that constraint. Across sessions the Architect presented a mean of 36.0 questions ($SD = 7.1$) over 3.75 rounds ($SD = 1.0$), or 9.6 questions per round, consistent with the design target described in Section~\ref{subsec:architect}.

Third, and most relevant to the question of whether a self-evaluating pipeline can be trusted to evaluate itself, the Judge scored the system's own first-stage output harshly. Across the four sessions the Forge's ideal-response draft, assessed against the 18-element rubric, received a mean of 3.97/10 ($SD = 0.75$), with the recurring criticism that the draft, read as a prompt, lacked an explicit role, task directive and output specification. The Scribe's polished prompt was then scored at 9.03, 9.32 and 9.47 in the three sessions for which the second-pass record was retained. P4's second-pass score was not preserved in the log; because the pipeline terminated after a single pass, it necessarily exceeded the 8/10 continuation threshold. The evaluation loop converged in one iteration in all four sessions, against a configured maximum of three. A multi-model judge that assigned a mean of 3.97/10 to its own pipeline's intermediate output is not behaving as a rubber stamp, which is a necessary---though not sufficient---condition for the quality-assurance loop to be meaningful.

Finally, the surface structure of the resulting prompts differed categorically rather than by degree. Each PAWNI-generated prompt was organised into explicitly labelled sections (mean 22.5, range 17--32), including named Role, Context, Goal, Task, Output Format and Constraints blocks. None of the four participants' naive prompt sequences contained any labelled section at all; they were continuous prose. This is the mechanism underlying the structural-completeness difference reported in Section~\ref{subsec:results_rq3}, observed directly rather than through rater judgement.

\begin{table*}[!htbp]
\centering
\caption{Component-level process measures extracted from the four complete PAWNI session logs. These describe agent behaviour; they are not outcome measures and no causal attribution to individual agents is implied. $^{\dagger}$P4's second-pass Judge record was not retained; the pipeline terminated after one pass, which requires a score above the 8/10 threshold.}
\label{tab:component}
\small
\begin{tabular}{l c c c c c}
\toprule
\textbf{Measure (agent)} & \textbf{P1} & \textbf{P2} & \textbf{P3} & \textbf{P4} & \textbf{M $\pm$ SD} \\
\midrule
Task-level knowledge points (Scout) & 56 & 63 & 54 & 54 & 56.8 $\pm$ 4.3 \\
Prompt-specific knowledge points (Scout) & 65 & 64 & 62 & 77 & 67.0 $\pm$ 6.8 \\
Knowledge categories returned (Scout) & 5 & 5 & 5 & 5 & 5.0 $\pm$ 0.0 \\
Q\&A rounds (Architect) & 3 & 5 & 4 & 3 & 3.8 $\pm$ 1.0 \\
Questions presented (Architect) & 32 & 46 & 36 & 30 & 36.0 $\pm$ 7.1 \\
Judge score, Forge draft (0--10) & 3.00 & 4.73 & 4.33 & 3.83 & 3.97 $\pm$ 0.75 \\
Judge score, Scribe prompt (0--10) & 9.03 & 9.32 & 9.47 & n/r$^{\dagger}$ & 9.27 $\pm$ 0.22 \\
Evaluation-loop iterations used (max 3) & 1 & 1 & 1 & 1 & 1.0 $\pm$ 0.0 \\
Labelled sections in final prompt (Scribe) & 19 & 32 & 17 & 22 & 22.5 $\pm$ 6.7 \\
Labelled sections in naive prompt sequence & 0 & 0 & 0 & 0 & 0.0 $\pm$ 0.0 \\
Seed prompt length (words) & 835 & 213 & 120 & 310 & 369.5 $\pm$ 319.9 \\
Final prompt length (words) & 2{,}009 & 2{,}530 & 1{,}998 & 2{,}466 & 2{,}251 $\pm$ 287 \\
\bottomrule
\end{tabular}
\end{table*}

% ============================================================
% BATCH 5: §7 Discussion + §8 Conclusion + References + Appendix
% Paper: Asking Questions the Right Way
% Target: CUI 2026 (ACM Conference)
% ============================================================

\section{Discussion}
\label{sec:discussion}

% ------------------------------------------------------------
\subsection{Better Questions Lead to Better Answers}
\label{subsec:disc_quality}

The results are consistent with---though at this sample size they cannot confirm---the hypothesis that structurally complete, context-rich prompts yield higher-quality LLM outputs. Across all four participants and all six quality dimensions measured, PAWNI-assisted outputs were rated higher than their naive counterparts. The effect sizes are uniformly large ($d > 2.0$), and the direction consistency is 4/4 for every output quality variable.

The mediating mechanism is prompt structural completeness. Naive prompts scored a mean of 42\% on structural completeness---meaning that over half of the essential prompt elements were absent even after multiple rounds of iterative refinement. PAWNI-generated prompts scored 91\%. This gap directly explains why the first LLM response in Condition~B was rated 6.5/7 for quality, compared to 3.8/7 in Condition~A: the LLM received a comprehensive specification in a single turn rather than an evolving, fragmented set of instructions across multiple turns.

This finding has a practical implication for conversational AI design. The current paradigm treats prompt engineering as the user's responsibility. PAWNI is an existence proof that this responsibility \emph{can} be offloaded to an agentic system that knows \textit{what to ask} and \textit{in what order to ask it}, guided by domain-specific knowledge accumulated over prior interactions. Whether it \emph{should} be, and for whom, is taken up in Section~\ref{subsec:beneficiaries}.

% ------------------------------------------------------------
\subsection{Cognitive Load Redistribution, Not Elimination}
\label{subsec:disc_cl}

The EEG and NASA-TLX data paint a nuanced picture of cognitive load under the two conditions. Subjective workload, as measured by NASA-TLX, showed a large and consistent reduction ($d = 2.60$; 4/4 participants). The EEG-derived $\theta/\alpha$ ratio, by contrast, showed a small session-level effect ($d = 0.48$; 3/4 participants), with P3 exhibiting an increase. This dissociation between subjective and neuro-physiological measures warrants examination.

Session-level means compress temporal dynamics into a single number, obscuring the within-session structure. The continuous cognitive load timelines (Figure~\ref{fig:cl_timeline}) reveal that PAWNI does not simply reduce cognitive load uniformly; it redistributes it. In Condition~A, the cognitive load profile follows a sawtooth pattern---high peaks during prompt writing and criteria evaluation, interspersed with low troughs during response waiting. This oscillatory pattern reflects the repeated cycle of effortful composition, passive waiting, and evaluative comparison that characterizes multi-turn LLM interaction.

In Condition~B, the profile is qualitatively different. The Architect Q\&A phase produces a moderate, sustained cognitive load---the user is actively thinking about their requirements, but the structured multiple-choice format reduces the formulation effort compared to free-form prompt writing. The subsequent pipeline processing and LLM execution phases produce lower cognitive load, as the user's role shifts from active composition to passive monitoring and verification. The phase-level analysis (Figure~\ref{fig:phase_cl_agg}) captures this distinction: the largest reductions occurred during passive waiting ($\downarrow$29\%) and composing content ($\downarrow$20\%), whilst the checking criteria phase---which requires similar evaluative effort regardless of how the prompt was generated---showed the smallest change ($\downarrow$7\%).

The 45\% reduction in NASA-TLX Overall Workload, despite the modest session-level EEG effect, suggests that subjective experience integrates not only the magnitude but also the \textit{quality} of cognitive effort. The effort invested during the Architect Q\&A phase is \textit{productive} effort---it results in a comprehensive prompt that reduces downstream evaluation work. By contrast, Condition~A's multi-turn cycle includes substantial \textit{wasteful} effort: correcting model misunderstandings, re-specifying forgotten requirements, and evaluating incomplete outputs. Participants appear to weigh this distinction when reporting subjective workload, even though the EEG captures both types of effort indiscriminately as frontal theta activity.

P3 presents an informative counter-pattern. This participant showed \textit{higher} $\theta/\alpha$ ratios with PAWNI ($M = 2.14$ vs.\ $1.55$), yet simultaneously reported lower NASA-TLX workload ($51.7 \rightarrow 27.5$). This dissociation suggests that the Architect's questions on quantum computing---a domain where P3 had limited prior knowledge---elicited genuine cognitive engagement (elevated theta) that was experienced as productive rather than burdensome. The participant's qualitative feedback confirmed this: the Architect's questions ``made me think about aspects of curriculum design I would not have considered.'' This aligns with the observation from Kosmyna et al.~\cite{kosmyna2025brain} that cognitive engagement during human-AI interaction is not inherently negative---what matters is whether the engagement supports learning and task comprehension or merely reflects effortful error correction.

% ------------------------------------------------------------
\subsection{The Single-Shot Advantage}
\label{subsec:disc_singleshot}

\begin{figure*}[!htbp]
\centering
\includegraphics[width=\textwidth]{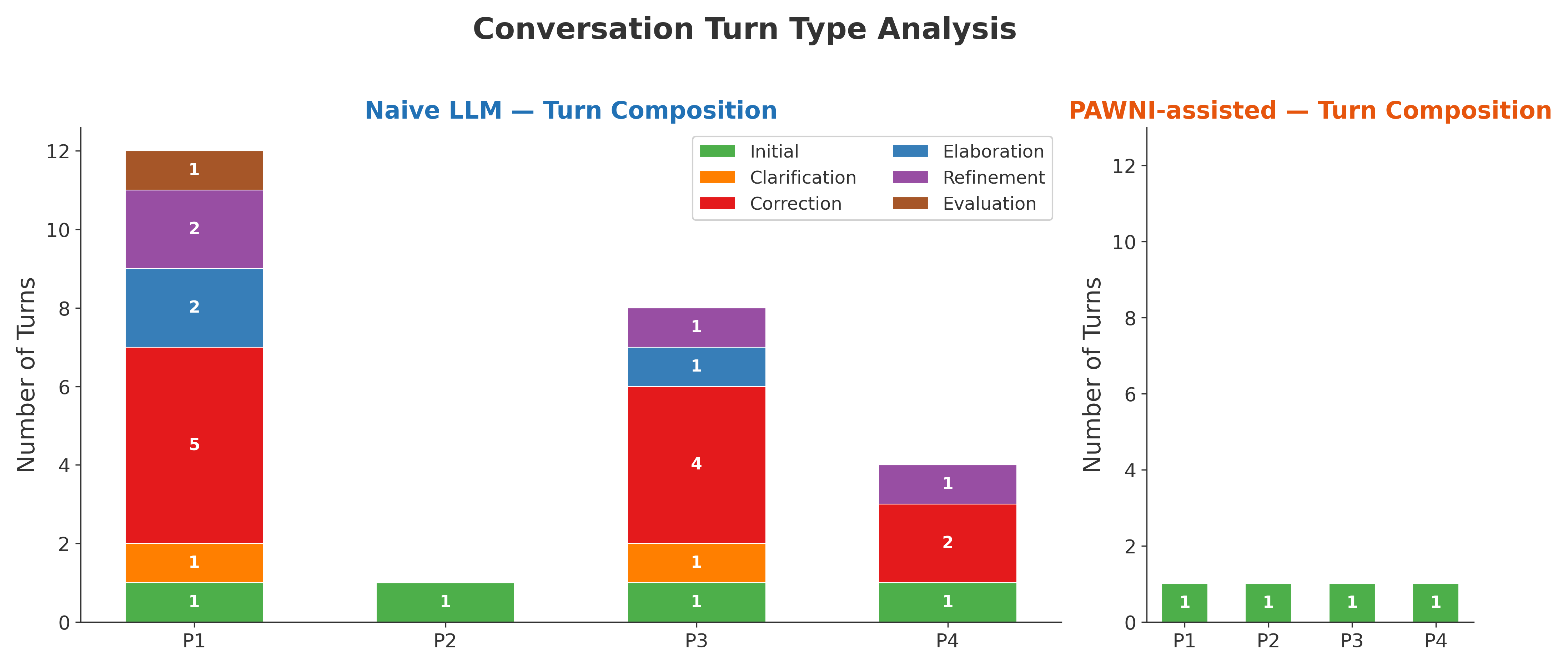}
\caption{Conversation turn type composition. Left panel: Condition~A (Naive LLM), showing stacked turn types per participant. Right panel: Condition~B (PAWNI-assisted), all participants reduced to a single initial turn. Correction turns (red) dominate the naive interaction.}
\Description{Two-panel stacked bar chart of conversation turn types by participant. The left panel (Naive LLM) shows multiple stacked segments (initial, clarification, correction, elaboration, refinement, evaluation) with correction segments prominent; the right panel (PAWNI-assisted) shows a single initial turn for each participant.}
\label{fig:app_conversation_flow}
\end{figure*}

PAWNI reduced all LLM interactions to a single turn, eliminating the iterative correction cycle entirely. Figure~\ref{fig:app_conversation_flow} decomposes each participant's conversational turns by type (Initial, Clarification, Correction, Elaboration, Refinement, Evaluation) for both conditions. This conversation flow analysis reveals that in Condition~A, correction turns (red) constituted the majority of interaction for P1 (5/12 turns), P3 (4/8 turns), and P4 (2/4 turns). These correction turns represent wasted interaction cycles where the user is repairing the model's misunderstandings rather than advancing toward the goal. In Condition~B (blue), all interactions consist of a single initial turn, as PAWNI's Architect agent front-loads the clarification process. In this sample, front-loaded intent clarification removed the need for correction turns entirely; whether it does so reliably is a question for a larger study. It is worth noting that the correction work does not vanish so much as change form: the clarification that P1 performed across five corrective turns in Condition~A was performed across three Architect rounds in Condition~B. The claim is not that less clarification occurred, but that it occurred before generation rather than after, and in a structured rather than an ad-hoc form.

Time-on-task did not show a consistent directional effect (1/4 participants were faster with PAWNI). This is expected and does not undermine the system's value. The PAWNI pipeline imposes an upfront time investment in the Architect Q\&A session (mean: 37.5 minutes), which is offset to varying degrees by reduced LLM interaction time (mean: 9 minutes in Condition~B vs.\ 44 minutes in Condition~A for the LLM-only component). For tasks where users would naturally iterate extensively (P2's travel planning), PAWNI saves time. For tasks where the pipeline Q\&A is itself lengthy (P1's technical specification), the net time increases. The critical point is that the \textit{quality} of time spent differs: PAWNI time is spent on productive requirement articulation, not on correcting model errors.

% ------------------------------------------------------------
\subsection{Limitations}
\label{subsec:limitations}

Several limitations constrain the interpretation of these findings and must be stated plainly.

\textit{Sample size and the status of the effect sizes.} With $N = 4$, statistical inference is not possible; all reported $p$-values exceed 0.05 by construction. More importantly, the effect sizes themselves must not be read as magnitude estimates. As set out in Section~\ref{subsec:effectsize}, a four-participant sample yields effect estimates that are upward-biased, estimator-dependent, and accompanied by intervals so wide that they are compatible with effects several times smaller. The values of $d = 9.69$ for structural completeness and $d = 4.24$ for output quality should therefore be read as ``all four participants moved in the same direction, and the gap was larger than the between-participant spread''---not as a claim that PAWNI produces an effect of that size. We report them because omitting them would be selective, not because we believe they will replicate at that magnitude.

\textit{What this study can and cannot support.} The study was designed as a formative evaluation with two purposes: to verify that the eight-agent pipeline behaves as specified on genuinely different complex tasks, and to validate a multi-method measurement protocol (EEG phase segmentation, NASA-TLX, criteria checklists, transcript coding) that had not previously been assembled for this interaction paradigm. Both purposes were served. Nothing in this study establishes that PAWNI improves outcomes for the general population of LLM users, and no claim to that effect is intended.

\textit{Order effects.} The fixed condition ordering (Condition~A always first) means that task familiarity from Day~1 carries into Day~2. Whilst this works against the PAWNI hypothesis, making the observed differences conservative, it precludes separating order effects from treatment effects. A future study should employ counterbalanced ordering.

\textit{Participant bias and demand characteristics.} No participant was a member of the PAWNI development team or had contributed to the system, but all four were recruited from the same institution and had been given a short practice run with the interface before the study. Familiarity of this kind, combined with the visible novelty of the PAWNI condition, creates a genuine risk of social-desirability bias and of demand characteristics: participants may have inferred which condition the researchers hoped would win. Section~\ref{subsec:bias} sets out the four procedural safeguards adopted and classifies each measure by its susceptibility to this bias. The residual risk is concentrated in the self-report measures. We therefore treat the SUS result (98.1 for PAWNI) as uninterpretable in isolation, and we do not rest any conclusion on it. The behavioural measures (turn counts, correction turns, prompt length), the blind-rated measure (structural completeness) and the physiological measure (EEG $\theta/\alpha$) are the ones on which the argument depends; notably, the EEG measure is also the one that showed the weakest and least consistent effect, which is what one would expect if the self-report measures were partly bias-driven. A study with participants who have never seen the system remains essential for external validity.

\textit{LLM confounding.} Each participant used a different commercial LLM. This was a deliberate choice---the claim under examination is that structured prompts help across models, not that they help with one particular model---and the within-subjects design holds the model constant within each participant, so no participant's comparison is confounded. It does, however, mean that the aggregate means in Table~\ref{tab:summary} are averages over four different systems and should not be read as characterising any one of them. It also means that the between-participant variance in Table~\ref{tab:summary} conflates task difficulty, participant, and model.

\textit{Task drift within one participant.} Inspection of the interaction logs revealed that P2's task specification was not identical across conditions. In Condition~A the participant described a trip for two travellers with no stated budget ceiling; in Condition~B, the Architect agent flagged the stated budget as infeasible for two travellers, and the participant revised the brief to a single traveller with a hard cap. The Architect's intervention is itself an instance of the behaviour the system is designed to produce---an infeasible premise surfaced before generation rather than after---but it does mean that P2's two outputs answer slightly different questions, and P2's output-quality ratings should be discounted accordingly. No comparable drift was found for P1, P3 or P4.

\textit{Whole-system, not component-level, evaluation.} PAWNI was evaluated as a single interactive pipeline. The user-facing outcome measures cannot be attributed to any individual agent, and this is a real limitation, discussed further in Section~\ref{subsec:attribution}.

\textit{EEG limitations.} Saline-based EEG sensors are susceptible to drying over sessions exceeding 60 minutes, potentially reducing signal quality in later portions of longer sessions. Movement artefacts from typing are partially mitigated by ICA but not entirely eliminated. The 32-channel density, whilst adequate for the cognitive load indices used, does not support source-level localisation. Because sessions differed in length between conditions, drying is partly confounded with condition for the longest sessions.

% ------------------------------------------------------------
\subsection{Implications for Conversational AI Design}
\label{subsec:implications}

The findings suggest three design principles for future conversational AI systems. First, \textit{question-asking should be a first-class interaction paradigm}. Current LLM interfaces are designed around the assumption that the user formulates the question and the model provides the answer. PAWNI inverts this: the system asks the questions, and the user provides the domain knowledge. This paradigm is more natural for complex tasks where the user knows \textit{what} they want but not \textit{how} to specify it comprehensively.

Second, \textit{front-loading intent clarification reduces total interaction cost}. The perceived overhead of a multi-step Q\&A process is outweighed by the elimination of multi-turn correction cycles. This suggests that commercial LLM interfaces would benefit from an optional ``deep prompt'' mode that guides users through structured requirements gathering before generating a response.

Third, \textit{domain knowledge should be dynamic, not static}. PAWNI's self-evolving pattern file system ensures that best practices are current, weighted by empirical confirmation, and tailored to each task type. Static prompt templates, by contrast, become outdated and lack task-specific granularity.

\subsection{Where Is Such a System Needed? Positioning Against Built-in Prompt Rewriting}
\label{subsec:positioning}

A reasonable objection to systems like PAWNI is that commercial LLM providers already perform prompt transformation internally: queries are expanded before retrieval, system prompts are auto-tuned, reasoning models restate the task before answering, and vendor tooling now offers one-click ``prompt improvement''. If the platform rewrites the prompt anyway, what is left for an external question engine to do?

The answer turns on a distinction between \emph{reformulating what the user said} and \emph{eliciting what the user did not say}. Automated prompt optimisation methods---APE~\cite{zhou2023ape}, OPRO~\cite{yang2024optimizers}, gradient-style textual optimisation~\cite{pryzant2023apo}, and compiled pipelines such as DSPy~\cite{khattab2024dspy}---search over the space of instruction phrasings to maximise a metric on a held-out set. They are extremely effective when a task distribution and a scoring function exist. They are inapplicable to a single user with a single novel task, because there is no held-out set and no metric: the criterion of success lives in the user's head. Query rewriting and expansion in retrieval systems face the same boundary; they enrich a query with terms the system can infer, not with constraints only the user knows.

The information PAWNI adds is, by construction, not recoverable by any amount of rewriting. In our four sessions the Architect elicited the target audience and accreditation context for a curriculum, a hard budget cap and a specific gastro-oesophageal condition for a travel plan, a state-management library and a BLE transmission contract for a mobile application, and a set of biomarker families and a regulatory submission target for a clinical protocol. None of these are latent in the original request. A rewriter that invented them would be hallucinating; a rewriter that omitted them produces exactly the generic output the participants rated 3.5--4.3 out of 7 in Condition~A.

This aligns the present work with a line of research that treats missing information as something to be requested rather than inferred. Clarifying-question generation has a long history in conversational information seeking~\cite{zamani2020clarifying}, and recent work trains models to ask before acting~\cite{andukuri2024stargate, wang2024learningtoask, zhang2024askbeforeplan, mazzaccara2024informative} or to ask well in high-stakes domains~\cite{li2025clinicalquestions}. PAWNI's contribution relative to this line is not the act of asking but the scaffolding around it: the questions are grounded in retrieved, weighted, task-specific domain knowledge, they are enumerated against an explicit 18-element completeness target, and the resulting prompt is evaluated by independent models before it is handed to the user. Built-in rewriting and PAWNI are therefore complements rather than competitors: rewriting improves the expression of a specification, whereas PAWNI improves the specification itself. A platform that did both would be strictly better than one that did either.

% ------------------------------------------------------------
\subsection{Who Benefits? Target Populations and a Training Function}
\label{subsec:beneficiaries}

A system that imposes a 27--55 minute structured interview is not for everyone, and it is worth being explicit about who it is for. Three characteristics of our task set point to the answer: the tasks were high-stakes relative to the user's effort budget, they had a deliverable whose adequacy the user could recognise but not fully specify in advance, and they sat in domains where the user had partial rather than expert knowledge. Where all three hold, the front-loaded interview is a good trade. Where any one fails---a familiar task, a low-stakes output, or a user who is already a domain expert with a mental checklist---the overhead is unlikely to be repaid.

This suggests several communities as natural early targets. First, \emph{domain experts without prompting expertise}: clinicians, lawyers, teachers and civil servants who know precisely what a good output looks like but have no vocabulary for output format, negative prompting or role assignment. Second, \emph{users writing in a second language}, for whom the cost of composing a 2{,}000-word specification in free text is far higher than the cost of answering multiple-choice questions. Third, \emph{people in regulated or audit-heavy settings}---protocol design, compliance drafting, procurement---where the cost of an incomplete specification is a rejected submission rather than a wasted afternoon. Fourth, \emph{novice and occasional LLM users}, who are documented to explore prompt designs opportunistically rather than systematically and to struggle to diagnose why an output failed~\cite{zamfirescu2023johnny}.

The fourth group points to a second use for the system that our data hint at but do not establish: PAWNI may function as a \emph{training} intervention rather than only a production tool. The Architect makes the structure of a good specification visible---it names the elements, asks about them in priority order, and shows the user which of their assumptions were never stated. P3, working on a curriculum for a topic he knew only partially, reported that the questions ``made me think about aspects of curriculum design I would not have considered,'' and this participant was also the one whose EEG $\theta/\alpha$ ratio \emph{rose} under PAWNI whilst his reported workload fell---a pattern consistent with productive engagement rather than reduced effort. Across all four sessions the participants answered a mean of 36 structured questions (range 30--46) over a mean of 3.75 rounds; that is a substantial amount of guided reflection on task requirements, and it is plausible that some of it transfers.

Whether it does transfer is an empirical question this study cannot answer, and a scaffolding tool that produces good prompts without teaching anything would be a legitimate but different contribution. The test is straightforward and we intend to run it: measure the structural completeness of a user's unaided prompts before PAWNI exposure, after several sessions with it, and after a wash-out period. If completeness rises and stays risen, the system is a tutor; if it reverts, it is a prosthesis. Both are useful; they imply very different deployment strategies.

% ------------------------------------------------------------
\subsection{Component Attribution and the Limits of Whole-System Evaluation}
\label{subsec:attribution}

An eight-agent pipeline raises two related concerns. The first is attribution: because the agents run in sequence and each consumes the previous one's output, no user-facing outcome measure can be traced to a particular agent. The second is validation: in principle each agent should be shown to do its job before the composite is evaluated. We accept both concerns and address them partially rather than pretending otherwise.

The pipeline logs reported in Section~\ref{subsec:component} provide component-level evidence for three of the eight agents. The Scout is observable in the number and category structure of the knowledge points it retrieved, which was stable across four unrelated domains (mean 56.8 task-level and 67.0 prompt-specific points, always across the same five categories). The Architect is observable in the number of clarification rounds and questions required to reach element coverage, which varied with task ambiguity (three rounds for the two most fully-specified initial requests, five for the least). The Judge is observable in the scores it assigned: crucially, it scored the system's \emph{own} first-stage output at a mean of 3.97/10, which is evidence that the multi-model evaluator is not simply endorsing whatever the pipeline produces. The Forge$\rightarrow$Scribe transition is observable as the difference between the two Judge passes.

What this does not provide is counterfactual evidence. We do not know what the final prompt would have looked like with the Scout disabled, with a single judge instead of four, or with the Architect replaced by a fixed 18-question form. Those are ablation questions, they require a dedicated study with many tasks and no human participants, and they are the appropriate subject of a separate paper. We note that this limitation is structural to the evaluation of composed LLM systems generally rather than specific to PAWNI: any pipeline whose stages are individually stochastic and jointly optimised for a subjective outcome resists clean decomposition. That is an argument for doing the ablation work, not for treating the composite result as if the decomposition had already been done. Until it is done, PAWNI's evaluated unit is the pipeline, and claims should be stated at that level.

% ============================================================
\section{Conclusion and Future Work}
\label{sec:conclusion}

This paper presented PAWNI, an eight-agent conversational AI system that transforms unstructured user queries into comprehensive, structured prompts for complex task resolution. The system is grounded in the thesis that asking the right questions leads to better answers. Through a three-tier prompt element framework (18 elements across Essential, Enhancement, and Elevation categories), a self-evolving pattern file system, and a reverse-reasoning evaluation loop with multi-model judges, PAWNI systematically addresses the structural deficiencies that characterize user-authored prompts.

An exploratory within-subjects evaluation ($N = 4$) across four domain-diverse tasks, using 32-channel EEG, NASA-TLX workload assessment, and behavioural metrics, produced the following observations. In this sample: (i) PAWNI-generated prompts produced LLM outputs rated approximately 2.5 points higher on a 7-point quality scale, with all four participants improving; (ii) all four participants reported lower subjective workload (NASA-TLX: $39.6 \rightarrow 21.7$), whilst the EEG-derived $\theta/\alpha$ ratio decreased in only three of four participants at the session level, with phase-level and temporal analyses revealing clearer separation between conditions; (iii) every participant reached a satisfactory output in a single LLM turn, against one to twelve turns unaided, and no correction turns were required; and (iv) prompt structural completeness rose from a mean of 42\% to 91\% of assessed elements, and prompt specificity from 0.29 to 0.79. Pipeline logs additionally show that the internal quality score of the prompt rose from a mean of 3.97/10 for the Forge draft to 9.27/10 for the Scribe output, with the evaluation loop converging in a single iteration in all four sessions.

We stress the limits of these observations. With four participants the study cannot support inference, and the accompanying effect sizes---though large---are unstable estimates whose magnitudes should not be carried forward. What the study does establish is that the measurement protocol works, that the system behaves as designed on four genuinely different complex tasks, and that the direction of effect was consistent across participants on every output-quality and prompt-quality variable. These are the preconditions for, not a substitute for, a confirmatory study.

Future work will address the limitations identified in this study. A confirmatory study is planned with counterbalanced condition ordering, a single controlled LLM across participants, and blinded outcome rating. The effect sizes reported here are deliberately not used to justify a sample size: because small-sample effect estimates are upward-biased, doing so would under-power the follow-up. We will instead power the confirmatory study on the smallest effect we would consider practically meaningful---a 15 percentage-point gain in structural completeness---which, at $\alpha = 0.05$ and 80\% power for a paired comparison, implies approximately 40 participants. A dedicated agent-ablation study, in which the Scout, Architect and Judge are removed or replaced in turn, is required before any causal claim can be attached to an individual agent. Longitudinal deployment studies will examine whether PAWNI's scaffolding effect persists over time---whether users internalize the questioning framework and produce better prompts even without PAWNI. Integration of PAWNI's Architect agent as an optional mode within existing LLM chat interfaces represents a natural pathway for deployment. Finally, the pattern file convergence dynamics and their relationship to task-type maturity warrant dedicated investigation.

%% ---------------------------------------------------------------------------
%% Acknowledgements and AI-assistance disclosure
%%
%% Uncomment before posting. arXiv has no ACM-style policy requirement here,
%% but declaring LLM assistance is good practice and costs nothing.
%% ---------------------------------------------------------------------------
% \section*{Acknowledgements}
%
% \subsection*{Disclosure of AI-Assisted Writing}
%
% LLM-based tools were used to support language editing, including grammar and
% clarity improvements, during the preparation of this manuscript. The authors
% remain solely responsible for the accuracy, originality, and integrity of the
% work. No AI system is listed as an author.

%% ---------------------------------------------------------------------------
%% Bibliography
%% ---------------------------------------------------------------------------
\bibliographystyle{ACM-Reference-Format}
\bibliography{1_PAWNI_bibliography_bibtex}

\clearpage

%% ============================================================================
%% APPENDIX
%% ============================================================================
\appendix
\section{Supplementary Figures}
\label{sec:appendix}

This appendix presents additional figures that provide per-participant detail
and complementary perspectives to the group-level analyses reported in the main
text.

\subsection{Per-Participant EEG Topographic Maps}

\begin{figure*}[!htbp]
\centering
\includegraphics[width=0.6\textwidth]{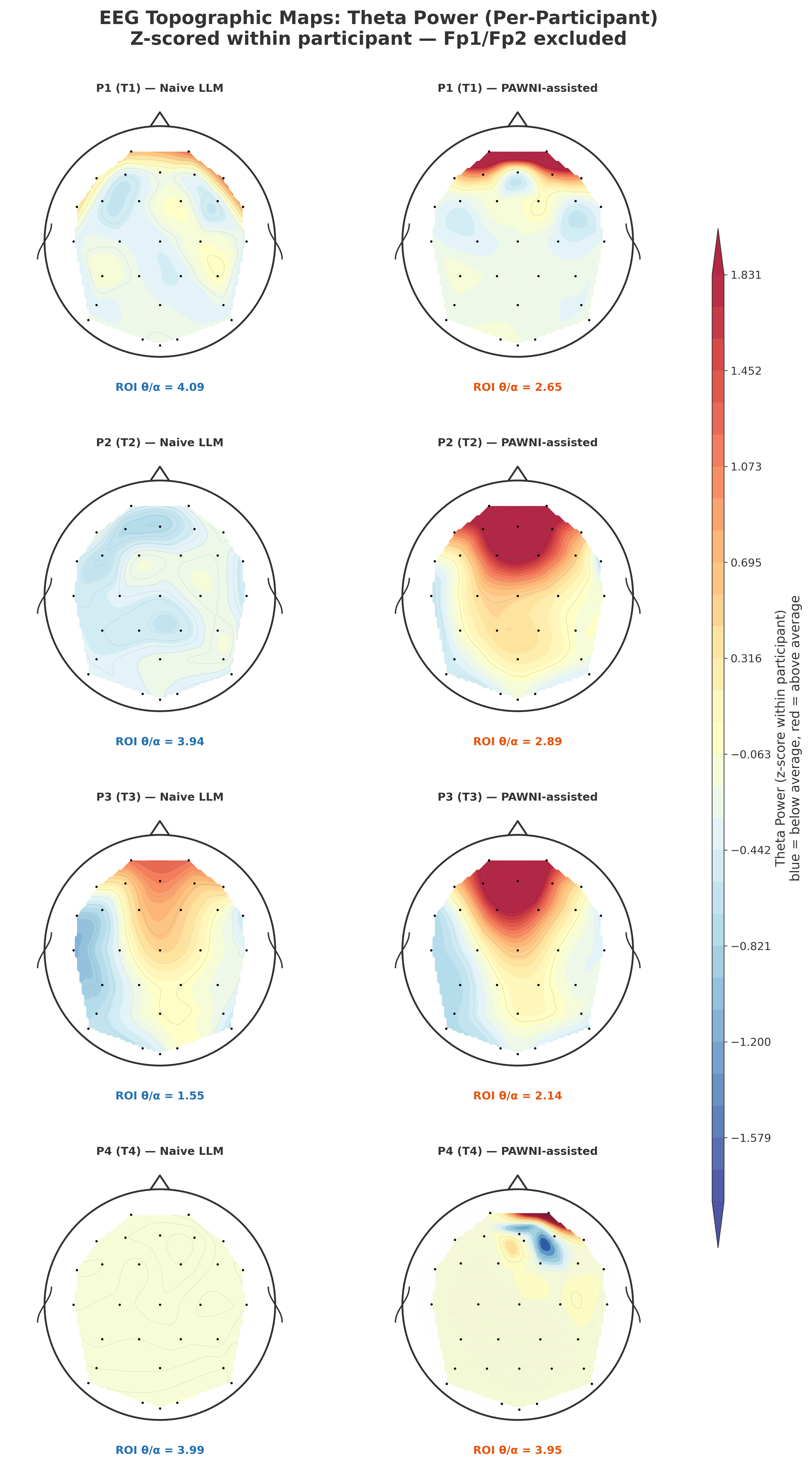}
\caption{Per-participant EEG topographic maps of theta power (z-scored within
participant). Left column: Naive LLM. Right column: PAWNI-assisted. Fp1/Fp2
excluded. ROI $\theta/\alpha$ ratios are shown below each map.}
\Description{Grid of scalp topography heatmaps (one row per participant)
showing theta-band power for two conditions: Naive LLM on the left and
PAWNI-assisted on the right. Each map is z-scored within participant, excludes
Fp1/Fp2 channels, and reports an ROI $\theta/\alpha$ ratio beneath the map.}
\label{fig:app_topomaps}
\end{figure*}

Figure~\ref{fig:app_topomaps} presents the per-participant theta power
topographic maps for both conditions. These individual maps reveal the
inter-participant variability underlying the group-average topomaps shown in
Figure~\ref{fig:topomaps_agg}. P1 and P2 show prominent frontal theta in the
PAWNI condition, consistent with focused task-directed processing. P3 shows
elevated theta across broader cortical regions in the PAWNI condition,
reflecting the higher engagement observed in the continuous timeline data. P4
shows relatively stable distributions across conditions.

\subsection{Criteria Coverage Trajectory}

\begin{figure*}[!htbp]
\centering
\includegraphics[width=\textwidth]{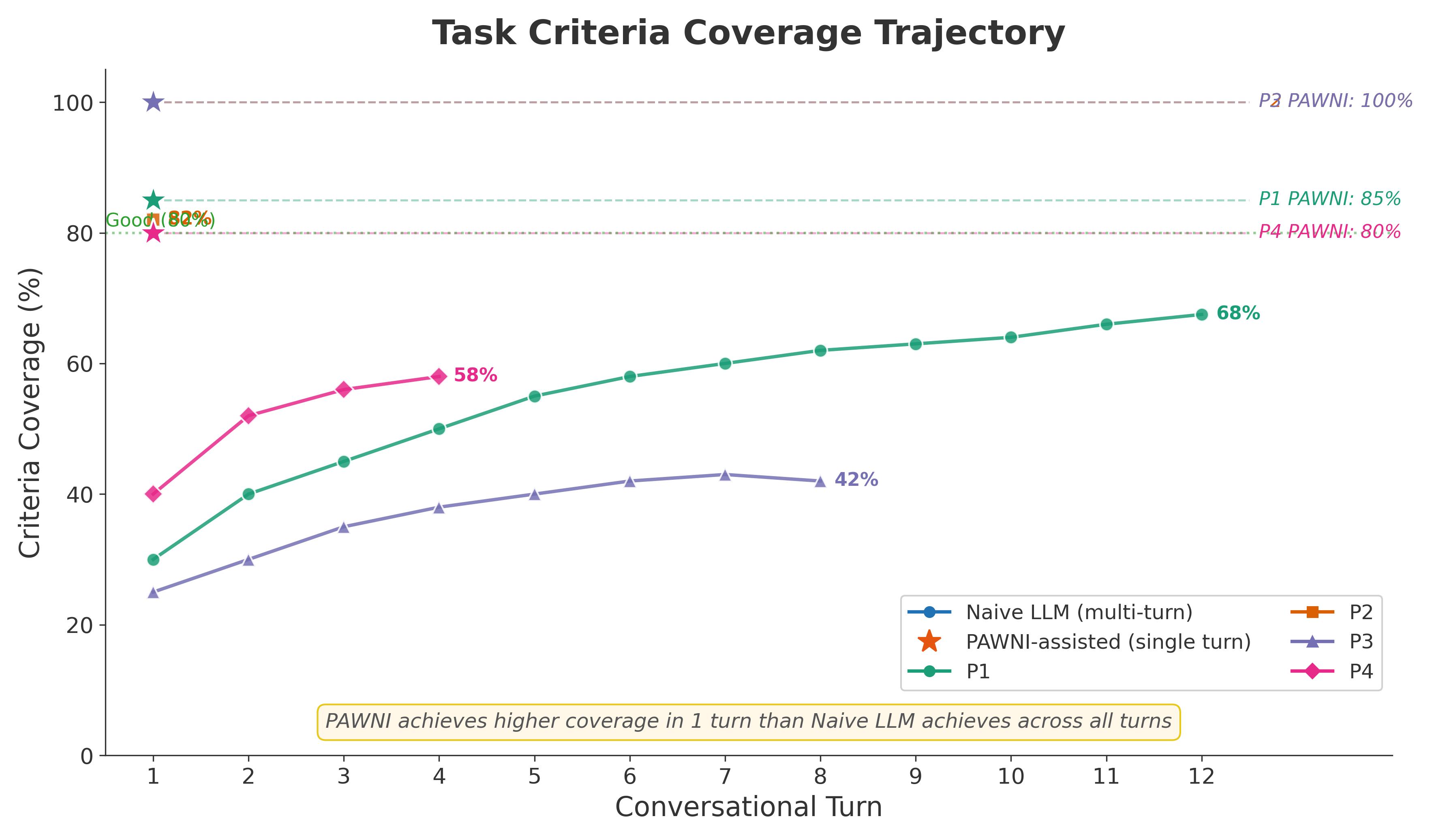}
\caption{Task criteria coverage trajectory. Lines show cumulative coverage
(\%) after each conversational turn in Condition~A. Stars indicate PAWNI
single-turn coverage. PAWNI achieves higher coverage in one turn than naive
interaction achieves across all turns.}
\Description{Line plot of cumulative task-criteria coverage percentage versus
conversational turn number for Condition~A (Naive LLM), with separate
trajectories for participants. Star markers indicate the single-turn coverage
achieved with PAWNI, which is higher than the final naive coverage for each
participant.}
\label{fig:app_criteria_trajectory}
\end{figure*}

Figure~\ref{fig:app_criteria_trajectory} shows how task criteria coverage
accumulates over conversational turns in Condition~A, compared against the
single-turn PAWNI coverage. For P1 and P4, the naive multi-turn trajectory
reaches a plateau well below the PAWNI single-turn coverage level, indicating
that additional turns yield diminishing returns. For P3, the naive trajectory
plateaus at 42\% after 8 turns, whilst PAWNI achieves 100\% in a single turn.
This demonstrates that the multi-turn correction cycle is not only slower but
also fundamentally less effective at achieving comprehensive criteria coverage.

\subsection{Summary Heatmap}

\begin{figure*}[!htbp]
\centering
\includegraphics[width=\textwidth]{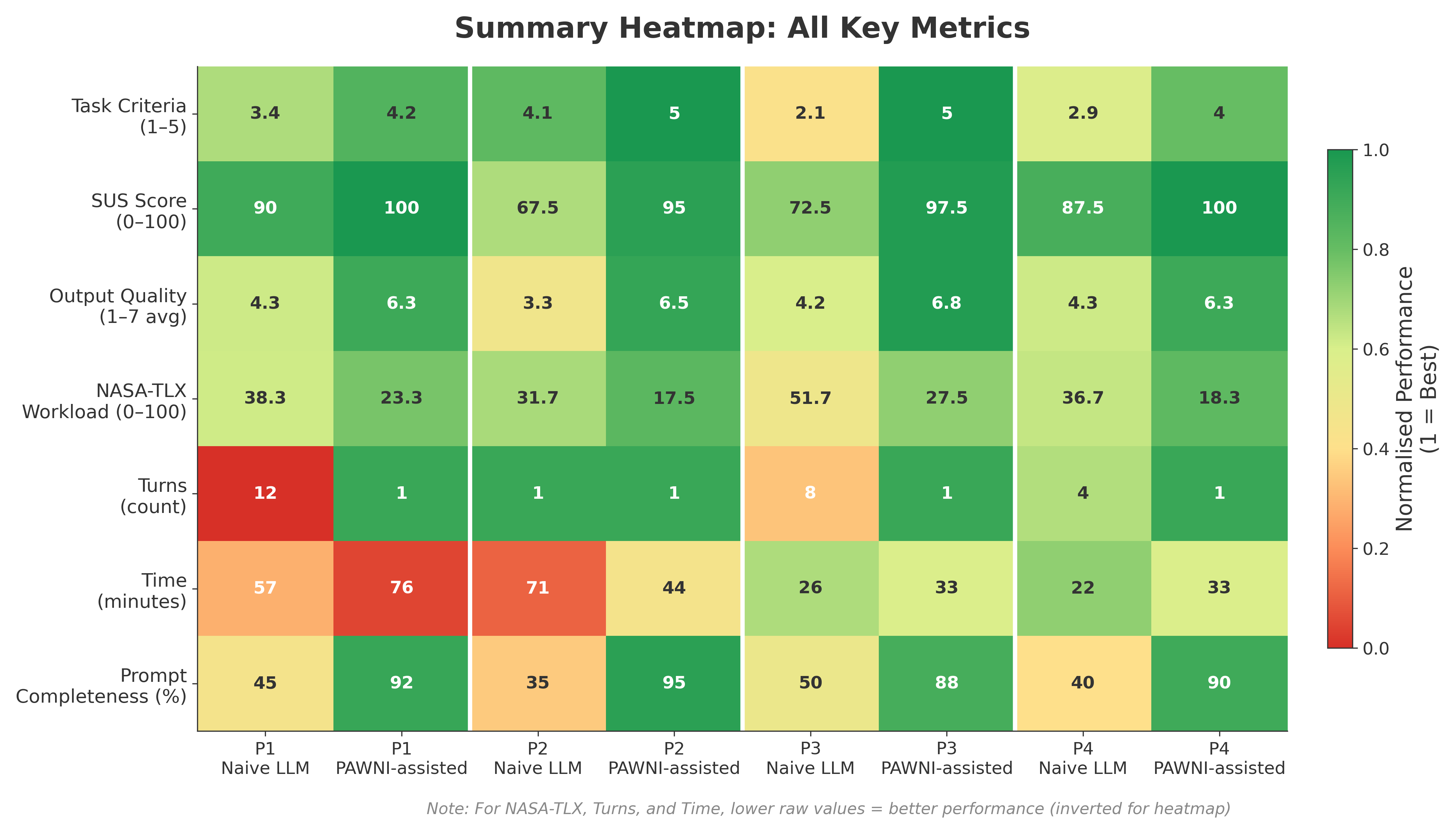}
\caption{Summary heatmap of all key metrics across participants and
conditions. Colour intensity reflects normalised performance (1 = best). For
NASA-TLX, Turns, and Time, lower raw values correspond to better performance
(inverted for display).}
\Description{Heatmap table summarising multiple normalised metrics for each
participant under Naive LLM and PAWNI-assisted conditions. Cell colour encodes
performance on a 0--1 scale (higher is better), with metrics where lower is
better inverted so that better performance consistently appears greener.}
\label{fig:app_heatmap}
\end{figure*}

Figure~\ref{fig:app_heatmap} consolidates all key metrics into a single
normalised heatmap, providing an at-a-glance comparison across all participants
and conditions. For metrics where lower is better (NASA-TLX, Turns, Time),
values are inverted for the heatmap so that green consistently indicates better
performance. The PAWNI columns are predominantly green across all participants
and metrics, with the notable exception of Time for P1 and P2.

\subsection{Master Comparison Dot Plot}

\begin{figure*}[!htbp]
\centering
\includegraphics[width=\textwidth]{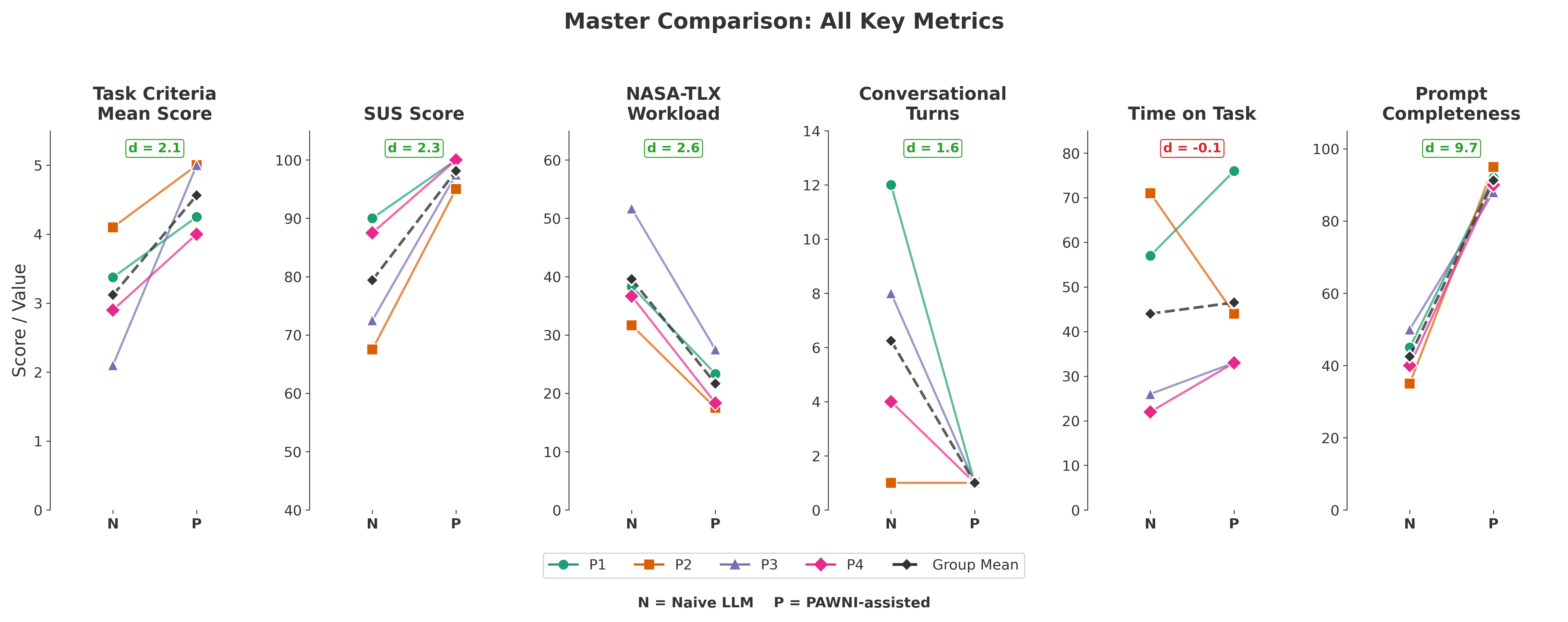}
\caption{Master comparison dot plots for six key metrics. Each line connects
one participant's Naive LLM (N) and PAWNI-assisted (P) scores. Dashed black
line shows the group mean. Green-boxed Cohen's $d$ values indicate large
effects; red-boxed values indicate negligible effects.}
\Description{Set of paired dot plots for six metrics, where each participant is
represented by two points (Naive LLM and PAWNI-assisted) connected by a line. A
dashed line indicates the group mean, and annotated Cohen's $d$ boxes highlight
which metrics show large versus negligible effects.}
\label{fig:app_master_dotplot}
\end{figure*}

Figure~\ref{fig:app_master_dotplot} presents paired dot plots for six key
metrics, showing individual participant trajectories from Naive LLM to
PAWNI-assisted conditions. This figure complements the summary table
(Table~\ref{tab:summary}) by showing per-participant variation and the
consistency of directional effects.

\subsection{Band Power Spectra}

\begin{figure*}[!htbp]
\centering
\includegraphics[width=\textwidth]{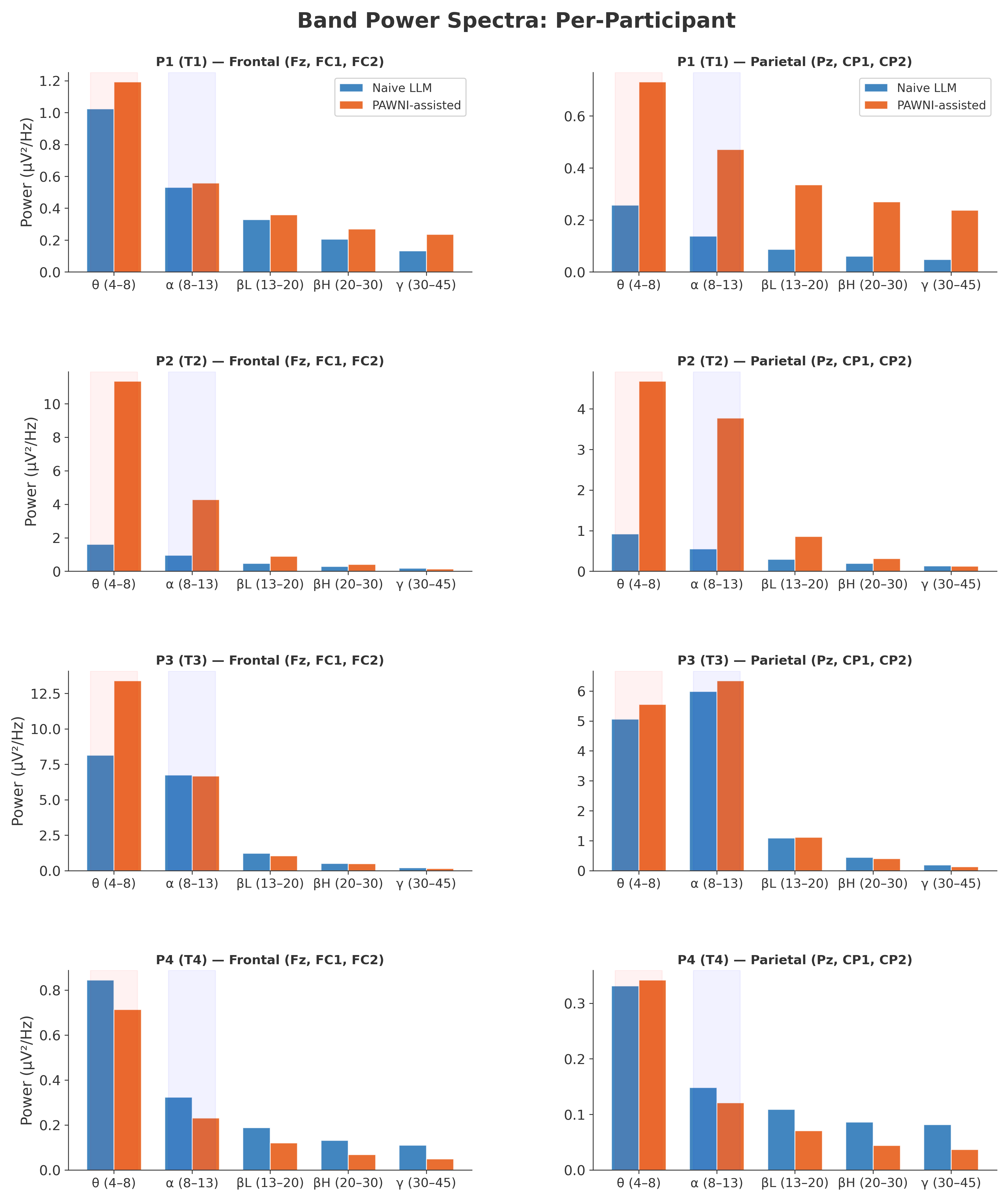}
\caption{Per-participant band power spectra at frontal (Fz, FC1, FC2) and
parietal (Pz, CP1, CP2) regions. Five frequency bands shown: $\theta$ (4--8
Hz), $\alpha$ (8--13 Hz), $\beta_L$ (13--20 Hz), $\beta_H$ (20--30 Hz),
$\gamma$ (30--45 Hz).}
\Description{Multi-panel frequency spectra showing EEG band power distributions
for each participant at frontal and parietal regions of interest. The figure
marks the standard frequency bands (theta, alpha, low beta, high beta, gamma)
across the 4--45 Hz range.}
\label{fig:app_band_power}
\end{figure*}

\begin{figure*}[!htbp]
\centering
\includegraphics[width=\textwidth]{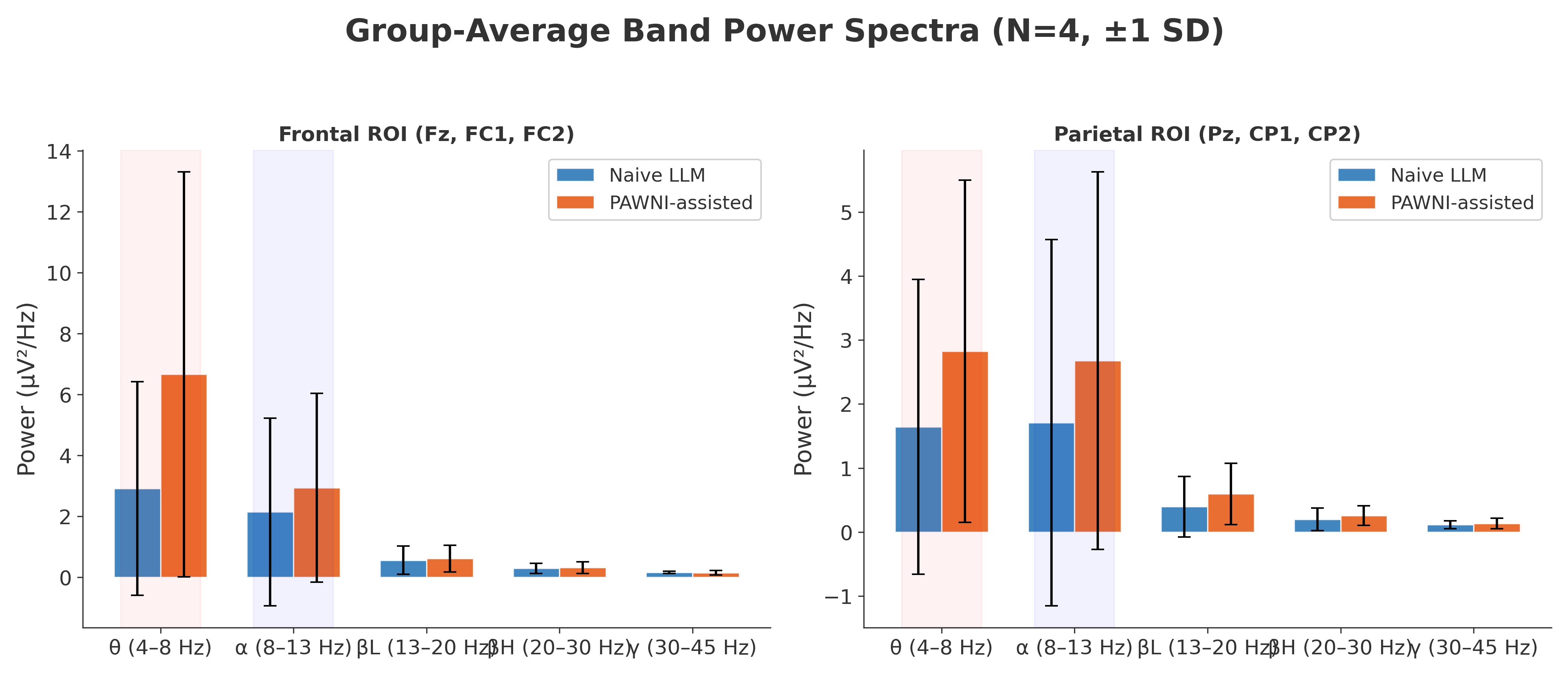}
\caption{Group-average band power spectra ($N = 4$, $\pm 1$ SD) at frontal and
parietal ROIs. Large error bars reflect inter-participant variability in
absolute power.}
\Description{Line plot of group-average EEG power spectra at frontal and
parietal regions of interest, with shaded regions or error bars indicating plus
or minus one standard deviation across four participants.}
\label{fig:app_band_power_agg}
\end{figure*}

Figures~\ref{fig:app_band_power} and~\ref{fig:app_band_power_agg} present the
frequency band power distributions at frontal and parietal regions of interest.
Per-participant spectra (Figure~\ref{fig:app_band_power}) reveal substantial
inter-participant variability in absolute power, which is expected given
individual differences in skull thickness, electrode impedance, and cortical
folding. The group-average spectra (Figure~\ref{fig:app_band_power_agg}) show
that frontal theta power was marginally higher in the PAWNI condition, whilst
parietal alpha power was comparable across conditions. This pattern is
consistent with the hypothesis that PAWNI elicits more focused frontal
engagement.

\subsection{Group-Average Difference Topomap}

\begin{figure*}[!htbp]
\centering
\includegraphics[width=0.7\textwidth]{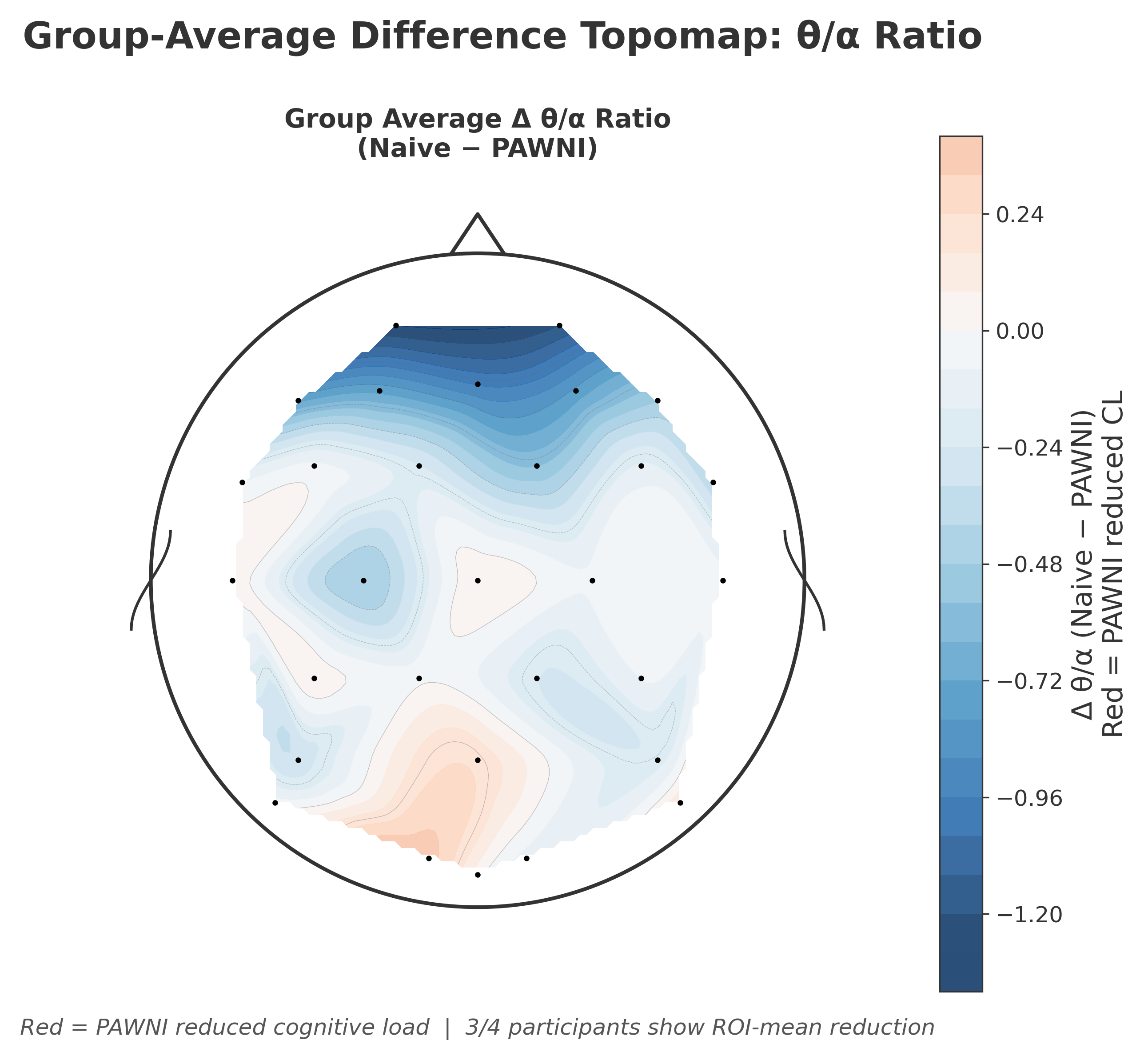}
\caption{Group-average difference topographic map ($\Delta\,\theta/\alpha$,
Naive $-$ PAWNI). Blue indicates PAWNI reduced cognitive load; red indicates
increase. The predominant frontal-central reduction is consistent across 3/4
participants.}
\Description{Single scalp topography difference heatmap showing the
group-average change in $\theta/\alpha$ ratio (Naive minus PAWNI). Most of the
frontal and central scalp regions are blue, indicating reduced cognitive load
under PAWNI on average.}
\label{fig:app_diff_topomap_agg}
\end{figure*}

Figure~\ref{fig:app_diff_topomap_agg} shows the group-average difference
topographic map for $\theta/\alpha$ ratio. The predominant blue colouration
across frontal and central regions indicates that, on average, PAWNI reduced
the $\theta/\alpha$ ratio across most of the scalp. A small parieto-occipital
region shows mild red colouration, suggesting marginally higher posterior
activity in the PAWNI condition, which may reflect greater visual engagement
with the structured Q\&A interface.

\subsection{Per-Participant Overlaid Cognitive Load Timelines}

\begin{figure*}[!htbp]
\centering
\includegraphics[width=\textwidth]{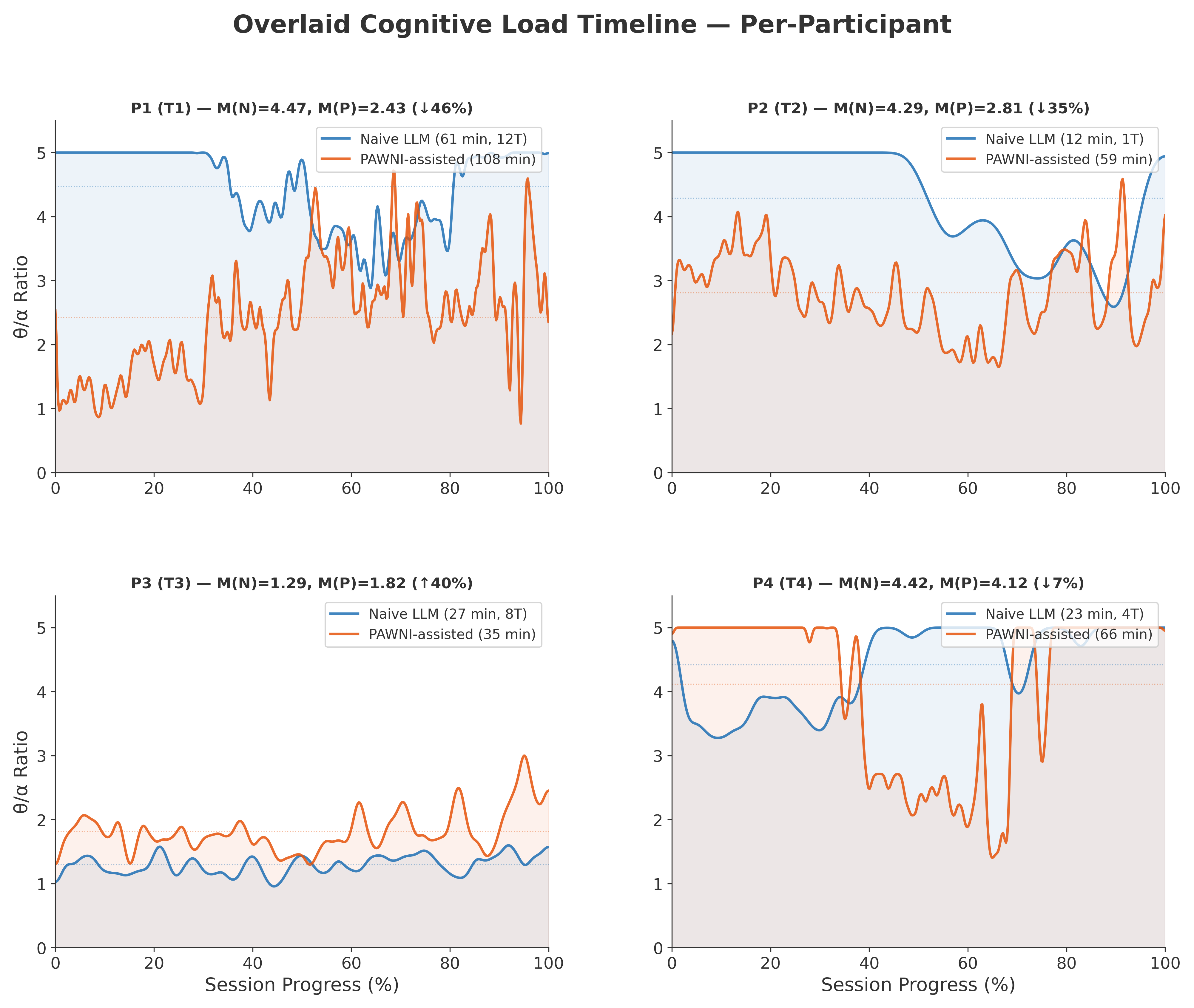}
\caption{Per-participant overlaid cognitive load timelines ($\theta/\alpha$
ratio) on normalised session progress (0--100\%). Blue: Naive LLM. Orange:
PAWNI-assisted. Shaded regions indicate the spread of each condition's values.
M(N) and M(P) denote session means.}
\Description{Overlaid cognitive load timelines for each participant, showing
Naive LLM and PAWNI-assisted conditions.}
\label{fig:app_overlaid_cl}
\end{figure*}

Figure~\ref{fig:app_overlaid_cl} overlays both conditions on a normalised
session-progress axis for each participant individually. This complements the
group-average timeline in Figure~\ref{fig:cl_overlay_agg} by revealing the
individual patterns. P1 and P2 show clear separation between conditions
throughout the session, with the PAWNI condition consistently lower. P3 shows
the conditions interleaving, consistent with the higher PAWNI engagement noted
previously. P4 shows early separation that converges towards the end of the
session.

\subsection{Per-Participant Phase-Averaged Cognitive Load}

\begin{figure*}[!htbp]
\centering
\includegraphics[width=\textwidth]{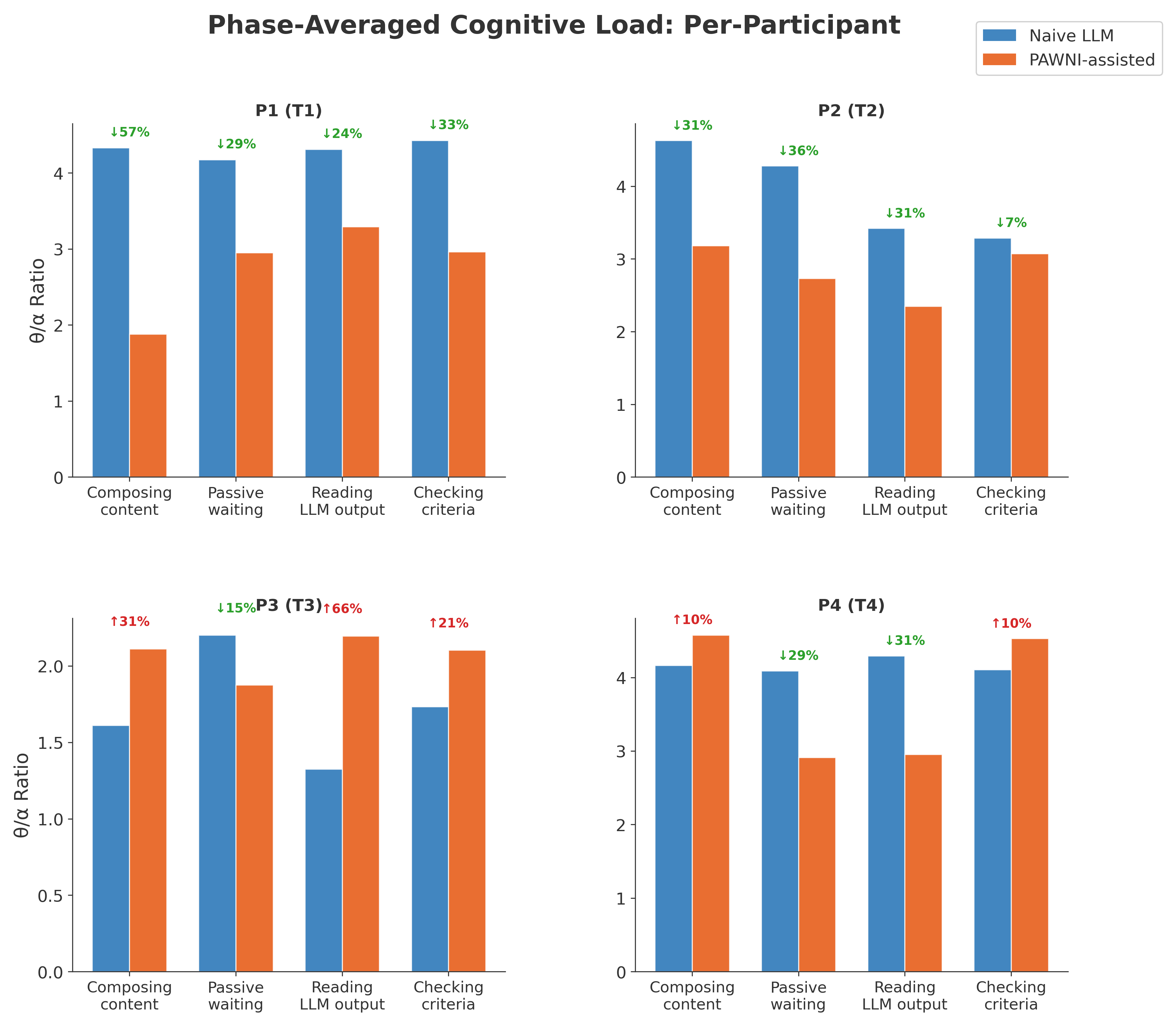}
\caption{Per-participant phase-averaged cognitive load ($\theta/\alpha$ ratio).
Four functionally comparable phases are compared between conditions. Percentage
labels indicate change (green: reduction, red: increase).}
\Description{Grouped bar charts by participant showing phase-averaged EEG
cognitive load ($\theta/\alpha$ ratio) across four interaction phases for Naive
LLM and PAWNI-assisted conditions. Percent change annotations are coloured
green for reductions and red for increases.}
\label{fig:app_phase_cl}
\end{figure*}

Figure~\ref{fig:app_phase_cl} presents the phase-averaged $\theta/\alpha$ ratio
for each participant across the four comparable interaction phases. P1 shows
reductions across all four phases (24--57\%). P2 shows reductions in all phases
(7--36\%). P3 shows increases in three of four phases, consistent with the
elevated engagement pattern. P4 shows a mixed pattern with reductions in two
phases and increases in two. The largest and most consistent reductions across
participants occur during the passive waiting and composing content phases.

\end{document}